\pdfoutput=1
\documentclass[twocolumn]{aastex62}
\listofchanges
\usepackage{blindtext}
\usepackage{textcomp}
\usepackage{afterpage}
\usepackage{natbib}
\usepackage{graphicx}
\usepackage{verbatim}
\usepackage{wrapfig}
\usepackage{mathrsfs}
\usepackage{nccmath}
\usepackage{amsmath}
\newcommand{\ssl}{\affiliation{Space Sciences Laboratory,University of California, Berkeley, CA 94720, USA}}
\newcommand{\pucb}{\affiliation{Physics Department,University of California, Berkeley, CA 94720, USA}}
\newcommand{\imp}{\affiliation{The Blackett Laboratory, Imperial College London, London, SW7 2AZ, UK}}
\newcommand{\qm}{\affiliation{School of Physics and Astronomy, Queen Mary University of London, London E1 4NS, UK}}
\newcommand{\sao}{\affiliation{Smithsonian Astrophysical Observatory, Cambridge, MA 02138, USA}}
\newcommand{\umich}{\affiliation{Climate and Space Sciences and Engineering, University of Michigan, Ann Arbor, MI 48109, USA}}

\newcommand{\cnrs}{\affiliation{LPC2E, CNRS and University of Orléans, Orléans, France}}
\newcommand{\gsfc}{\affiliation{Code 695, NASA, Goddard Space Flight Center, Greenbelt, MD 20771, USA}}
\newcommand{\uofa}{\affiliation{Lunar and Planetary Laboratory, University of Arizona, Tucson, AZ 85721, USA}}
\newcommand{\cu}{\affiliation{Astrophysical and Planetary Sciences Department, University of Colorado, Boulder, CO, USA}}
\newcommand{\lasp}{\affiliation{Laboratory for Atmospheric and Space Physics, University of Colorado, Boulder, CO 80303, USA}}
\newcommand{\umin}{\affiliation{School of Physics and Astronomy, University of Minnesota, Minneapolis, MN 55455, USA}}
\newcommand{\swri}{\affiliation{Space Science and Engineering, Southwest Research Institute, San Antonio, TX 78238, USA}}

\newcommand{\V}[1]{\mathbf{#1}}  %Bold for vectors 
\newcommand\Alfven{Alfv\'en } %Proper names
\newcommand\Alfvenic{Alfv\'enic }
\newcommand{\figref}[1]{Fig.~\ref{#1}}   
\newcommand{\secref}[1]{\S\ref{#1}}   
\newcommand{\T}[1]{{\tt #1}}
\graphicspath{{./}{figs/}}

\shorttitle{PSP beams and waves}
\shortauthors{Verniero et al.}
\begin{document}
\title{Parker Solar Probe observations of proton beams simultaneous with ion-scale waves}

\correspondingauthor{J. L. Verniero}
\email{jlverniero@berkeley.edu}

\author[0000-0003-1138-652X]{J. L. Verniero}\ssl
\author[0000-0001-5030-6030]{D. E. Larson}\ssl
\author[0000-0002-0396-0547]{R. Livi}\ssl
\author[0000-0003-0519-6498]{A. Rahmati}\ssl
\author[0000-0001-6077-4145]{M. D. McManus}\ssl\pucb
\author[0000-0003-4832-7638]{P. Sharma Pyakurel}\ssl
\author[0000-0001-6038-1923]{K. G. Klein}\uofa
\author[0000-0002-4625-3332]{T. A. Bowen}\ssl
\author[0000-0002-0675-7907]{J. W. Bonnell}\ssl
\author[0000-0001-6673-3432]{B. L. Alterman}\umich\swri
\author[0000-0002-7287-5098]{P. L. Whittlesey}\ssl
\author[0000-0003-1191-1558]{David M. Malaspina}\cu\lasp
\author[0000-0002-1989-3596]{S. D. Bale}\ssl\pucb\imp\qm
\author[0000-0002-7077-930X]{J. C. Kasper}\umich\sao
\author[0000-0002-3520-4041]{A. W. Case}\sao
\author[0000-0003-0420-3633]{K. Goetz}\umin
\author[0000-0002-6938-0166]{P. R. Harvey}\ssl
\author[0000-0001-6095-2490]{K. E. Korreck}\sao
\author[0000-0003-3112-4201]{R. J. MacDowall}\gsfc
\author[0000-0002-1573-7457]{M. Pulupa}\ssl
\author[0000-0002-7728-0085]{M. L. Stevens}\sao
\author[0000-0002-4401-0943]{T. Dudok de Wit}\cnrs

\begin{abstract}
Parker Solar Probe (PSP), NASA's latest and closest mission to the Sun, is on a journey to investigate fundamental enigmas of the inner heliosphere.
This paper reports initial observations made by the Solar Probe Analyzer for Ions (SPAN-I), one of the instruments in the Solar Wind Electrons Alphas and Protons (SWEAP) instrument suite.
We address the presence of secondary proton beams in concert with ion-scale waves observed by FIELDS, the electromagnetic fields instrument suite.
We show two events from PSP's 2\textsuperscript{nd} orbit that demonstrate signatures consistent with wave-particle interactions.
We showcase 3D velocity distribution functions (VDFs) measured by SPAN-I during times of strong wave power at ion-scales.
From an initial instability analysis, we infer that the VDFs departed far enough away from local thermodynamic equilibrium (LTE) to provide sufficient free energy to locally generate waves.
These events exemplify the types of instabilities that may be present and, as such, may guide future data analysis characterizing and distinguishing between different wave-particle interactions.
\end{abstract}

%%%%%%%%%%%%%%%%%%%%%%%%%%%%%%%%%%%%%%%%%%%%%%%%%%%%%%%
\section{Introduction} \label{sec:intro}

Parker Solar Probe (PSP) is NASA's latest mission to the Sun.
It is on a journey to approach the Sun at a distance never before achieved so that we can answer many outstanding questions in heliophysics.
These include:
\begin{enumerate}
\item Why is the solar corona hotter by several orders of magnitude than expected? 
\item What is the nature of the fundamental physical processes that permeate in the young solar wind? 
\item Based on (2), which plasma heating mechanisms dominate to answer (1)? 
\end{enumerate}
\citet{Fox:2016} provides more information about the mission and other science goals. 

The solar wind is a weakly collisional plasma \citep{Alterman:2018,Maruca:2011,Verscharen:2019,Kasper:2008}. As such, kinetic-scale wave-particle interactions play a significant role in solar wind thermalization \citep{Kasper:2002,Hellinger:2006,Klein:2018}, thereby returning the plasma velocity distributions (VDF) to local thermodynamic equilibrium (LTE). Jointly analyzing particle VDFs and electromagnetic fields at small scales comparable to the ion gyroscale can reveal and characterize these wave-particle phenomena.

The most stable state of a plasma is a Maxwellian distribution and sufficient deviation from LTE provides free energy sources capable of wave generation. A secondary peak that is markedly distinct from the bulk or core of the proton distribution is one such example of a LTE departure \citep{Feldman:1973,Feldman:1974}. Commonly referred to as proton beams \citep{Alterman:2018}, these non-Maxwellian features are symmetry breaking properties that can potentially cause the plasma to become unstable and subsequently undergo an energy transfer process that returns it to an equilibrium state.

In this paper, we report PSP's first observations of proton beams simultaneous with ion-scale waves, specifically during PSP's second orbit around the sun (hereafter called Encounter 2) collected by the Solar Probe Analyzer for Ions (SPAN-I). This unique instrument is a key component on the particle instrument suite, Solar Wind Electrons Alphas and Protons (SWEAP). We show 3D VDFs at SPAN-I's fastest measurement cadence and PSP's closest proximity to the Sun. Using linear Vlasov theory, we also provide initial kinetic instability results based on these PSP observations. These measurements illustrate SWEAP's ability to resolve kinetic processes in a manner that may answer presently unresolved questions about heating and acceleration mechanisms in the inner heliosphere \citep{Fox:2016,Kasper:2016,Bale:2016}.

\secref{sec:bac} reviews (1) results characterizing the instabilities associated with non-Maxwellian VDFs and (2) recent observations of ion-scale waves in the solar wind.
Section \secref{sec:meth} describes the instruments used for data collection and the methodology we applied to analyze the PSP data.
In \secref{sec:obs}, we present two example events that demonstrate a striking correlation between proton beams and ion-scale wave storms. An unexplained nonlinear phenomenon that occurred during one of these events is presented in \secref{sec:static}.
Section \secref{sec:stab} presents an initial instability analysis \citep{Verscharen:2018,Klein:2017c} of these events.
Finally, Section \secref{sec:con} reports our discussion and conclusions. 

\subsection{Background}\label{sec:bac}

Ion-cyclotron waves (ICWs) and \Alfven waves share the same branch on the left-hand side of the dispersion relation diagram \citep{Belcher:1969,Belcher:1971}.
Those that are left-hand (LH) polarized are called \Alfven/ICW (A/IC) waves \citep{Gary:1993}.
At a fixed wave number, waves whose dispersion curves lie on the magnetosonic branch have a higher frequency, are right-hand (RH) polarized, and are referred to as fast magnetosonic (FM) waves \citep{Gary:1993}. FM waves are characterized as whistler waves at sufficiently high frequency (i.e. several times the ion gyrofrequency) \citep{Gary:1993}.
The mechanism for generating these waves depends on both the plasma environment and the shape of the ion VDFs in the solar wind \citep{Leubner:1986,Gary:1993,Gary:2000,Tu:2002,Hellinger:2006,Bale:2009,Maruca:2011,Kasper:2013}.
For example, temperature anisotropies with $T_{\perp}<<T_{\parallel}$ in a high $\beta$ plasma drive the RH magnetosonic instability, while $T_{\perp}>>T_{\parallel}$ in a low $\beta$ plasma drives the LH ion cyclotron instability \citep{Kennel:1966,Gary:1993}. Mechanisms such as ion-cyclotron wave dissipation \citep{Isenberg:2007,Isenberg:2009,Isenberg:2011} and stochastic heating \citep{Chandran:2010a,Chandran:2013} have been proposed to explain this preferential ion heating. However, the origin of such anisotropies still remains an open question.

Throughout the solar wind, relative drift speeds between core protons, secondary proton beams, and $\alpha$-particles have been observed \citep{Feldman:1973,Feldman:1974,Marsch:1982,Marsch:1987,Neugebauer:1996,Steinberg:1996,Kasper:2006,Podesta:2011a}.
These differential flows provide a source of free energy that can be tapped to excite waves through resonant wave-particle interactions \citep{Daughton:1998,Hu:1999,Gary:2000,Marsch:2006,Maneva:2013,Verscharen:2013a,Verscharen:2013b}.

The nature of proton beam driven instabilities in the solar wind, and subsequent wave generation, still remains a topic of fervent research. Early work by \citet{Montgomery:1975,Montgomery:1976} predicted strong growth rates for the parallel magnetosonic mode, which highly influenced the direction of research until \citet{Daughton:1998} showed that the ion-cyclotron mode has the largest growth rate under typical solar wind conditions and is dominant over the previously thought parallel magnetosonic mode.
 
\citet{Marsch:1982} performed the first large scale \textit{in situ} statistical studies of non-thermal features in ion VDFs. The observations were made by Helios at 0.3 AU, which was home to the first 3D plasma instrument enabling field-aligned tracing of proton beams. Using 1D cuts through the distribution (along the magnetic field direction), they derived relative beam-to-core drift velocities that increased closer to the sun. They also demonstrated that statistically significant resolved beams were observed 20\%-30\% of the time. Consistent with \citet{Montgomery:1975,Montgomery:1976}, \citet{Marsch:1982} suggested that magnetosonic instabilities were responsible for beam drift speed and density.

Inclusion of $\alpha$-particles has been shown to significantly affect instability calculations \citep{Quest:1996,Gary:2000,Li:2000,Araneda:2002,Gary:2003,Hellinger:2003,Rosin:2011,Podesta:2011b,Verscharen:2013a,Verscharen:2013b}. For example,
\citet{Marsch:1987} used 1 AU Helios data to show that $\alpha$-particle beams themselves do not generate waves, but $\alpha$-proton core differential flow tends to stabilize the proton beam.
\citet{Verscharen:2013c} found that when $\alpha$-particles propagate in the same direction as the waves, drift-anisotropy plasma instabilities are maximized.
To drive ICWs unstable via an $\alpha$ resonance, enough $\alpha$-particles must exist in the $\alpha$ velocity distribution function \citep{Verscharen:2013c}.

Although ICWs can theoretically propagate at any angle with respect to the magnetic field, $\V{B}$, they are mostly observed while propagating parallel (or anti-parallel) to $\V{B}$.
ICWs that propagate parallel to the magnetic field are circularly polarized, intrinsically left-handed (LH) waves \citep{Stix:1992}.
ICWs were first observed in planetary magnetospheres associated with pick-up ions \citep{Russell:1990,Kivelson:1996,Brain:2002,Delva:2011}.
Studies from later missions suggested that other generation mechanisms were possible, including solar origin, local plasma instabilities, and interstellar pick-up ions \citep{Tsurutani:1994,Murphy:1995}.

Using STEREO measurements collected at 1 AU, \citet{Jian:2009} combined visual inspection and spectral analysis of magnetometer data to obtain a 1 week long statistical sample of ICWs far from planetary sources and other influences to differentiate between these various generation mechanisms.
They established the following criteria to identify intrinsically LH ICWs \citep{Stix:1962}:
\begin{enumerate}
\item the total power was well above the noise floor of the magnetometer, with dominant power transverse to $\V{B}$ rather than parallel,
\item the absolute value of the ellipticity was $>$ 0.7, corresponding to $>$ 70\% circular polarization, and
\item The angle of wave propagation was nearly parallel ($\pm 10^\circ$) to $\V{B}$.
\end{enumerate} 
Waves that match criteria (1)-(3) when the magnetic field is predominantly radial appear both LH and right-hand (RH) polarized in the spacecraft frame.
\citet{Jian:2009} identified these waves in 64\% of measurements.
By assuming that ICWs propagate at the \Alfven speed, \citet{Jian:2009} obtained an approximate expression for the Doppler-shifted wave frequencies in the plasma frame.
This revealed that all of the waves were intrinsically LH polarized.
As such, the waves that were LH polarized in the spacecraft frame were moving outward from the spacecraft and the RH waves in the spacecraft frame where moving inwards towards it.
Since the spacecraft-frame RH waves appeared weaker than the LH waves and the wave frequencies in the spacecraft frame were larger than the local proton gyrofrequency, \citet{Jian:2009} asserted that these waves were not locally generated.
As such, the authors concluded that these ICWs were generated at or near the sun.
Building on the results of \citet{Jian:2009,Jian:2010}, \citet{Jian:2014} combined \textit{in situ} plasma measurements and reported increased $\alpha$-proton drift speeds, temperature, and density ratios during a 4 hour ``Low Frequency Storm" period, suggesting that these plasma parameter changes could have played a role in wave generation.

Recent multi-year studies using Wind and STEREO observations have elucidated the connection between ion-scale waves with the behavior of simultaneously measured ion VDFs \citep{Wicks:2016,Gary:2016,Zhao:2018,Zhao:2019,Woodham:2019}. Using linear Vlasov stability analysis, it has been widely shown that temperature anisotropy drives the ion-cyclotron instability, but both proton beam and $\alpha$-particle differential flows can drive either the magnetosonic instability or the ion-cyclotron instability. The mechanisms for intrinsic LH ICW wave generation are well-interpreted, while the mechanisms for generating RH ICWs are found to be more complex and remains an open question.

\citet{Bale:2019} reported the first observations of circularly polarized ion-scale waves during PSP's first perihelion at 36-54 $R_s$, suggesting that kinetic-scale plasma instabilities existed with the associated measured waves.
In a statistical study of wavelet spectra, \citet{Bowen:2020inpress} found that 30\%-50\% of circularly polarized ion-scale waves were present when the magnetic field was radial. 
However, they claim that these waves could also be present when the field is non-radial, but are masked by perpendicular turbulent fluctuations and cannot be seen with single-spacecraft measurements.
 
The ubiquitous generation of ion-scale waves found in the inner heliosphere suggests that they may play a dominant role in solar wind thermalization \citep{Bale:2019,Bowen:2020inpress}. 
We provide a complementary detailed case study for individual ion-scale wave events.

\section{Methodology} \label{sec:meth}
This paper utilizes data from PSP's particle and electromagnetic fields instrument suites. 
In \secref{sec:inst}, we briefly describe those instruments. \secref{sec:coord} details the SPAN-I coordinate system. In \secref{sec:anal}, we elucidate the techniques used for data analysis.

\subsection{Instruments} \label{sec:inst}
Non-Maxwellian features, such as beams and high-energy shoulders, have been observed by SPAN-I, as part of the SWEAP instrument suite. 
\citet{Kasper:2015} describes SWEAP's primary objectives.
In brief, they are to shed light on solar wind sources, coronal and solar wind heating mechanisms, and energetic particle transport.
In addition to SPAN-I, SWEAP consists of two electron electrostatic analyzers, SPAN-E \citep{Whittlesey:2020}, and a sun-pointing Faraday cup, Solar Probe Cup (SPC) \citep{Case:2019inpress}. 
The combined field-of-view (FOV) of all four SWEAP instruments cover the full sky with minor exceptions for ions in the anti-ram direction of the spacecraft orbital path \citep{Kasper:2015}. 
\citet{Kasper:2019} provides a brief preview of SWEAP's first scientific results.

This paper specifically utilizes data measured by SPAN-I, which \citet{Livi:2020inpress} describes in detail.
SPAN-I is a top-hat electrostatic analyzer (ESA) and mass discriminator that measures 3D velocity distribution functions of the ambient ion populations in the energy range 2 eV-30 keV at a maximum cadence of 0.216 s. However, the observations in this paper are from lower cadence 7 s downlinked data.
SPAN-I's FOV covers 247.5$^\circ$ x 120$^\circ$.
To determine the mass per charge of the incoming particles, the analyzer employs a time-of-flight section at the exit of the ESA that enables separation of protons, $\alpha$-particles and other heavy ions in the solar wind. 
SPAN-I's main caveat is that the Sun-instrument line of sight is obscured by PSP's thermal protection shield (TPS).
As such, the observed VDFs are partially obscured and the resulting moments are considered ``partial moments.''
It is expected that in later Encounters, the plasma will be observed more in SPAN-I's FOV. 
Additionally, a combined analysis of SPC and SPAN-I data may also lead to full, un-obscured measurements of the ion VDFs in the solar wind.
The benefit of observing data from SPAN-I is that the sides of the velocity distributions are the most useful for beam detection. This is because non-thermal features of VDFs tend occur at higher velocities than the bulk flow, which is more likely to appear in SPAN-I's FOV.

Electric and magnetic fields onboard PSP are measured by the FIELDS investigation.
FIELDS's instruments include electric antennas and fluxgate magnetometers (MAG) \citep{Bale:2016}.
\citet{Bale:2019} reports initial FIELDS observations from PSP's first two Encounters.
The MAG data has a bandwidth of $\approx$140 Hz \citep{Bale:2016}.
Unless otherwise noted, we use MAG data downsampled to 16 Hz resolution in \secref{sec:obs} so as to improve computational efficiency and simultaneously capture ion-scale physics at approximately 1 Hz. 

\subsection{Coordinates} \label{sec:coord}

Top-hat ESAs such as SPAN-I measure particle velocities in 3D phase-space using coordinates $(E,\theta,\phi)$. The cylindrically symmetric configuration, shown in Figure 2 of \citet{Livi:2020inpress}, consists of two nested hemispheres below a particle entrance aperture. The instrument FOV is the cylindrical figure of revolution defined by the entrance aperture which rotates $360^\circ$ about the azimuthal symmetry axis of the hemispheres, $\phi$.  Particles with energy, $E$, determined by the voltage difference between the two nested hemispheres, arrive at the entrance aperture between two deflecting electrodes at an angle of incidence, $\theta$. The instrument sweeps through a full range of $\theta$, and then steps to the next lower energy step. Separate anodes simultaneously measures the distribution as a function of $\phi$.  An entire distribution is measured in 216 ms to assemble the 3D particle velocity information.

The velocity-space coordinates of the SPAN-I instrument, $(v_x,v_y,v_z)$, are dictated by the geometry of the ESA and the particle travel direction. The $v_z$-axis is defined as the rotational symmetry axis of instrument. The $v_x$-direction is aligned to the sunward look direction such that a particle traveling from the direction of the sun will be measured in the $-v_x$ direction. The subsequent $v_y$-axis is then defined orthogonal to $v_x$ and $v_z$ in a right-handed coordinate system. The reader may refer to \citet{Livi:2020inpress} for further details.

\subsection{Data Analysis} \label{sec:anal}

We apply 1D Levenberg-Marquardt fits \citep{Levenberg:1944,Marquardt:1963} to the L2 SPAN-I data to extract a number density ($n$), velocity ($v$), and a temperature ($T$) for two proton populations \citep{Livi:2014}. As applied to SPAN-I, the main assumption is that the radial component of the plasma is fully in the instrument's FOV.
Therefore, we select the FOV bin that contains the highest number of particle counts for a single interval.
Using data from this bin, we assume that the ion population being measured is isotropic in temperature and that the velocity component is fully radial.
We define the ``core" population to be the distribution that fits to the peak in phase-space density. The ``beam" represents a shoulder or separate resolved peak on the tail of the distribution.
While others have specifically defined beams to be the particle population that has the smaller of the two densities, \secref{sec:obs} makes clear that, along the dimension of highest energy flux, this criterion does not apply in the events we report. For clarity, we have included the superscript, *, to all fit parameters to indicate that these values were obtained along the direction of highest particle count flux, corresponding to highest energy flux. Throughout this paper, the terms ``beam" and ``core" refer to the fits procured from 1D cuts in this direction. We reiterate that these fits were constrained by both SPAN-I's limited FOV and the 1D approximation.

We apply a Morlet wavelet transform to the FIELDS data to extract polarization and wave power as a function of frequency.
A wavelet is a common alternative to the Fourier transform in signal processing that 
extracts localized information from a signal in both time and frequency, whereas the Fourier transform can only yield local frequency information.
The Morlet basis is one particular basis for a wavelet transform that is mathematically equivalent to a Gaussian-windowed Short-Time Fourier transform.
\citet{Farge:1992} provides the first cohesive review of wavelet transforms and their physical applications.
\citet{Farge:2015} review their application to plasma physics.
For a step-by-step guide to the wavelet analysis technique, the reader may also consult \citet{Torrence:1998}.

We extracted the magnetic field ellipticity and wave normal vector properties using Minimum Variance Analysis (MVA).
\citet{Sonnerup:1967} performed an early application of MVA to spacecraft data so as to identify vectors perpendicular to the surface of the magnetopause current layer in  Explorer 12 data.
\citet{Means:1972} provides a canonical description of the method.
\citet{Dunlop:1995} describes applications to the Cluster mission.
Following the introduction of \citet{Dunlop:1995}, we summarize the general procedure. Let $\V{B}=\V{B}_n(t_n)$ be the time series of a magnetic field sampled at $N$ points indexed by $n$. The variance, $\sigma^2$, of $\V{B}$ in some direction, $\V{\hat{x}}$, is
\[
\sigma^2 = \frac{1}{N}\sum_n^N \left(\V{B}_n \V{\cdot \hat{x}} -  \langle \V{B} \rangle \V{\cdot \hat{x}} \right)^2
\]
where $\langle \V{B} \rangle = \frac{1}{N}\sum_n^N B_n$ is the mean value of $\V{B}$. The goal is to find $\V{\hat{x}}$ such that $\sigma^2$ is minimized. This is equivalent to solving
\begin{equation}
\V{M \hat{x}} = \lambda \V{\hat{x}}
\label{eq:eig}
\end{equation}
for eigenvalues $\lambda$, where $M_{ij} = \langle B_i B_j \rangle - \langle B_i \rangle \langle B_j \rangle$ and $\langle B_i B_j \rangle =  \frac{1}{N} \sum_n^N B_i^n B_j^n$ for $i,j=x,y,z$. Then, the eigensolutions of Eq. \ref{eq:eig} define the principle axes of an ellipsoid. The eigenvector, $\V{\hat{x}_{min}}$, associated with the minimum eigenvalue is the solution that minimizes $\sigma^2$. Then, the angle between the wavevector, $\V{k}=\V{\hat{x}_{min}}$, and the mean magnetic field unit vector, $\V{\hat{b}}_0$, is
\[
\theta_{kb} = \cos^{-1}\left(\V{k \cdot \hat{b}}_0 \right )
\]
Hence, the minimum eigensolution to Eq. \ref{eq:eig} finds the angle of wave propagation with respect to the magnetic field direction with minimum variance. The other two eigensolutions that make up the ellipsoid yields information about the ellipticity of the magnetic field perturbations (waves). If two eigenvalues $\lambda_1,\lambda_2$ are approximately equal, with one, $\lambda_3$, being sufficiently subdominant, then the wave is near-circularly polarized. The difference $\lambda_{diff}=|\lambda_1 - \lambda_2|$ defines the measure of ellipticity. In the limit $\lambda_{diff} \rightarrow 0$, the ellipticity is 1 and the wave is circularly polarized. On the other hand, if $\lambda_{diff} \rightarrow \infty$, then the wave is linearly polarized and the ellipticity is 0. 

For the MVA analysis performed in \secref{sec:obs}, we used a frequency range of 1-10 Hz and the mean magnetic field was smoothed over a 193 sample window, corresponding to approximately 12 seconds for the downsampled 16 Hz magnetic field data. As discussed throughout the rest of the paper, the information gleaned from MVA is essential for characterizing the observed ion-scale waves.

\section{Observations} \label{sec:obs}
We present two events from Encounter 2 that, by visual inspection, presented themselves as the most noteworthy wave-particle interaction event candidates, as seen by SPAN-I in conjunction with FIELDS measurements. These events show compelling evidence that wave-particle interactions occurred, but characterizing the energy transfer mechanisms remains an open question.

\subsection{Event 1: A perfect storm} \label{sec:e1}
\begin{figure*}
\centering
\includegraphics[trim=5 0 5 0,clip,scale=1]{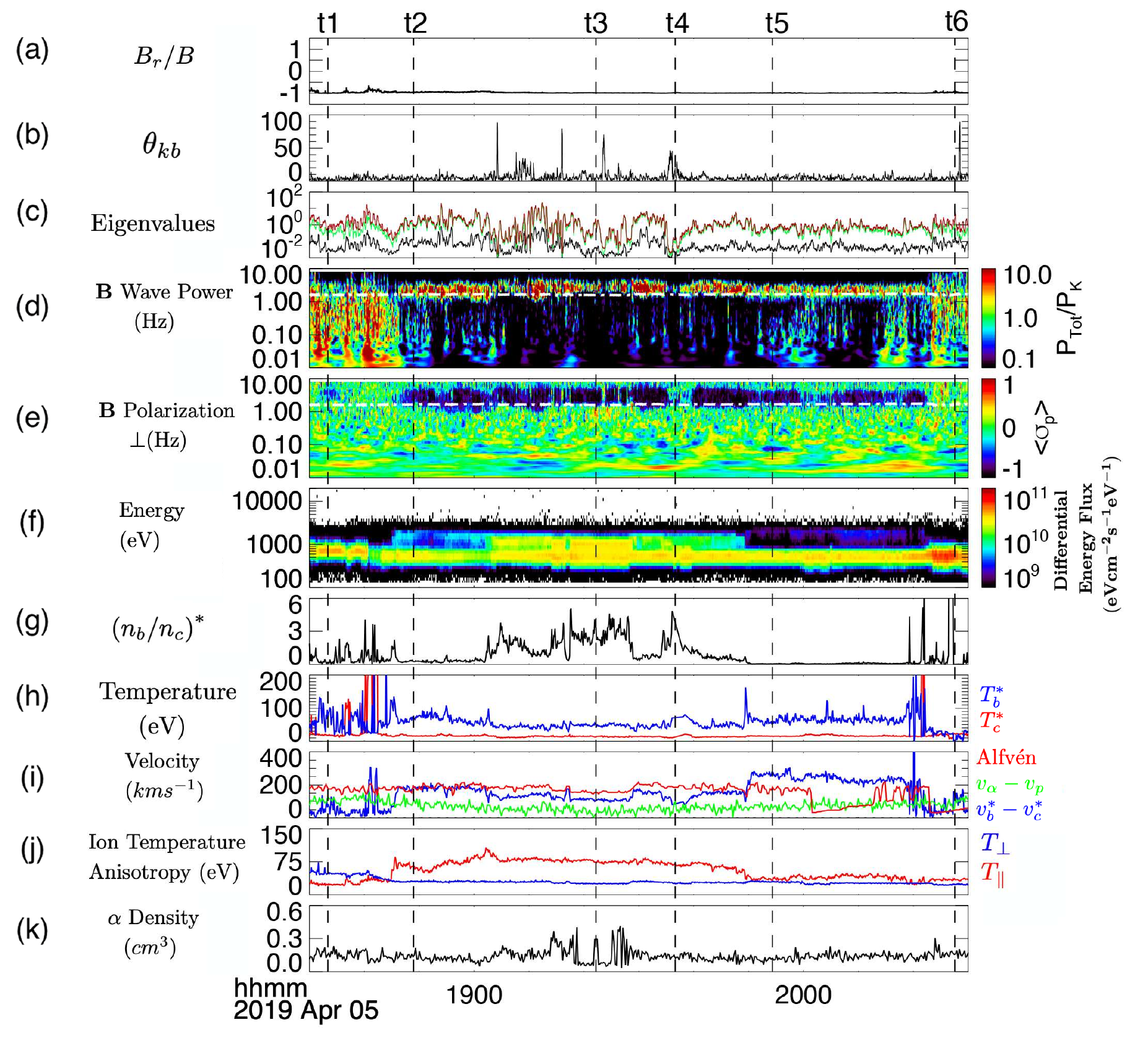}
\caption{Example event on 2019-04-05 (Event $\#$1) featuring a strong correlation between a proton beam and an ion-scale wave storm. Shown is the (a) radial magnetic field component, (b) angle of wave propagation w.r.t. $\V{B}$, (c) eigenvalues from MVA, (d) wavelet transform of $\V{B}$, (e) perpendicular polarization of $\V{B}$, (f) SPAN-I measured moment of differential energy flux, (g) proton beam-to-core density ratio fits, (h) temperature fits of proton beam (blue) and core (red), (i) proton beam-core differential velocity fits (blue) and $\alpha$-proton (green) differential velocity SPAN-I moments compared to the \Alfven velocity (red), (j) SPAN-I measured moments of temperature anisotropy, and (k) SPAN-I measured $\alpha$ density moments. In panels (d) and (e), the white dashed-dotted line represents the local proton gyrofrequency.
\label{fig:4_5_allvars}}
\end{figure*}

 \begin{figure*}
 \centering
\hspace{.05in} (a) t1 = 2019-04-05/18:33:22 
\vfill
\includegraphics[trim=60 10 10 48,clip,scale = .17]{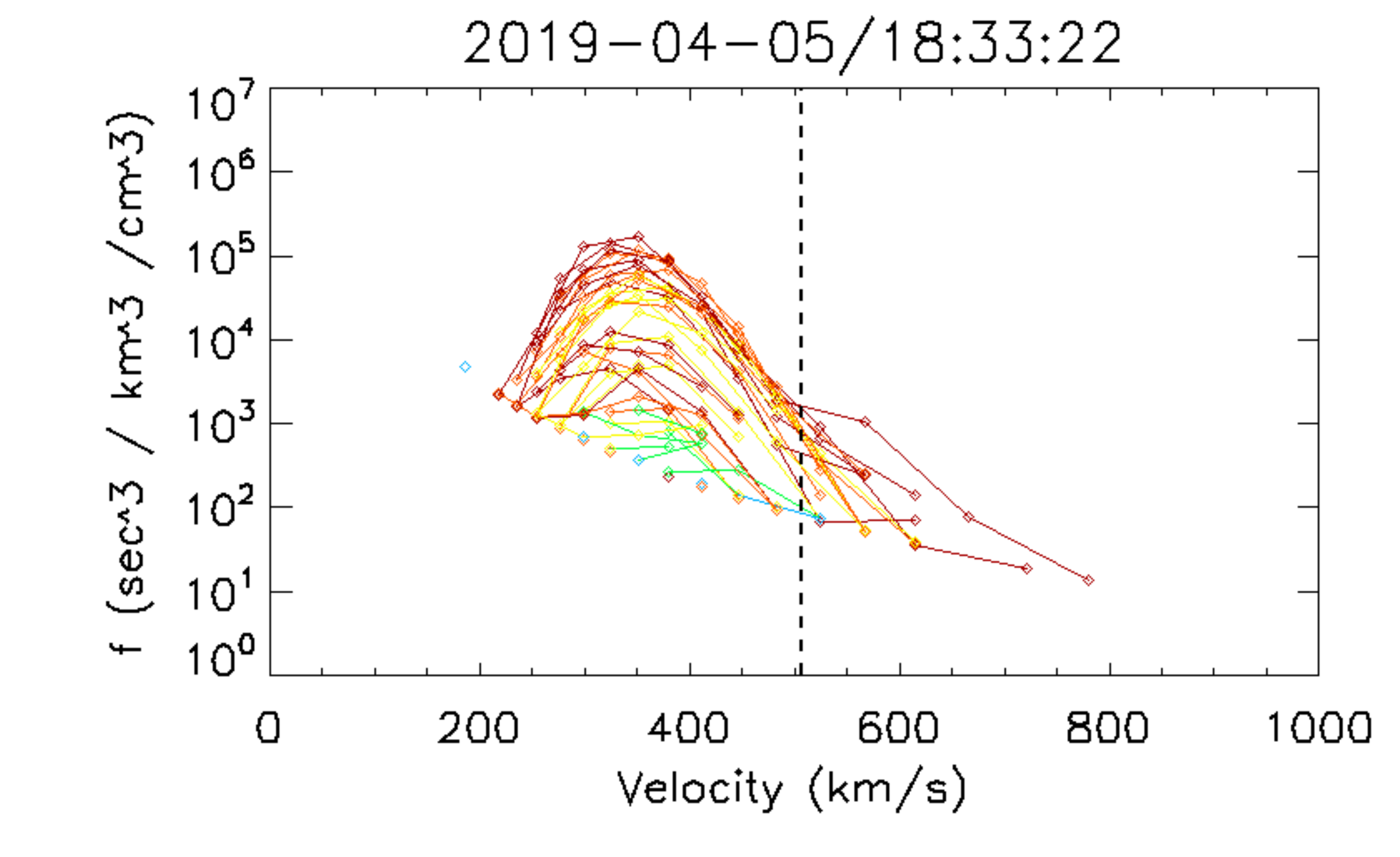} \hspace{.02in}
\includegraphics[trim=30 10 300 52,clip,scale=.17]{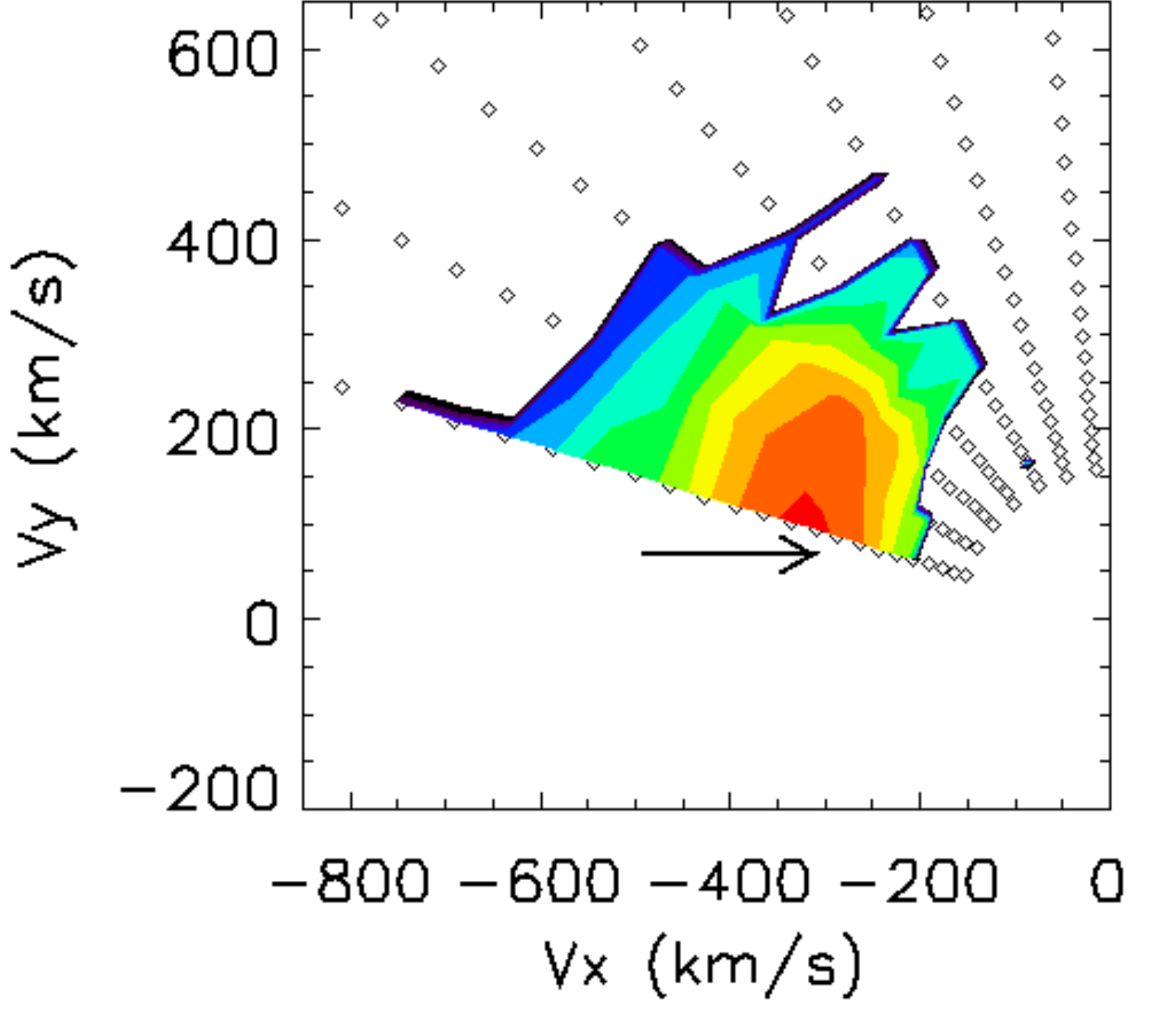}
\hspace{.02in}
\includegraphics[trim=30 10 290 52,clip,scale=.17]{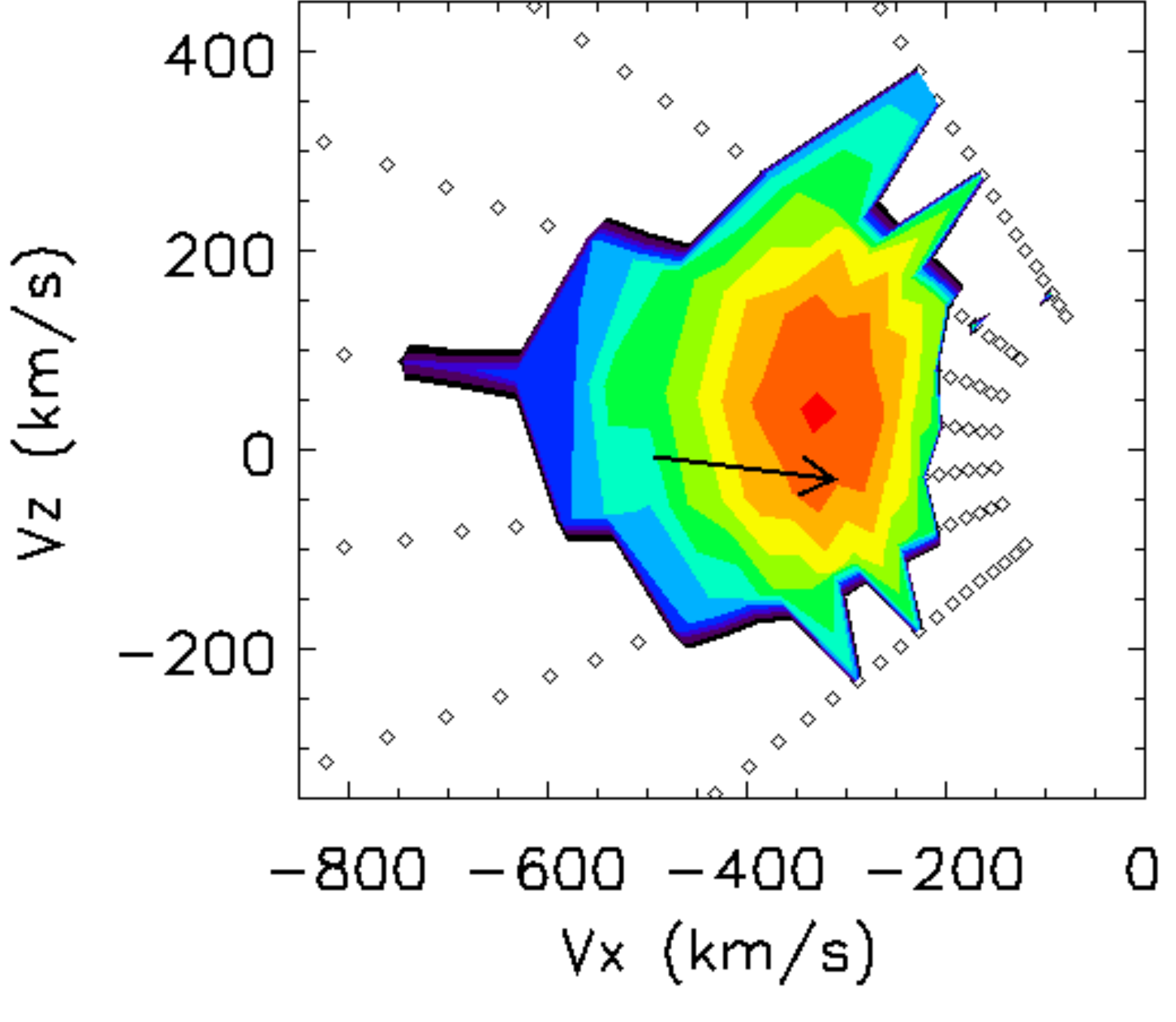}
\includegraphics[trim=20 0 0 0,clip,scale=.14]{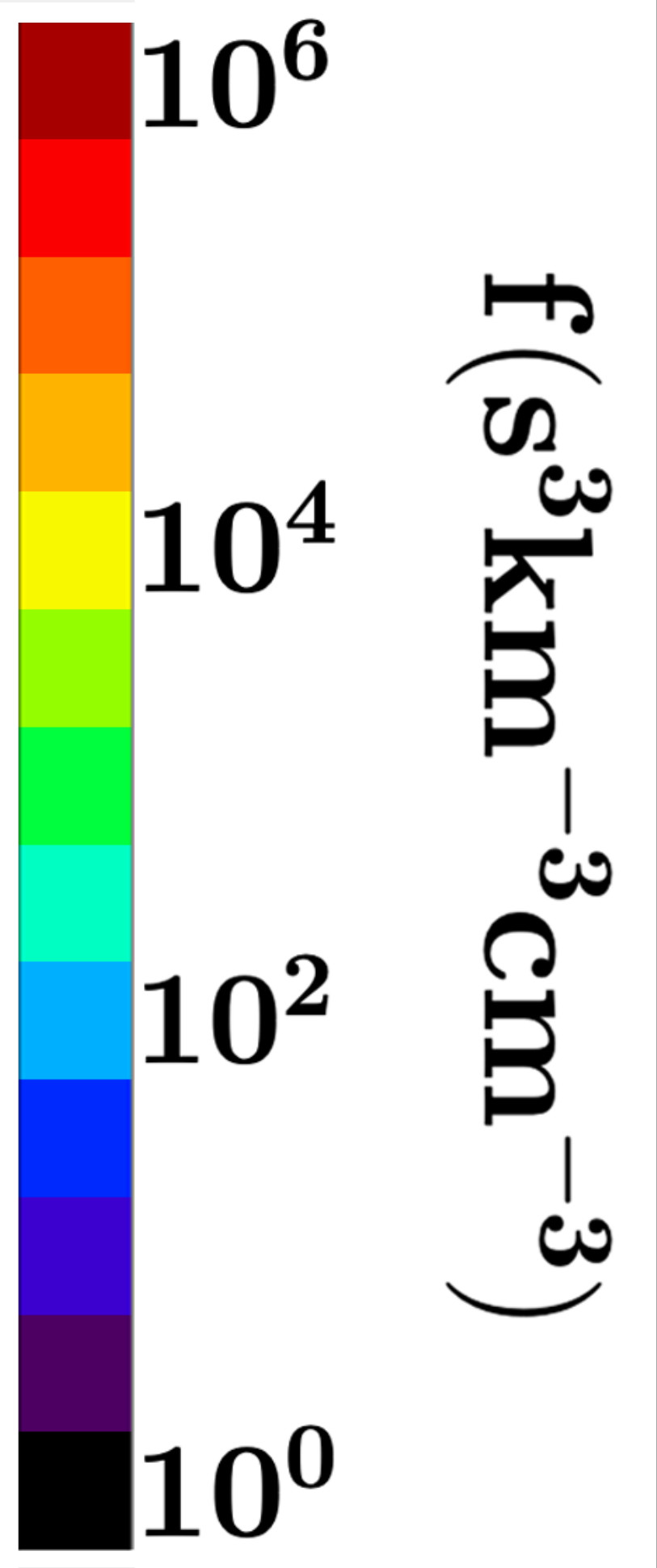}

\hspace{.05in} (b) t2 = 2019-04-05/18:48:59
\vfill
\includegraphics[trim=60 10 10 48,clip,scale = .17]{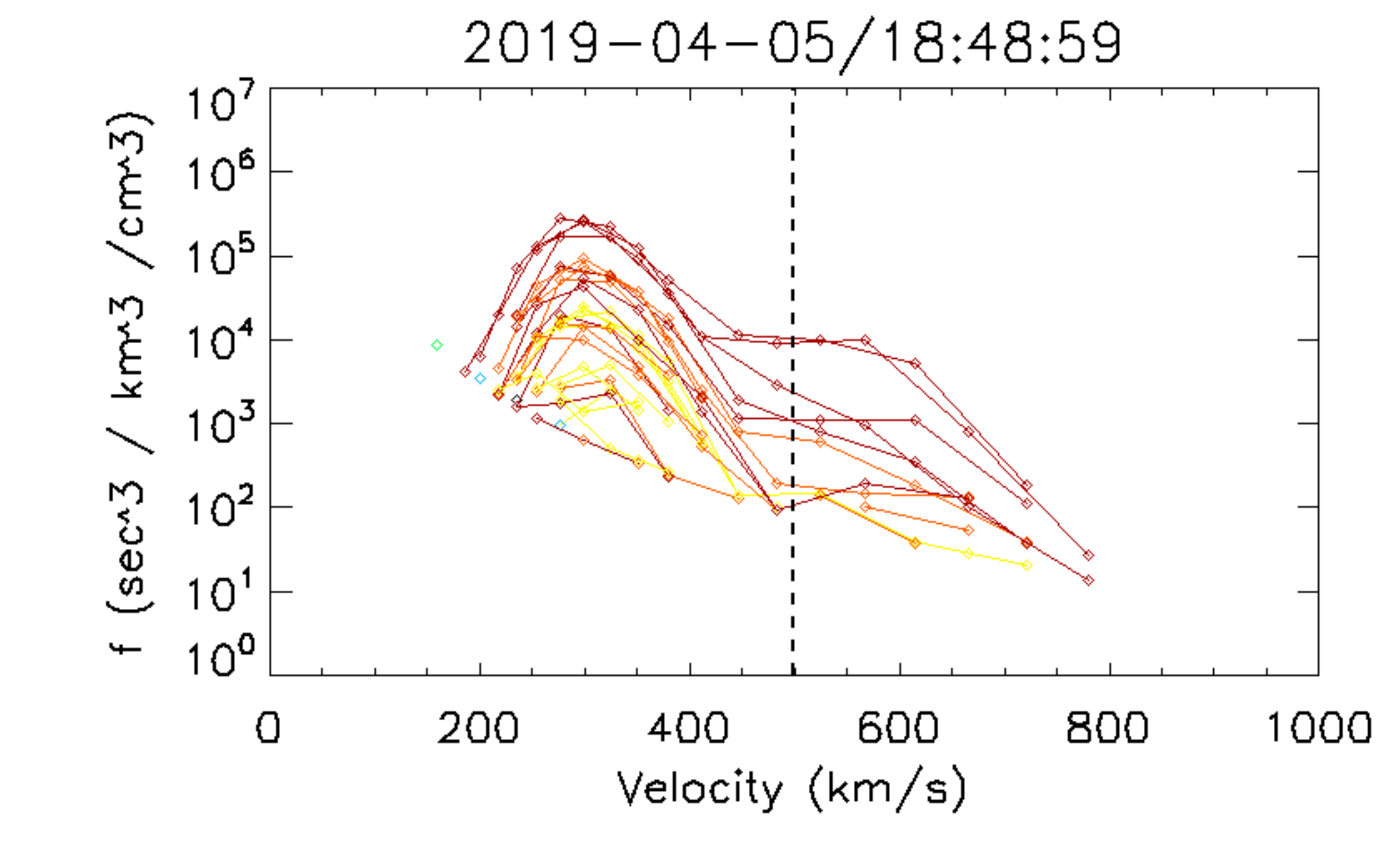} \hspace{.02in}
\includegraphics[trim=30 10 300 52,clip,scale=.17]{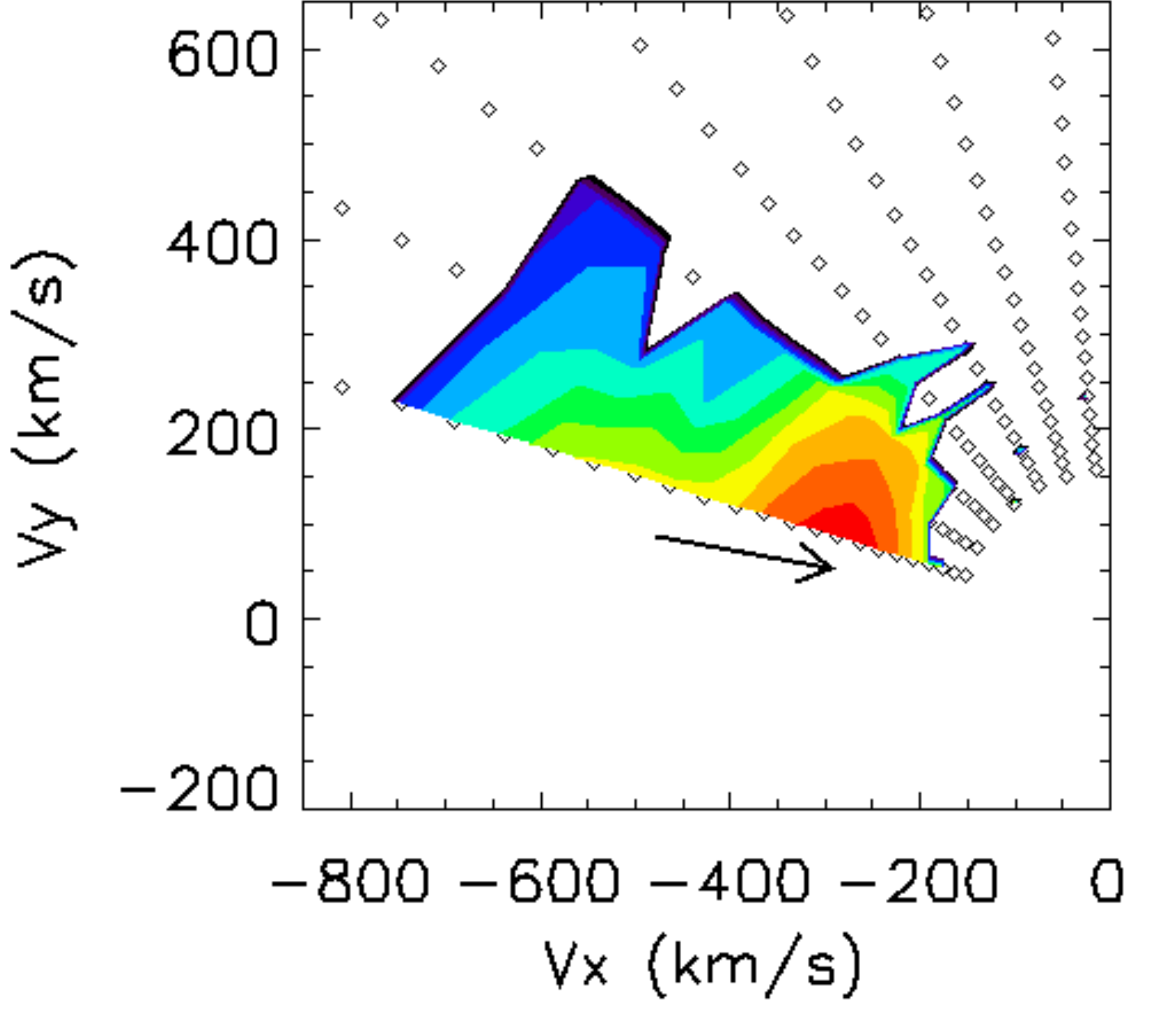}
\hspace{.02in}
\includegraphics[trim=30 10 290 52,clip,scale=.17]{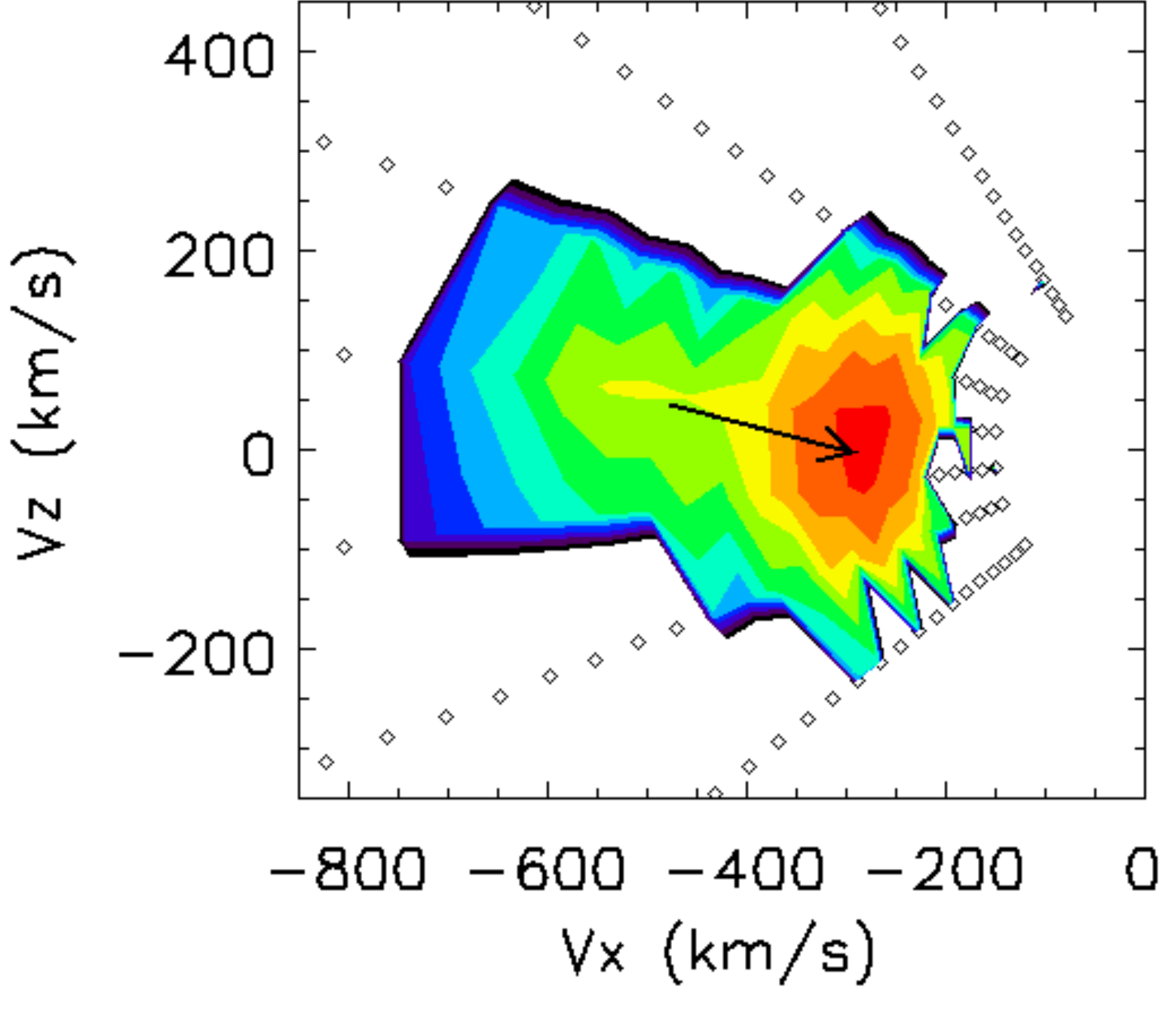}
\includegraphics[scale=.14]{fig2_legend.pdf}

\hspace{.05in}(c) t3 = 2019-04-05/19:22:11
\vfill
\includegraphics[trim=60 10 10 48,clip,scale = .17]{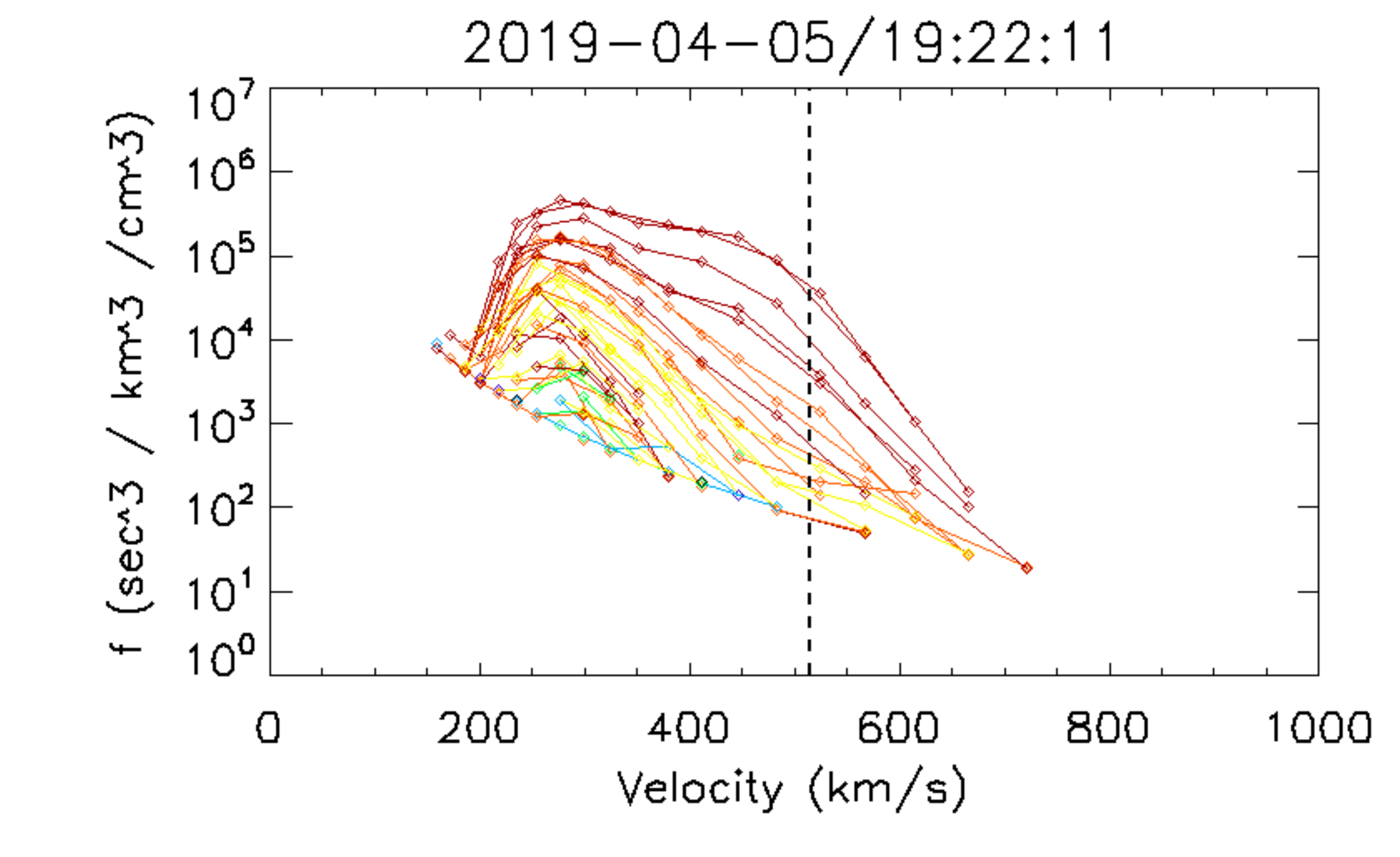} \hspace{.02in}
\includegraphics[trim=30 10 300 52,clip,scale=.17]{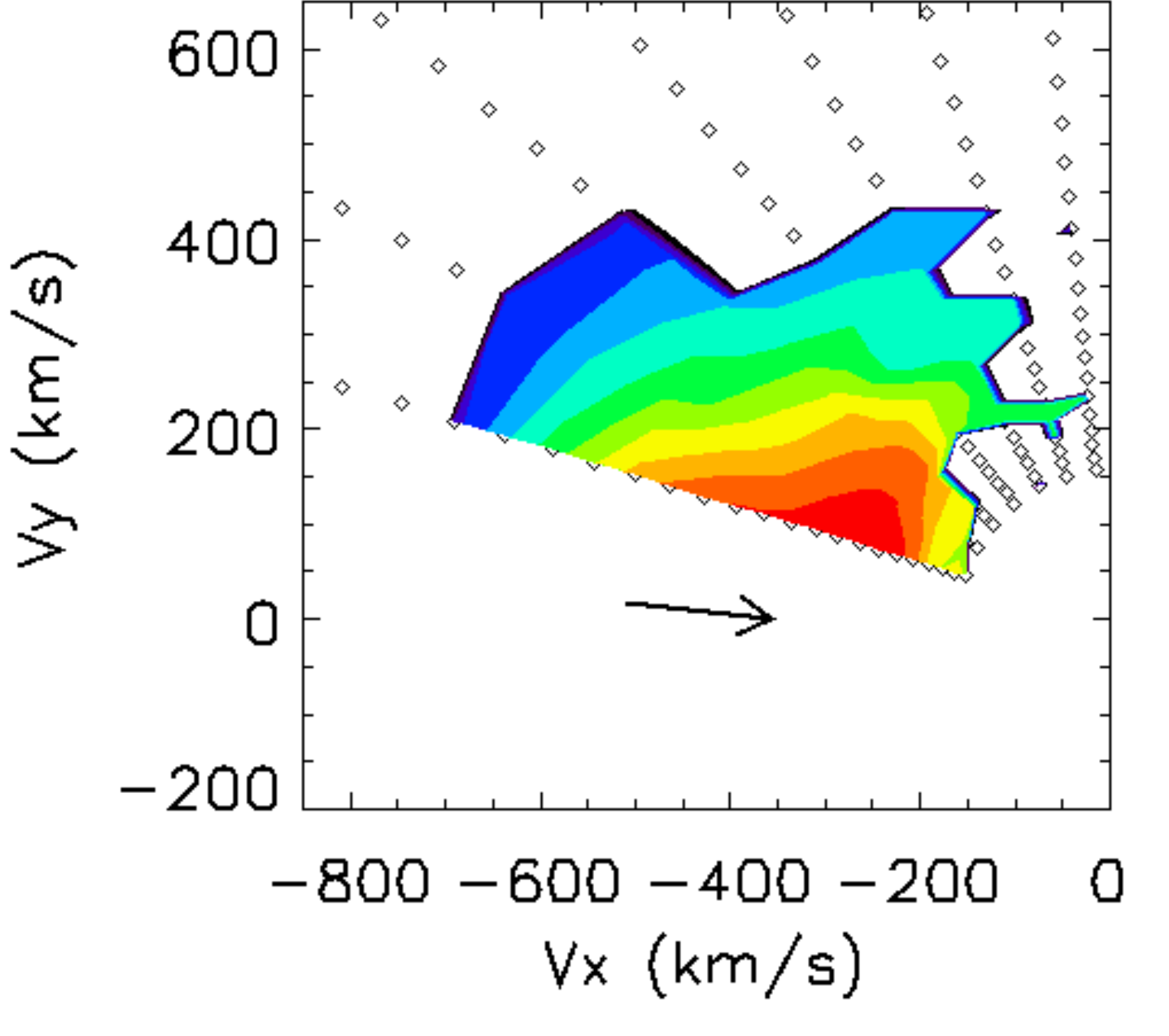}
\hspace{.02in}
\includegraphics[trim=30 10 290 52,clip,scale=.17]{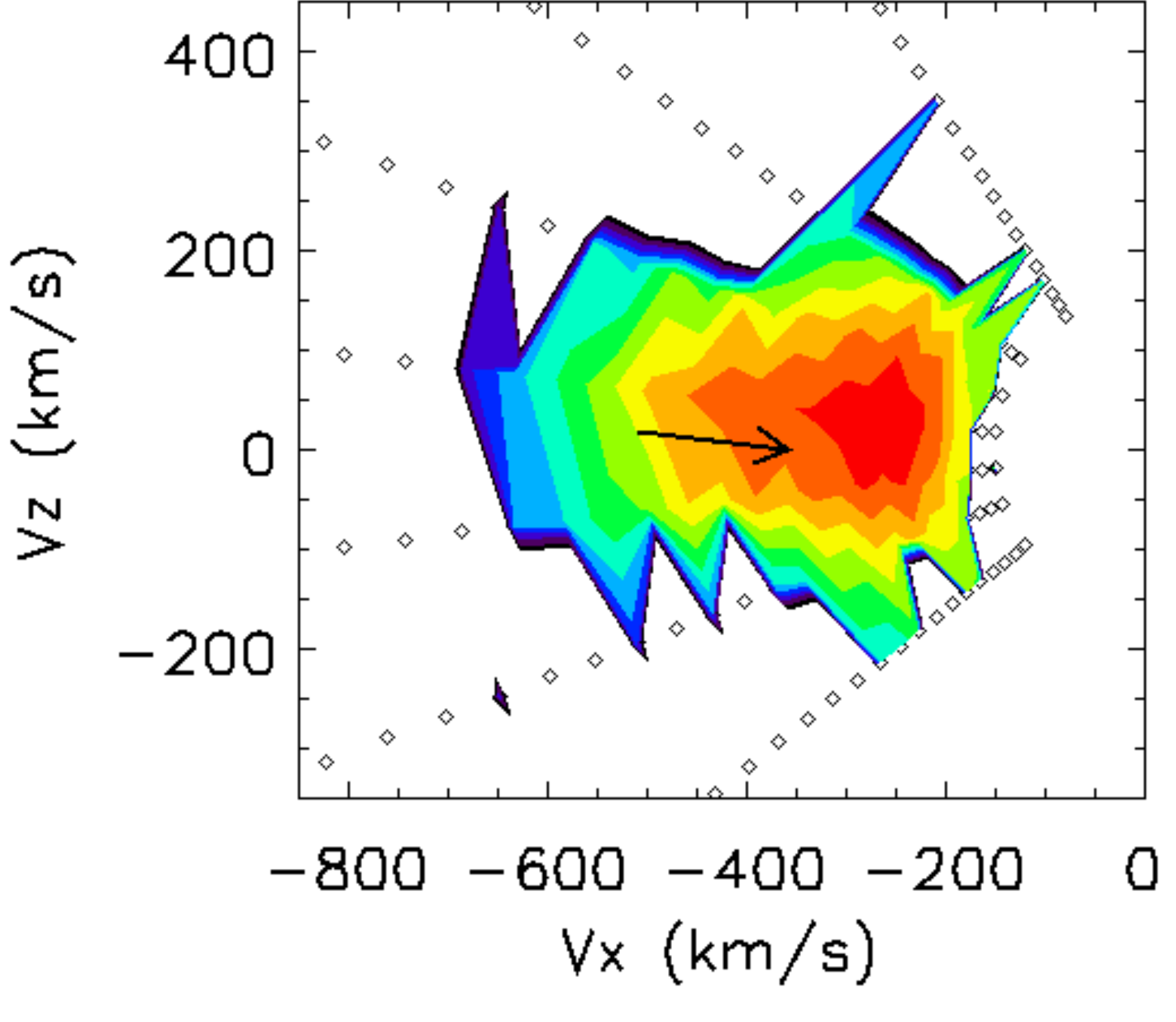}
\includegraphics[scale=.14]{fig2_legend.pdf}
\vfill

\hspace{.05in}(d) t4 = 2019-04-05/19:36:37
\vfill
\includegraphics[trim=60 10 10 48,clip,scale = .17]{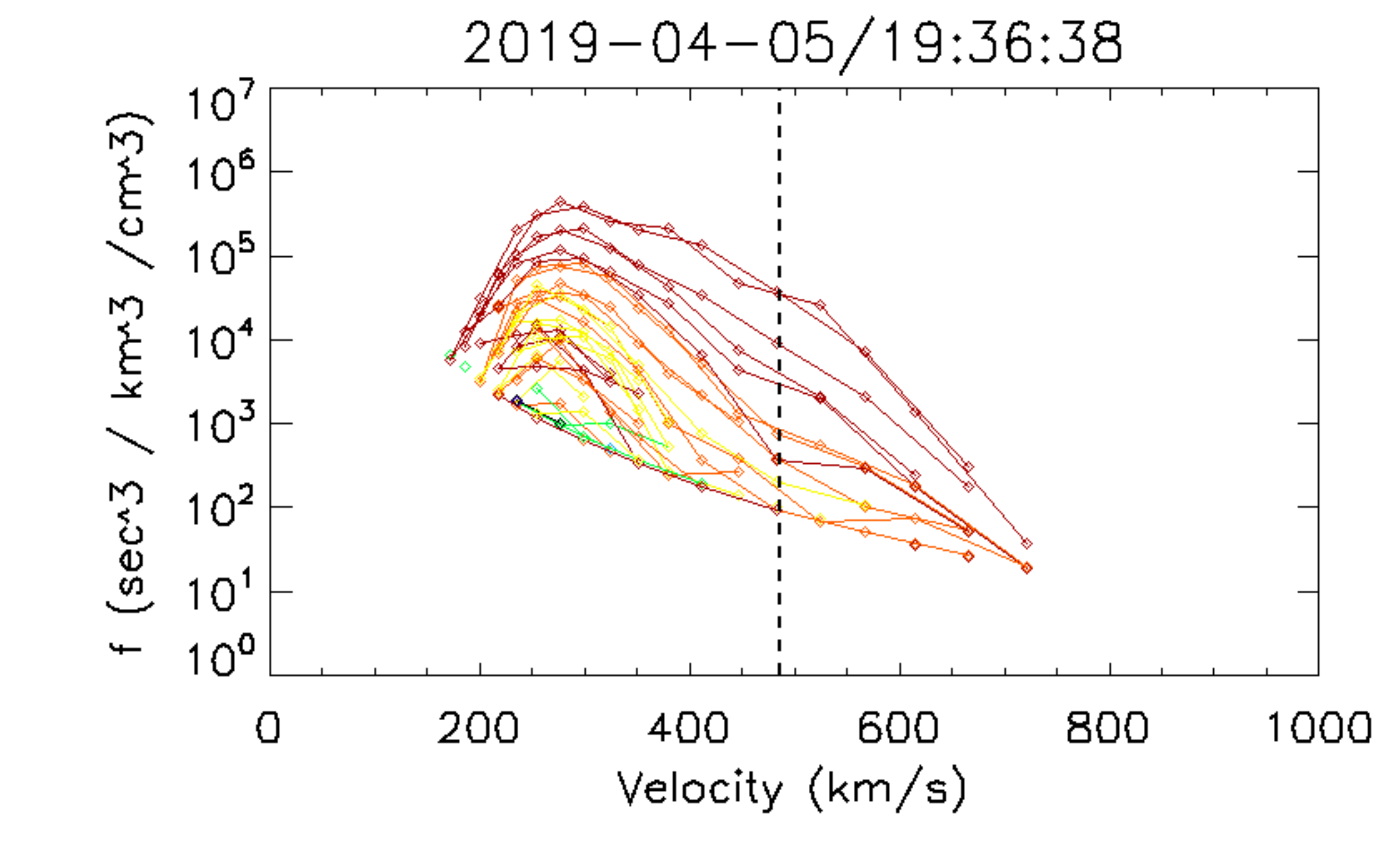} 
\hspace{.02in}
\includegraphics[trim=30 10 300 52,clip,scale=.17]{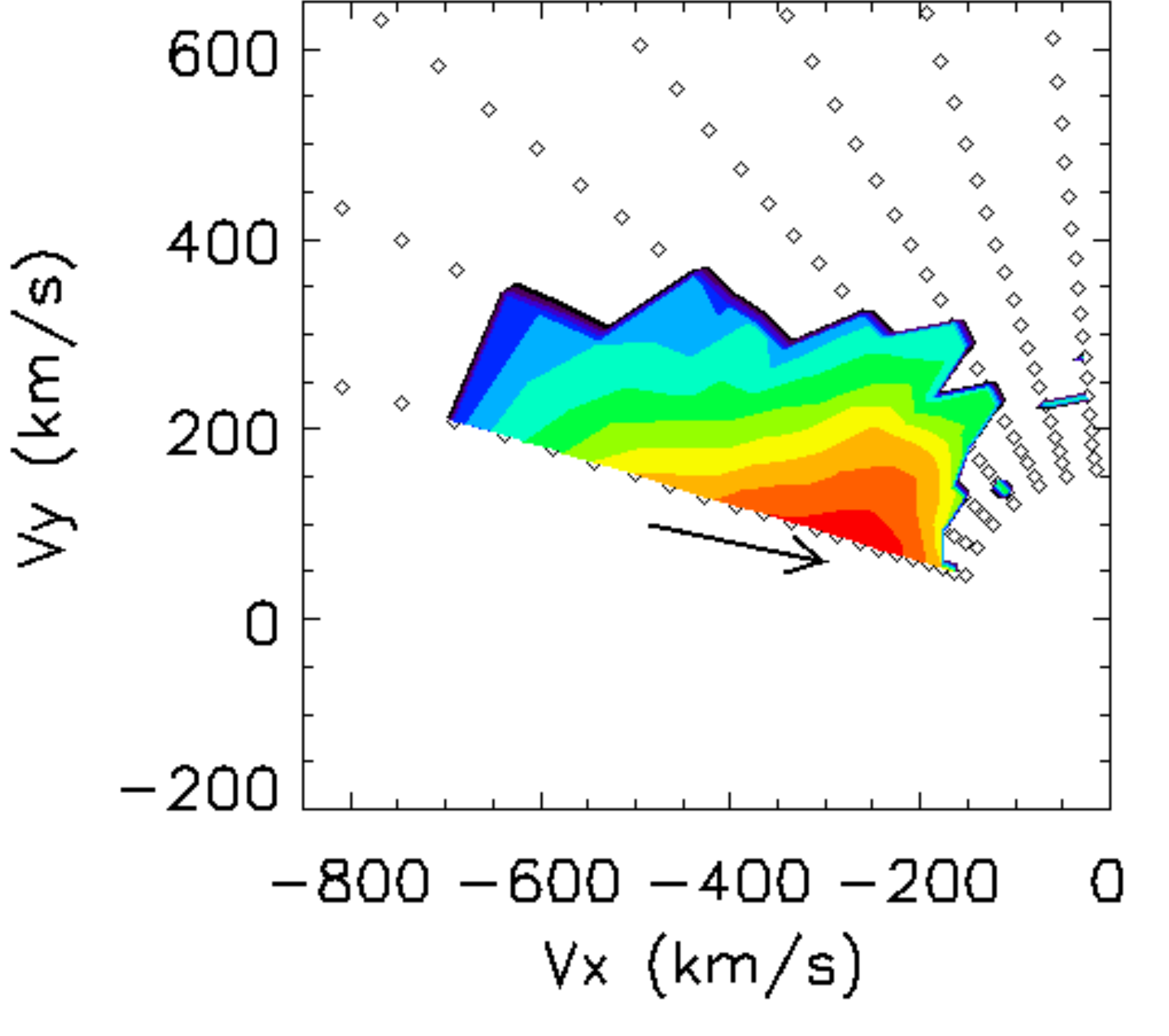}
\hspace{.02in}
\includegraphics[trim=30 10 290 52,clip,scale=.17]{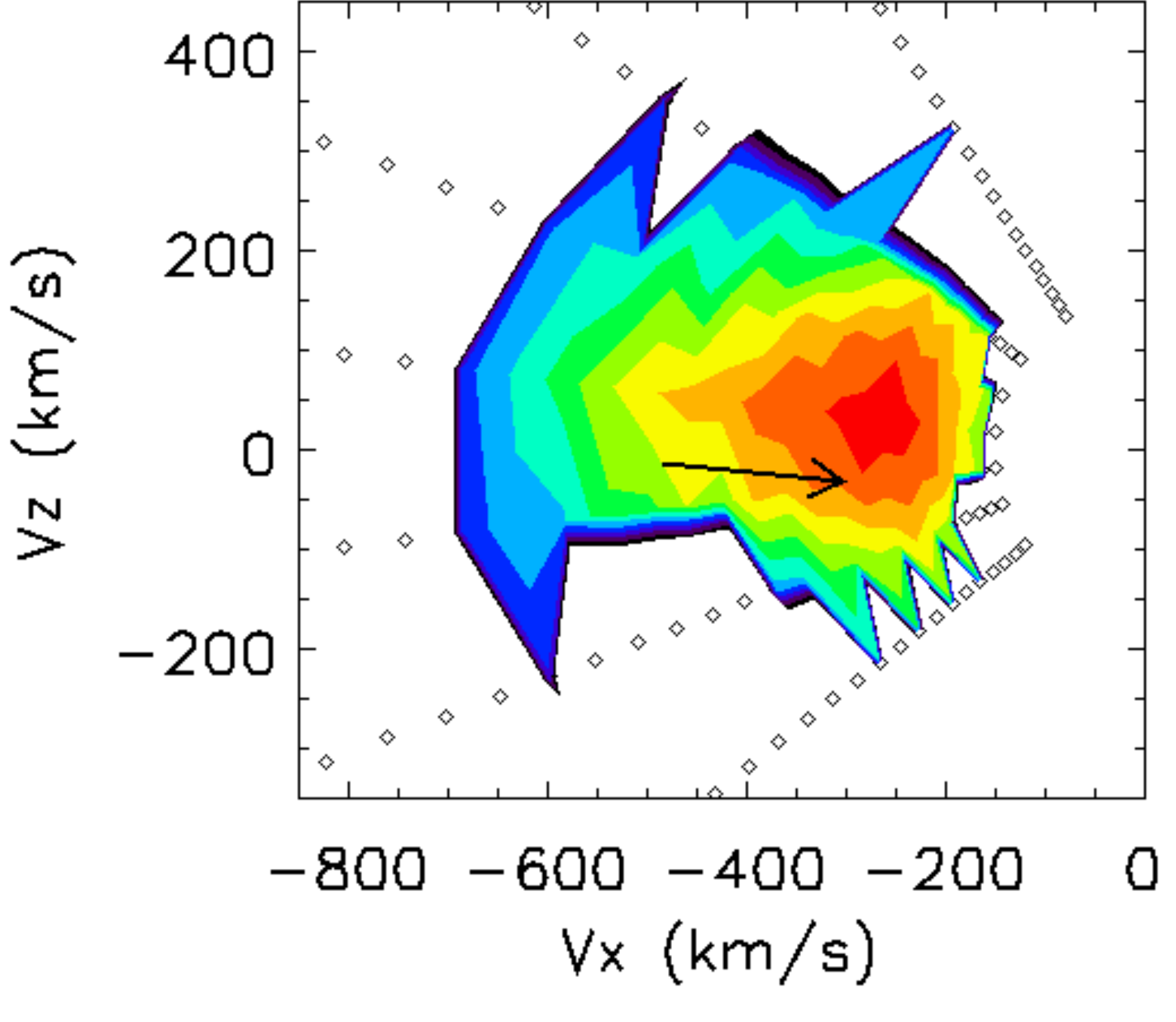} 
\includegraphics[scale=.14]{fig2_legend.eps} 

 \hspace{.05in}(e) t5 = 2019-04-05/19:54:20
\vfill
\includegraphics[trim=60 10 10 48,clip,scale = .17]{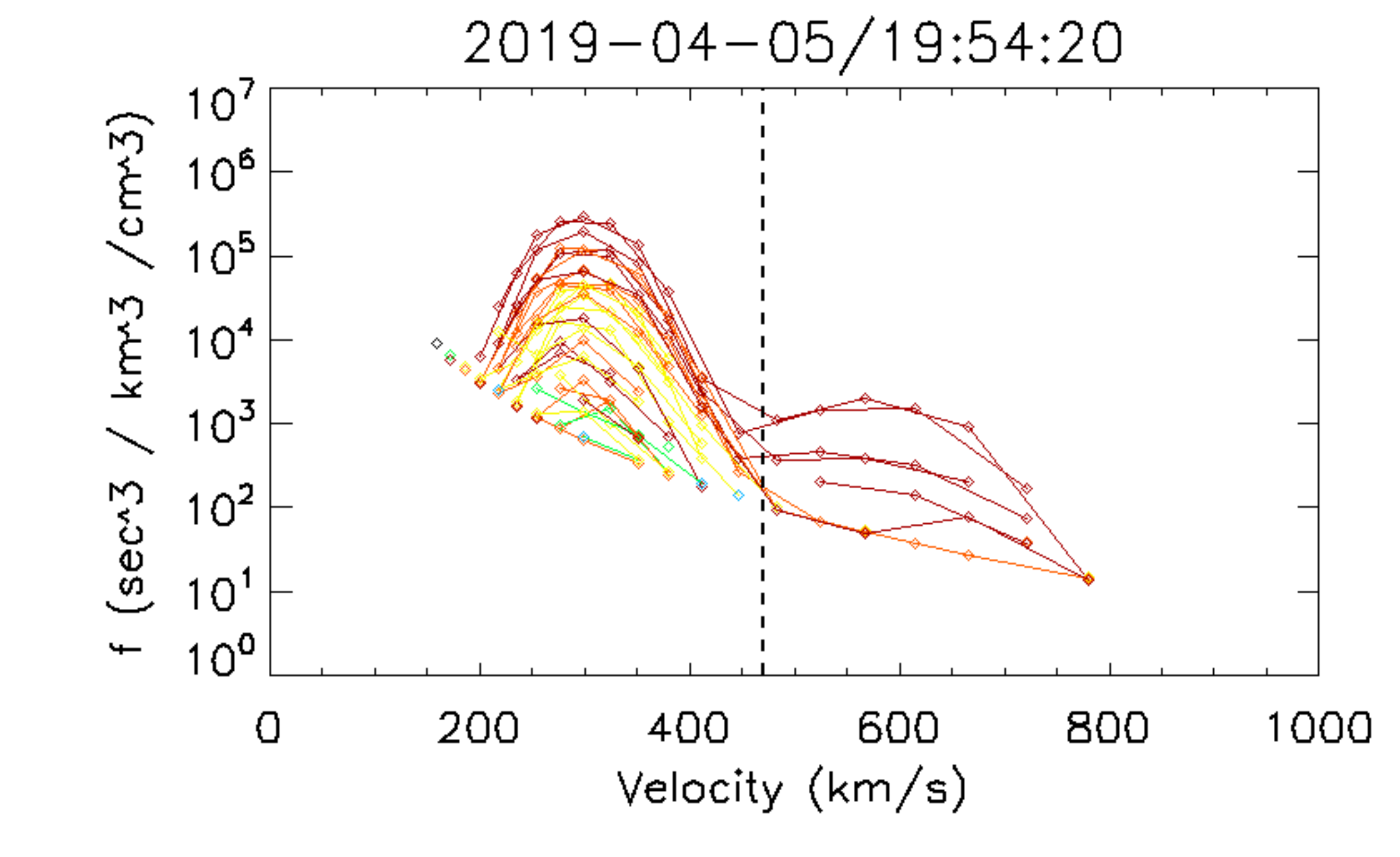} \hspace{.02in}
\includegraphics[trim=30 10 300 52,clip,scale=.17]{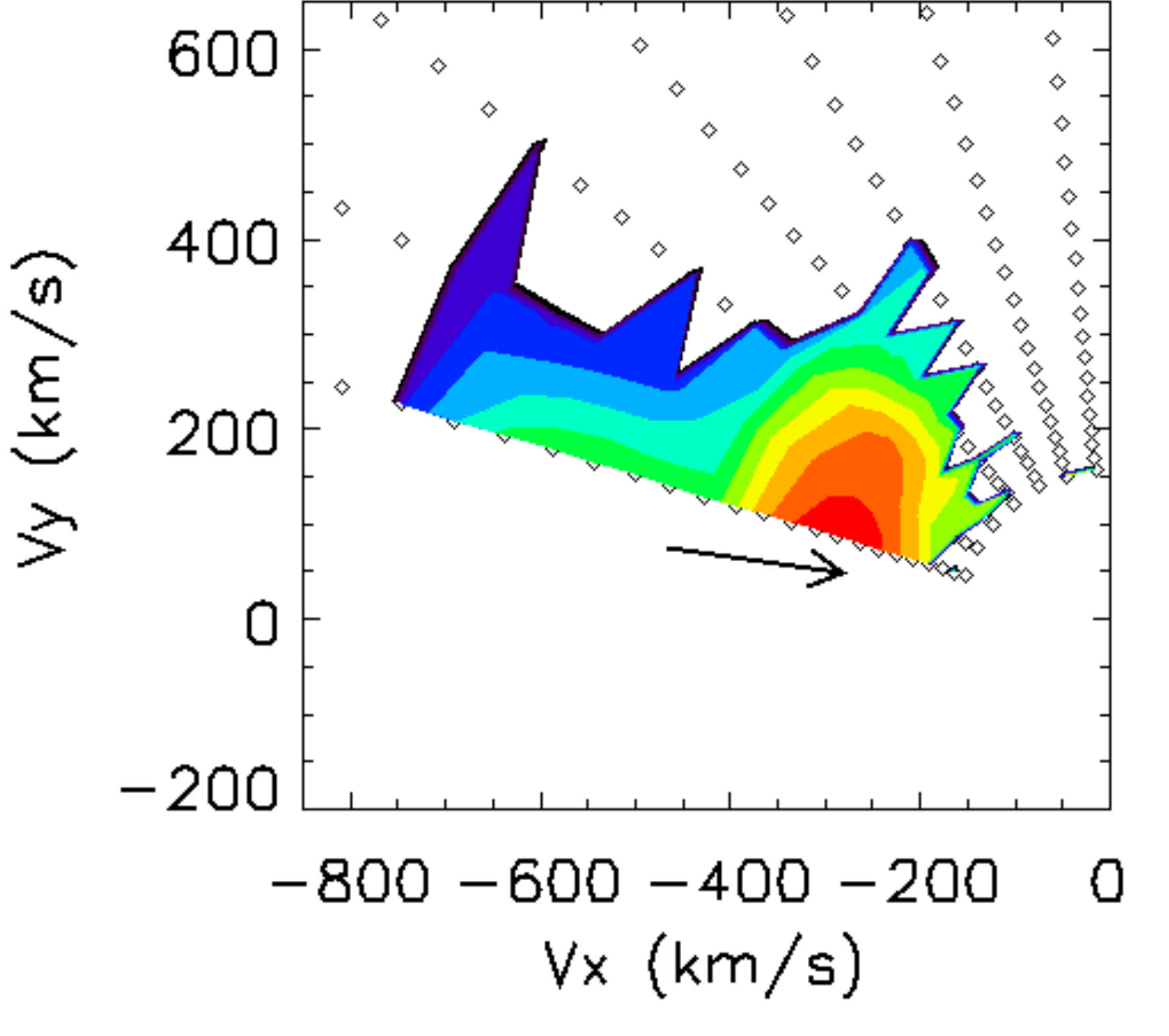}
\hspace{.02in}
\includegraphics[trim=30 10 290 52,clip,scale=.17]{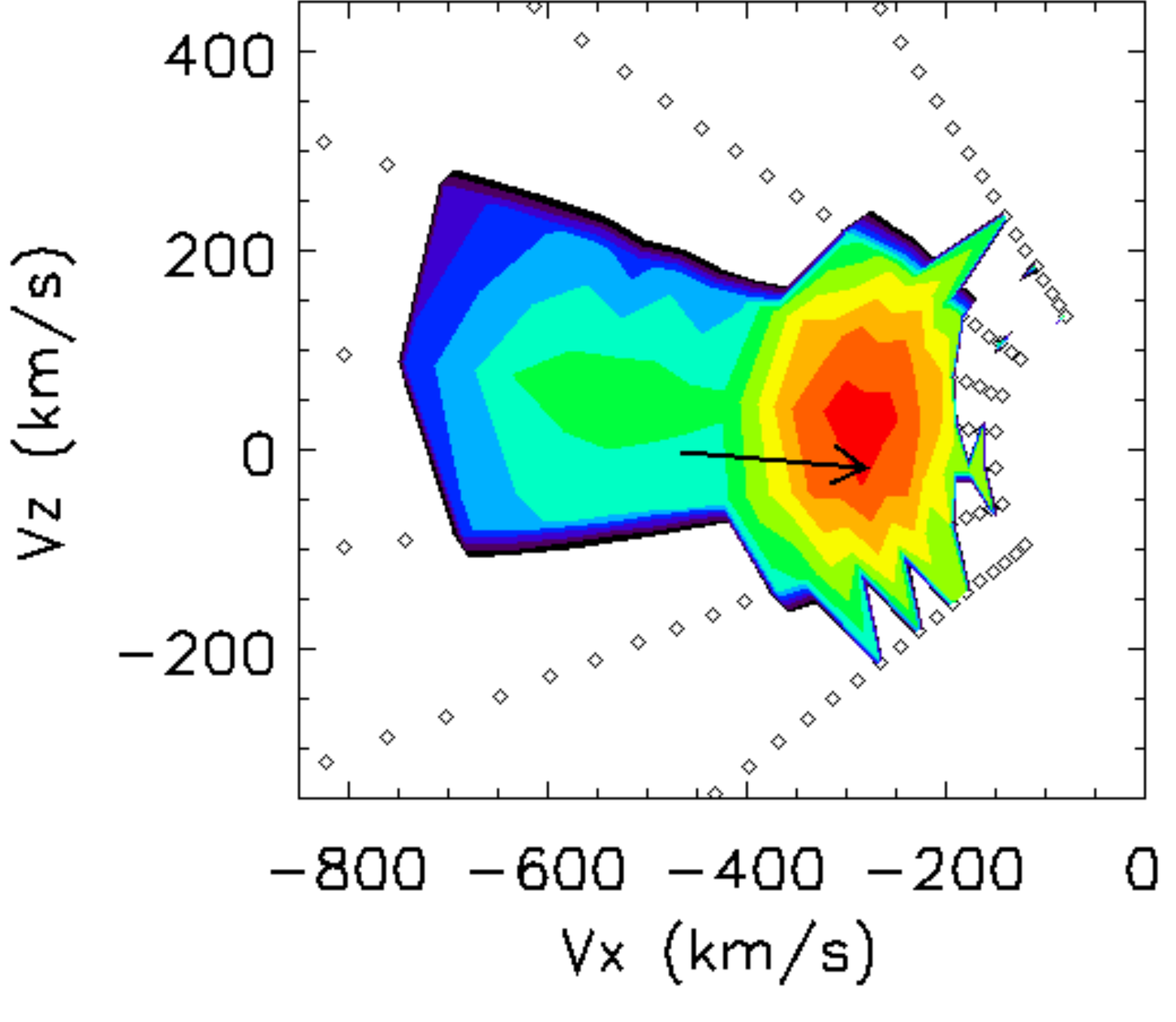}
\includegraphics[scale=.14]{fig2_legend.pdf} 

  \hspace{.05in}(f) t6 = 2019-04-05/20:27:33
\vfill
 \includegraphics[trim=60 10 10 48,clip,scale = .17]{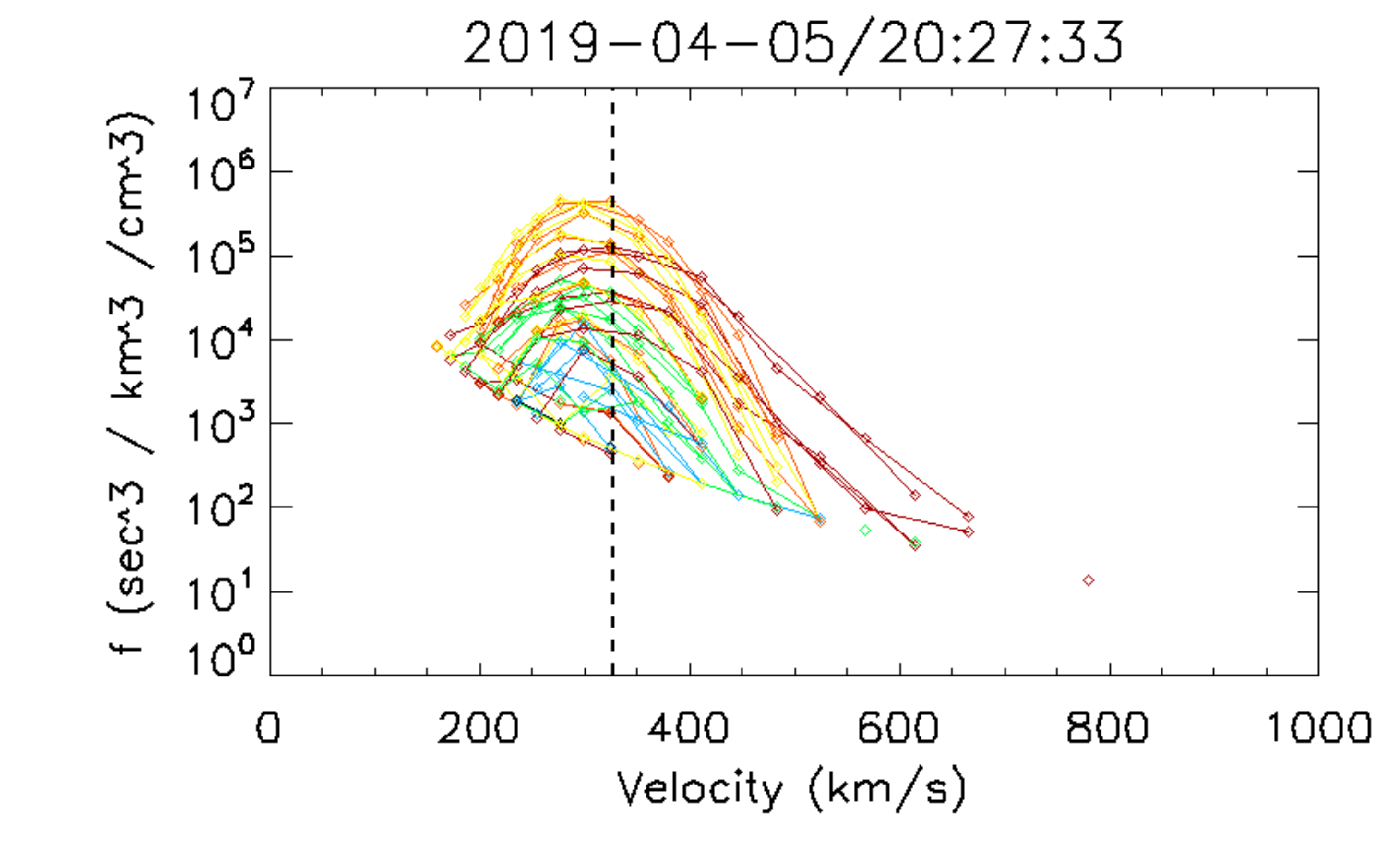} \hspace{.02in}
\includegraphics[trim=30 10 300 52,clip,scale=.17]{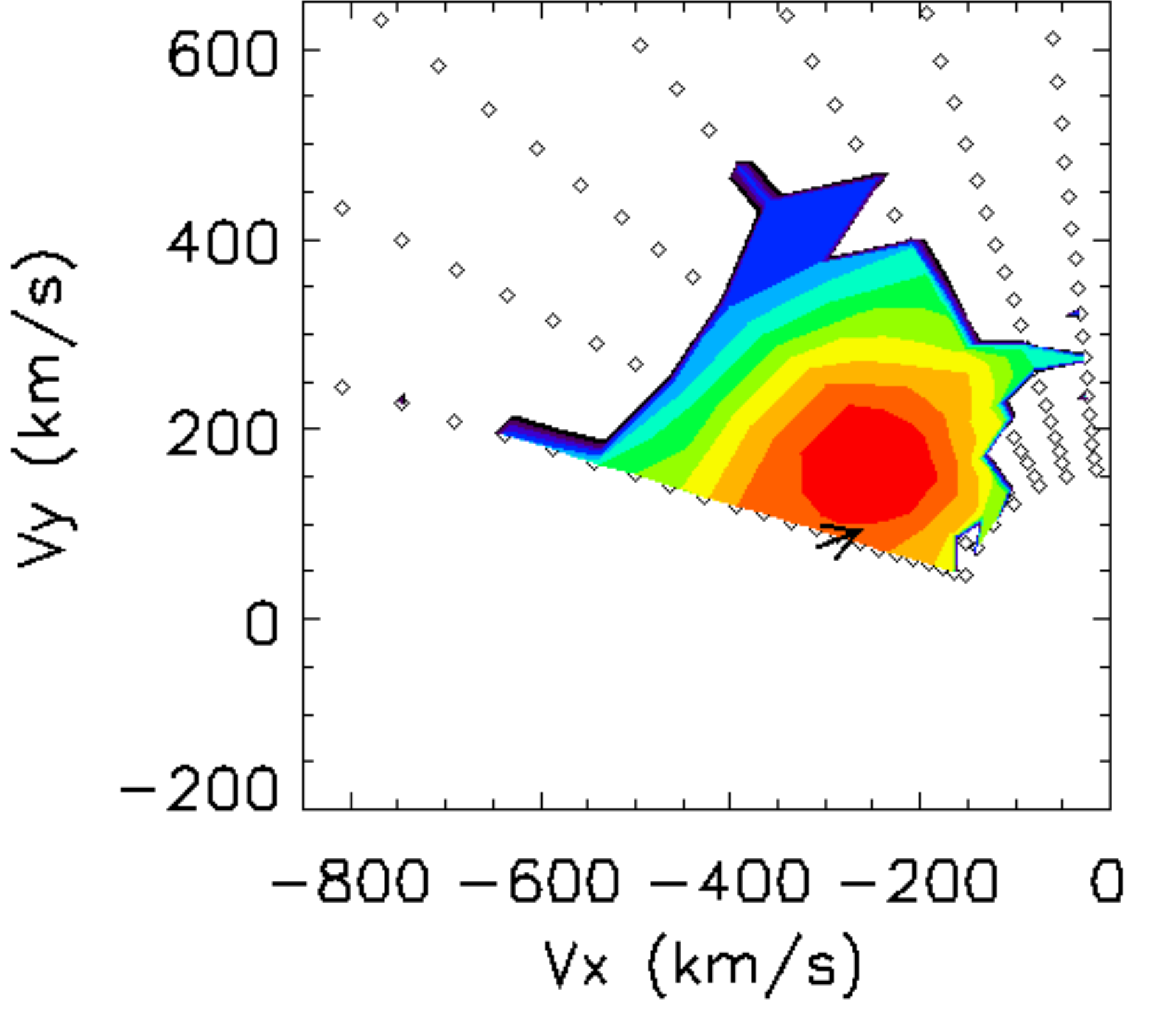}
\hspace{.02in}
\includegraphics[trim=30 10 290 52,clip,scale=.17]{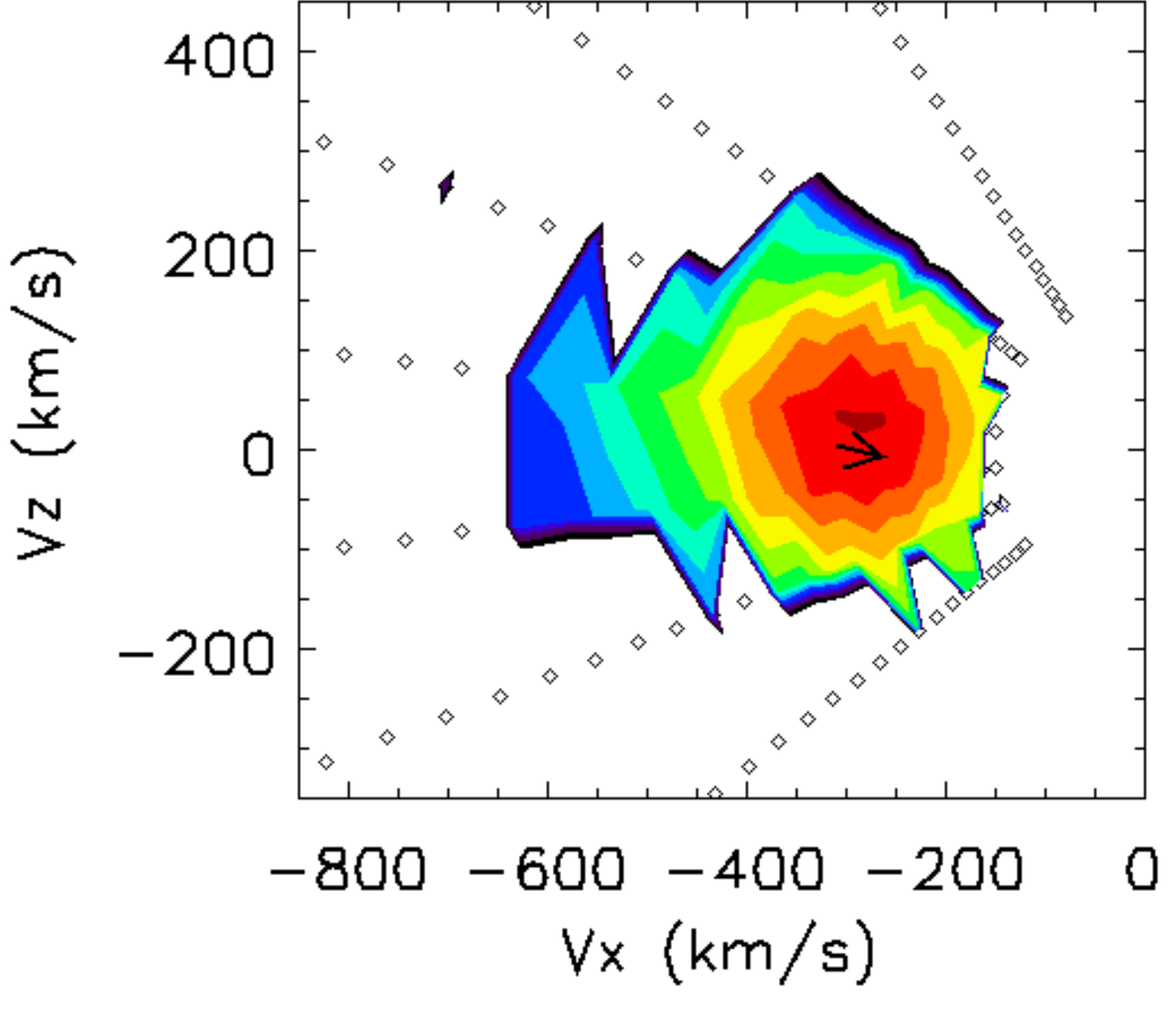}
\includegraphics[scale=.14]{fig2_legend.pdf}
\caption{Beam evolution for times indicated by the dashed black lines in \figref{fig:4_5_allvars}, with coordinates defined in \secref{sec:coord}. Left: Proton VDFs, where each line refers to an energy sweep at different elevation angles. The dashed vertical line represents \Alfven speed. Middle: VDF contour elevations which are summed and collapsed onto azimuthal plane. Right: VDF contour elevations which are summed and collapsed onto $\theta$ plane. The black arrow represents the magnetic field direction in SPAN-I coordinates, where the head is at the solar wind velocity (measured by SPC) and the length is the \Alfven speed.
\label{fig:4_5_vdfs}}
\end{figure*}

%[trim=left bottom right top, clip
\begin{figure*}[ht]
\hspace{.5in} (a) t3 = 2019-04-05/19:22:11
\hspace{2in} (b) t4 = 2019-04-05/19:36:37
\vfill
\includegraphics[trim=30 50 30 80, clip,scale=.35]{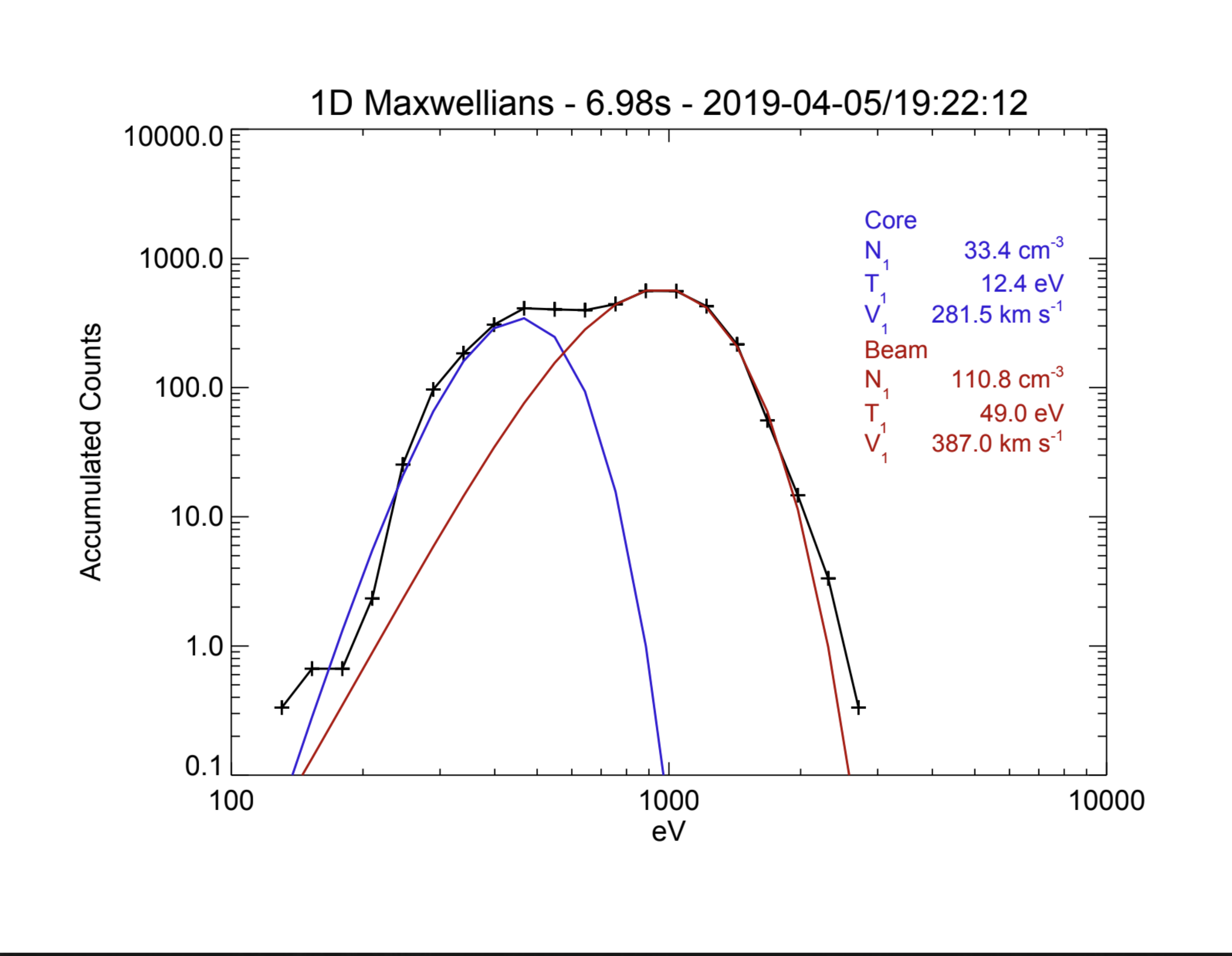}
\hfill
\includegraphics[trim=30 50 30 80, clip,scale=.35]{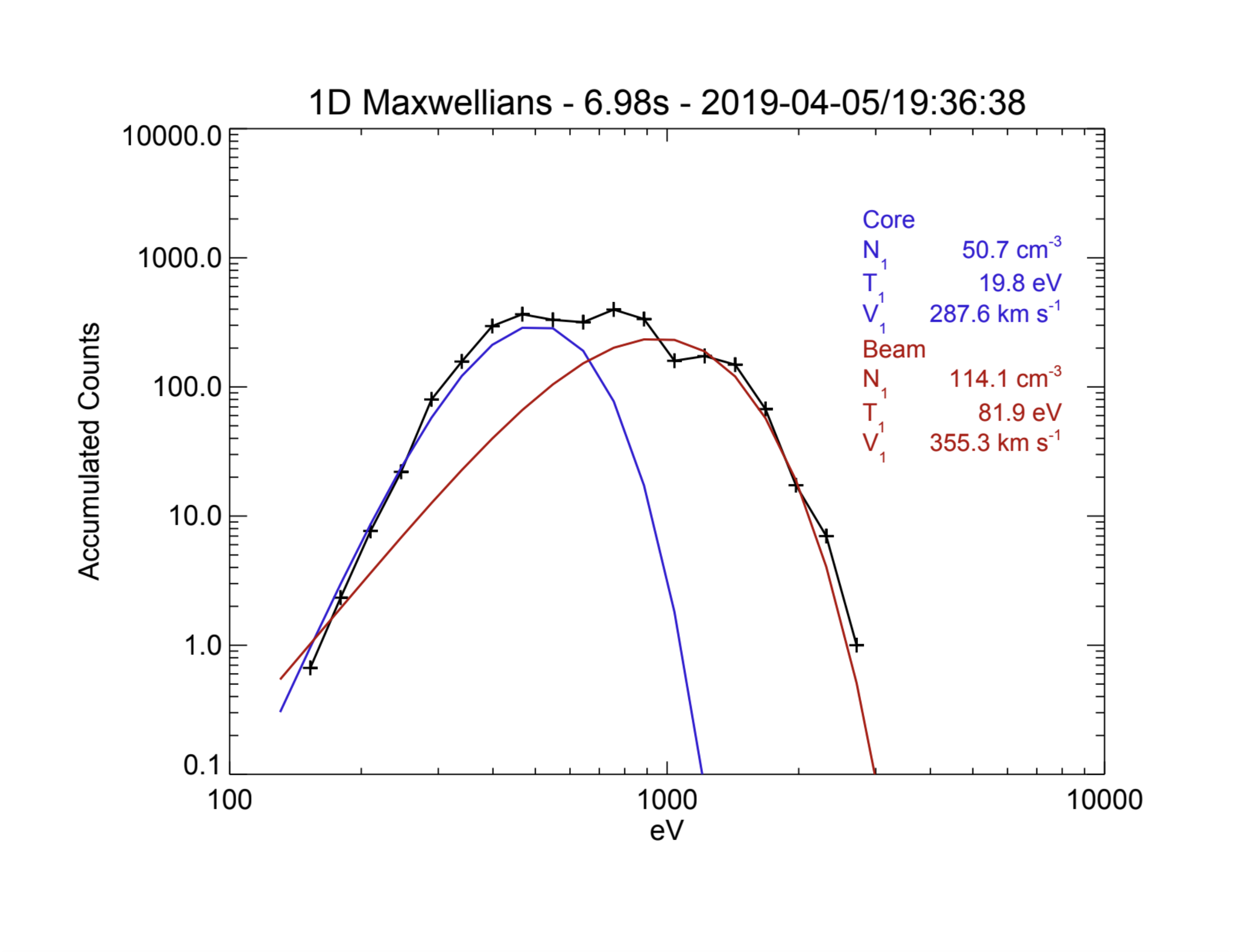}
\caption{Fits from 1D cuts of SPAN-I VDFs, along the direction of highest energy flux, at (a) t3 = 2019-04-05/19:22:11 and (b) t4 = 2019-04-05/19:36:37.}
\label{fig:fits}
\end{figure*}

The most prominent correlation event between a proton beam and an ion-scale wave storm occurred on 2019-04-05, which was during near-Perihelion of Encounter 2 at a distance of approximately 36.3 $R_s$. This event, hereafter called Event $\#$1, is demonstrated by \figref{fig:4_5_allvars}. The magnetic field information depicted in \figref{fig:4_5_allvars}(a)-(e) shows that the (a) quiet magnetic field was radially aligned, (b) wave vector, $\V{k}$ was propagating at angles $\theta_{kb}< 10^\circ$ with respect to $\V{B}$, (c) minimum variance coordinates depicted two nearly-identical eigenvalue components, with one being subdominant by 1-2 orders of magnitude, indicating ellipticity values close to 1 for near-circular polarization, (d) ion-scale wave signature as a narrowband frequency between 1 and 5 Hz (0.59$\Omega_i$ - 3.34$\Omega_i$), with peak wave power ($P_{Tot}$) of 7.98 dominant over the background magnetic field ($P_k$) at 2.4 Hz (1.41$\Omega_i$), and (e) LH polarization (blue) in the spacecraft frame. Here, $\Omega_i$ refers to the local ion gyrofrequency. \figref{fig:4_5_allvars}(f) shows a strong enhancement of differential energy flux, measured by SPAN-I, that was present simultaneous with the wave event. Note that the intermittent values of $\theta_{kb}$ above $20^\circ$ in \figref{fig:4_5_allvars}(b) correspond to intermittent periods of zero wave power in \figref{fig:4_5_allvars}(d). The frequencies and polarizations in \figref{fig:4_5_allvars}(d) and (e), respectively, are not Doppler-shifted to the plasma frame (see \secref{sec:stab}), and the white dashed-dotted line displayed in \figref{fig:4_5_allvars}(d) and (e) corresponds to the local proton gyrofrequency at $\approx$1.7 Hz. 

To obtain initial proton beam and core plasma parameters by SPAN-I, we perform 1D Maxwellian fits along the direction of highest energy flux, described in \secref{sec:anal}; one population is fitted to the peak phase-space density of the VDF, interpreted as the core; the second population moving outside this region is fitted to the tail of the VDF, interpreted as the beam. \figref{fig:4_5_allvars}(g)-(i) show results of these fits: (g) beam-to-core density ratio $\left(n_b/n_c\right)^*$, (h) temperature of the beam $T_b^*$ (blue) and core $T_c^*$ (red), and (i) relative beam-to-core drift velocity $v_D^*=v_b^*-v_c^*$ (blue). Note that we omitted  $\chi$ values greater than 5. 

We see from \figref{fig:4_5_allvars}(g) that $\left(n_b/n_c\right)^*$ was well-correlated with the enhanced proton differential energy flux (interpreted as the proton beam), increasing up to $\approx$5. \figref{fig:4_5_allvars}(h) shows that the beam temperature $T_b^*$ (blue) increased in both the beginning and end of the event, with decreased values during the middle time period. Meanwhile, the core temperature $T_c^*$ (red) stayed roughly the same. \figref{fig:4_5_allvars}(i) shows the \Alfven speed (red), SPAN-I measured moments of $\alpha$-proton differential speed (green), along with the $v_D^*$ fits (blue). Note that the sudden decrease in the \Alfven speed after t5 was most likely due to a type III radio burst obstructing the density measurement from quasithermal noise. The $\alpha$ parameters were obtained by SPAN-I moments after appropriately removing contamination \citep{McManus:2020inprep,Livi:2020inpress}. \figref{fig:4_5_allvars}(j) shows the total temperature anisotropy (obtained by SPAN-I measured temperature moments), with $T_\perp$ in blue and $T_\parallel$ in red.  We see that at t1, right before the differential energy flux enhancement, $T_\perp/T_\parallel\approx 2$. Then, once the differential energy flux began to increase significantly at t2, $T_\parallel$ grew larger than $T_\perp$, with the maximum parallel temperature anisotropy at $T_\perp/T_\parallel \approx 0.35$. While this could be seen as parallel thermalization correlated with the beam, our interpretation is limited by SPAN-I measured partial moments of total temperature anisotropy, rather than separate fitted beam and core components. \figref{fig:4_5_allvars}(k) shows the $\alpha$ density, indicative of an anti-correlation with $n_b/n_c$ during the midpoint of the most intense period of the differential energy flux. This suggests that it was indeed protons that were the dominant species involved in the possible wave-particle interaction event.

By inspecting the time series of the proton VDF in the left panels of \figref{fig:4_5_vdfs}, we can see that at the beginning (\figref{fig:4_5_vdfs}(b)) and end (\figref{fig:4_5_vdfs}(e)) times of the event, the beam was clearly resolvable and well-separated from the core. As the beam intensity increases in \figref{fig:4_5_allvars}(c) and (d), the beam-to-core drift velocity, $v_D^*$ decreases. Note that the dashed black vertical line indicates the \Alfven velocity. As shown in both panel \figref{fig:4_5_allvars}(i) and the left panels of \figref{fig:4_5_vdfs}(b)-(d), $v_D^*$ is moving approximate to the \Alfven speed at times t2-t4. Near the end of the event at t5, \figref{fig:4_5_vdfs}(e) shows that $v_D^*$ well-surpasses $v_A$, with $\left(v_D/v_{A,c}\right)^* \approx 1.67$, where $\left(v_{A,c}\right)^*$ is the \Alfven velocity associated with the fitted core proton population.

The middle panels of \figref{fig:4_5_vdfs} show contours of the VDFs, averaged over the azimuthal $(\phi)$ dimension, at their respective times (see \secref{sec:coord} for coordinate definitions). These time series plots best represent how much of the VDF was in SPAN-I's FOV. The right panels of \figref{fig:4_5_vdfs} show contours of the VDFs, averaged over the elevation angle ($\theta$) direction. The black arrow shows the direction of the magnetic field, where the head is placed at the solar wind velocity (measured my SPC) and the length is the \Alfven speed. \figref{fig:4_5_vdfs}(a) shows that at t1, before the wave event when $T_\perp/T_\parallel > 1$, we observed a small super-\Alfvenic beam from the left panel, with a mostly sub-\Alfvenic core distribution. As the beam becomes present, the time series depicted in \figref{fig:4_5_vdfs} confirms that the beam is nearly field-aligned, moving about the \Alfven speed during t2-t4 (\figref{fig:4_5_vdfs}(b)-(d)), and becomes super-\Alfvenic by t5 (\figref{fig:4_5_vdfs}(e)). Note that the drift speed between the core and beam is sufficiently large to appear as two separate peaks in the right contour plot in \figref{fig:4_5_vdfs}(e). Directly after the wave storm event at t6, there was a strong increase in differential energy flux at lower energy values, shown in \figref{fig:4_5_allvars}(f). Inspecting the VDF at this time, \figref{fig:4_5_vdfs}(f) shows the plasma returning to a more equilibrium state and most of the core came into SPAN-I's FOV. This is seen by the concentric circles at $v_x\approx 330$ $kms^{-1}$ in the contour plot in the middle panel. 

The unexpectedly large beam-to-core density ratios are best justified from \figref{fig:fits}, which show 1D cuts of the two populations in the direction of highest energy flux. At t3 = 2019-04-05/19:22:11 (also corresponding to \figref{fig:4_5_vdfs}(c)), the 1D fits are shown in \figref{fig:fits}(a), indicating that $n_b^* = 110.8$ $cm^{-3}$ and $n_c^* = 33.4$ $cm^{-3}$, yielding $\left(n_b/n_c\right)^* \approx 3.3$. At t4=2019-04-05/19:36:37 (also coinciding to \figref{fig:4_5_vdfs}(d)), one can observe a third defined peak centered around 1000 eV. This can also be seen in the left panel of \figref{fig:4_5_vdfs}(d). It is unclear whether this third peak is part of the secondary proton beam, or should be treated as an entirely distinct tertiary ion beam. For the present time, we treat it as part of the secondary proton beam, although this leads to some error in overcompensating $n_b^*$. In future Encounters, we will look for other tertiary peaks to determine the best treatment for conducting the fits. 

\subsection{Event 2: April Fools} \label{sec:e2}
The event presented in \secref{sec:e1} showed a large wave storm with exclusively left-handed polarization in the spacecraft frame. In this section, we show a second event, which occurred on 2019-04-01 (hereafter called Event $\#$ 2) at approximately 44.6 $R_s$, highlighting observations of both right and left-handed polarization periods.
\begin{figure*}
\centering
\includegraphics[trim=5 0 5 0,clip,scale=1]{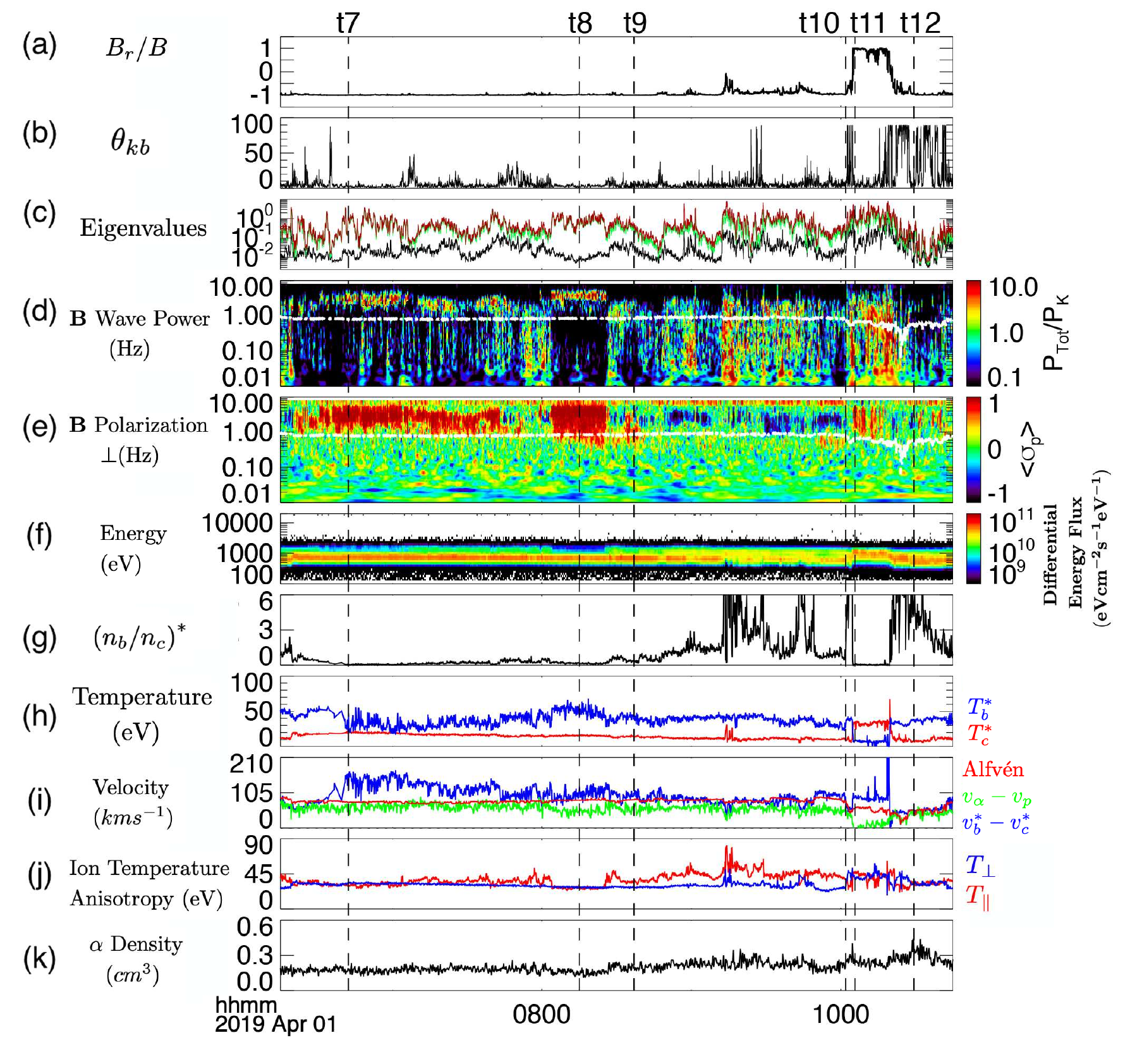}
\caption{Example event on 2019-04-01 (Event $\#$ 2) featuring both LH (blue) and RH (red) ion-scale wave polarizations in the spacecraft frame. Shown is the (a) radial magnetic field component, (b) angle of wave propagation w.r.t. $\V{B}$, (c) eigenvalues from MVA, (d) wavelet transform of $\V{B}$, (e) perpendicular polarization of $\V{B}$, (f) SPAN-I measured moment of differential energy flux, (g) proton beam-to-core density ratio fits, (h) temperature fits of proton beam (blue) and core (red), (i) proton beam-core differential velocity fits (blue) and $\alpha$-proton (green) differential velocity SPAN-I moments compared to the \Alfven velocity (red), (j) SPAN-I measured moments of temperature anisotropy, and (k) SPAN-I measured $\alpha$ density moments. In panels (d) and (e), the white dashed-dotted line represents the local proton gyrofrequency.\label{fig:4_1_allvars}}
\end{figure*}

\begin{figure*}
    \centering
    \hspace{.05in} (a) t7 = 2019-04-01/06:42:20 
\vfill
\vspace{.05in}
\includegraphics[trim=60 10 10 48,clip,scale = .17]{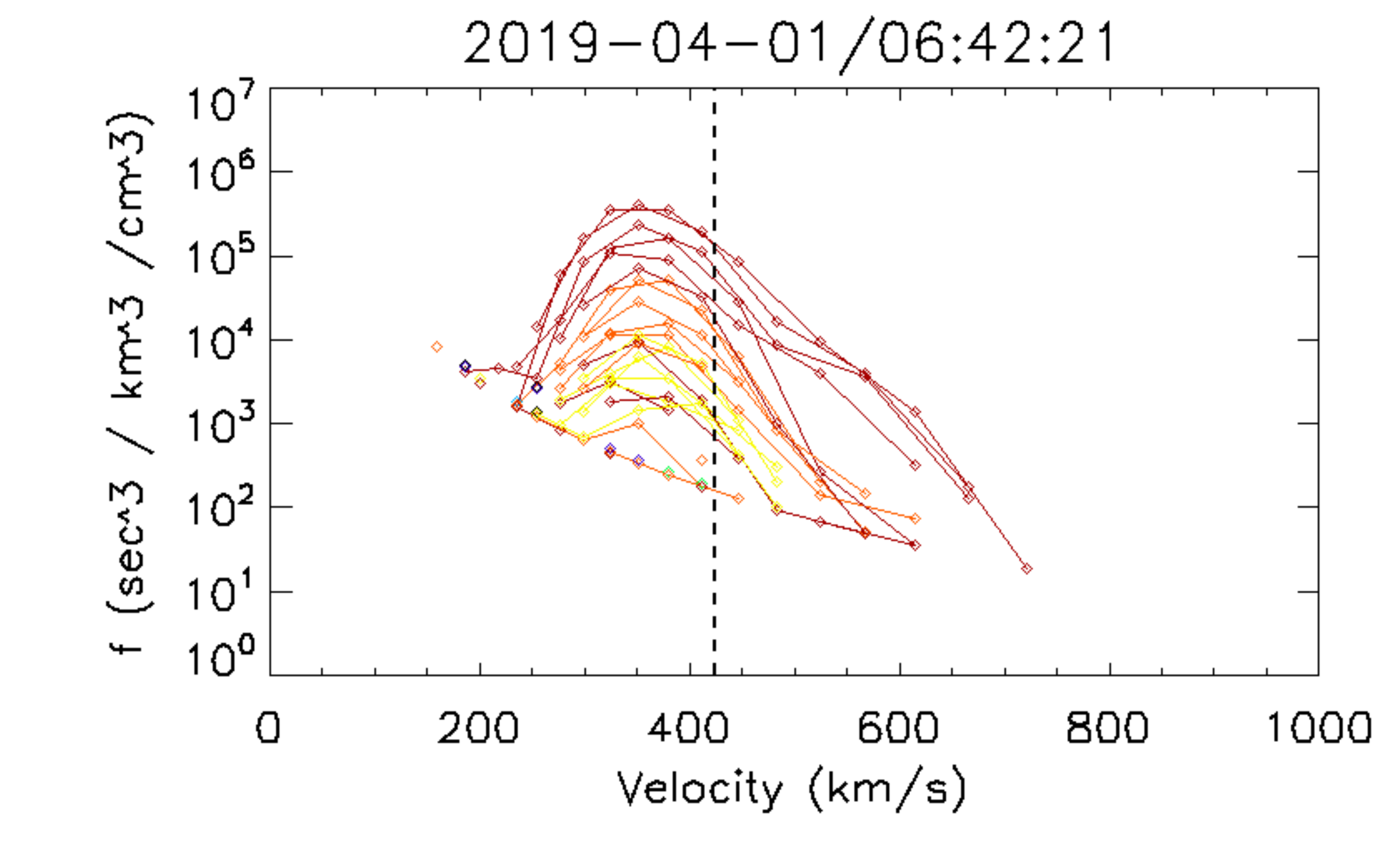} \hspace{.02in}
\includegraphics[trim=30 10 300 52,clip,scale=.17]{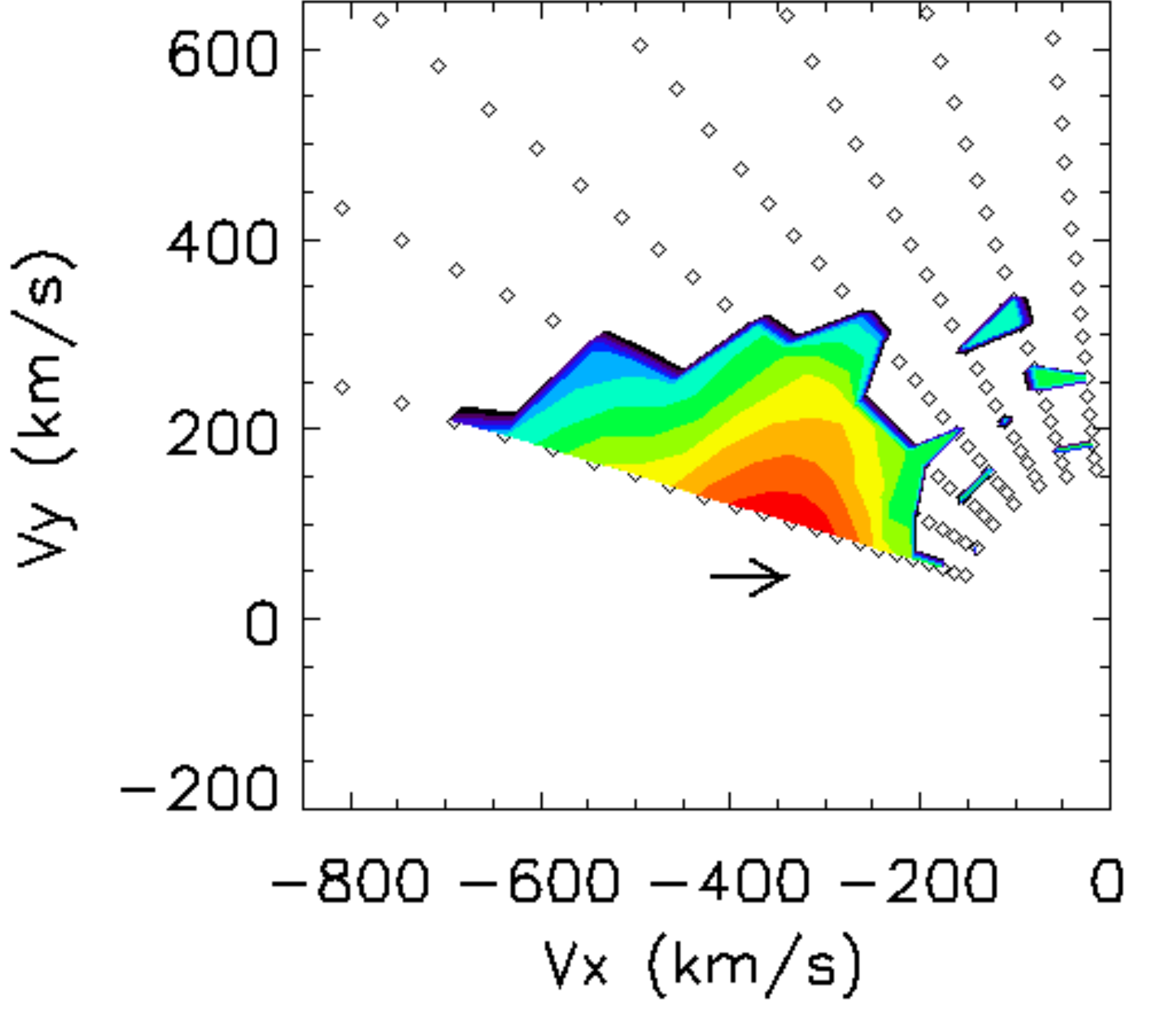}
\hspace{.02in}
\includegraphics[trim=20 10 290 52,clip,scale=.17]{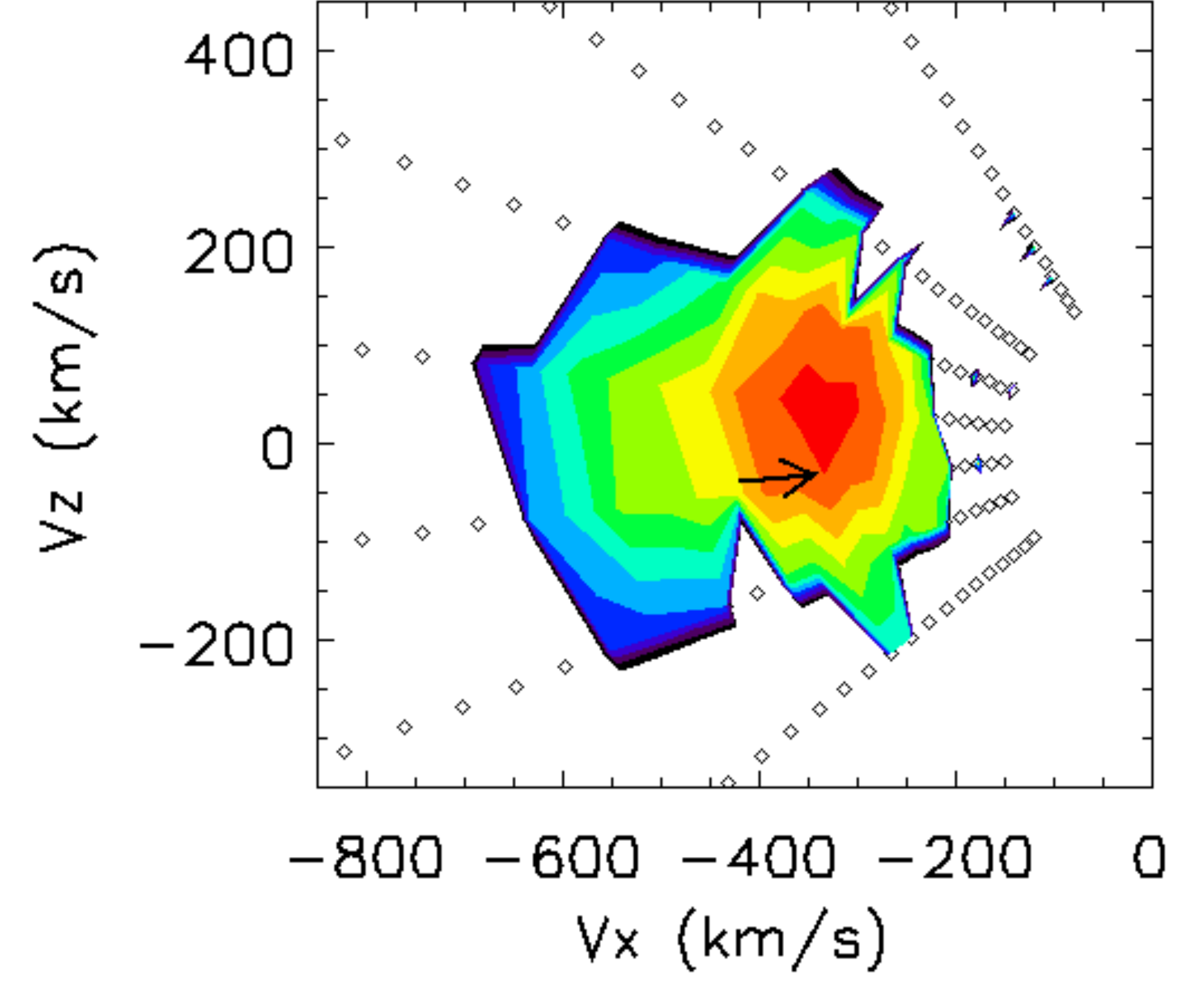}
\includegraphics[scale=.14]{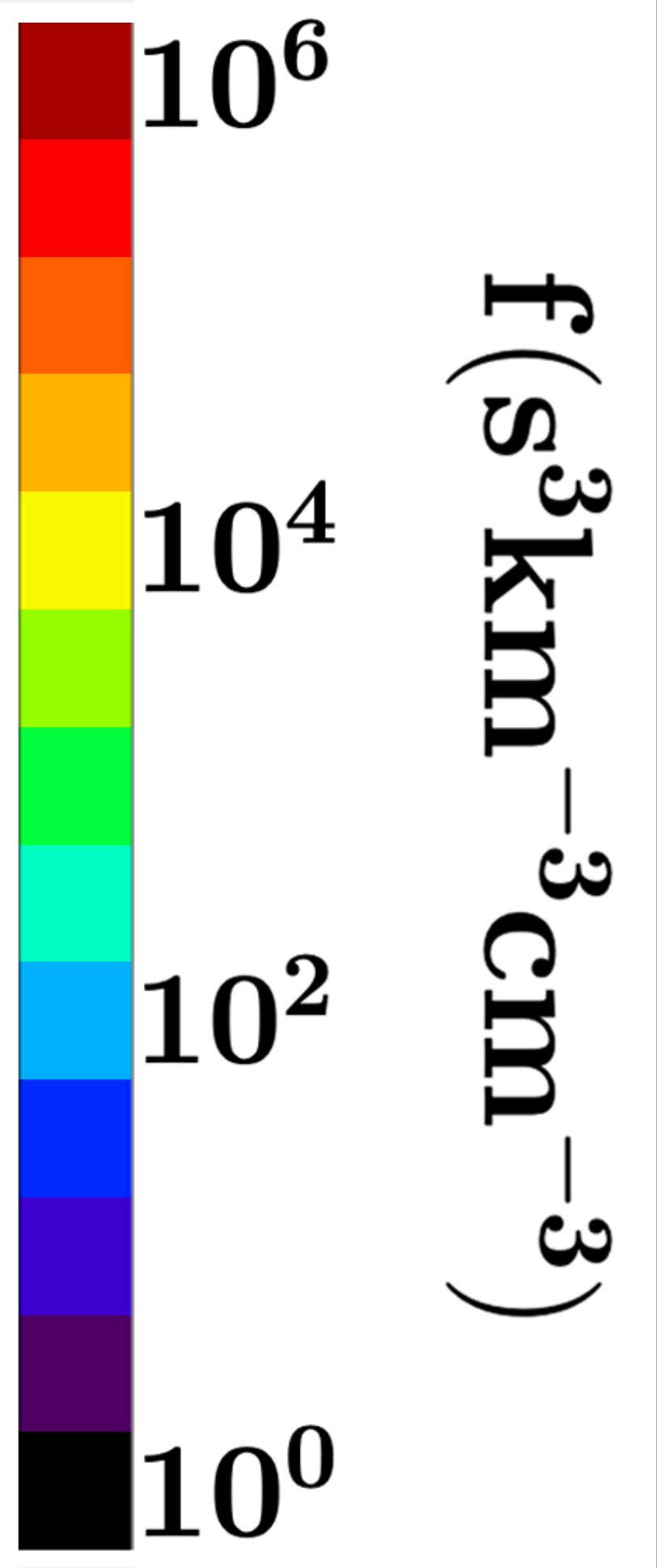}
\vfill

\hspace{.05in} (b) t8 = 2019-04-01/08:14:59 
\vfill
\includegraphics[trim=60 10 10 48,clip,scale = .17]{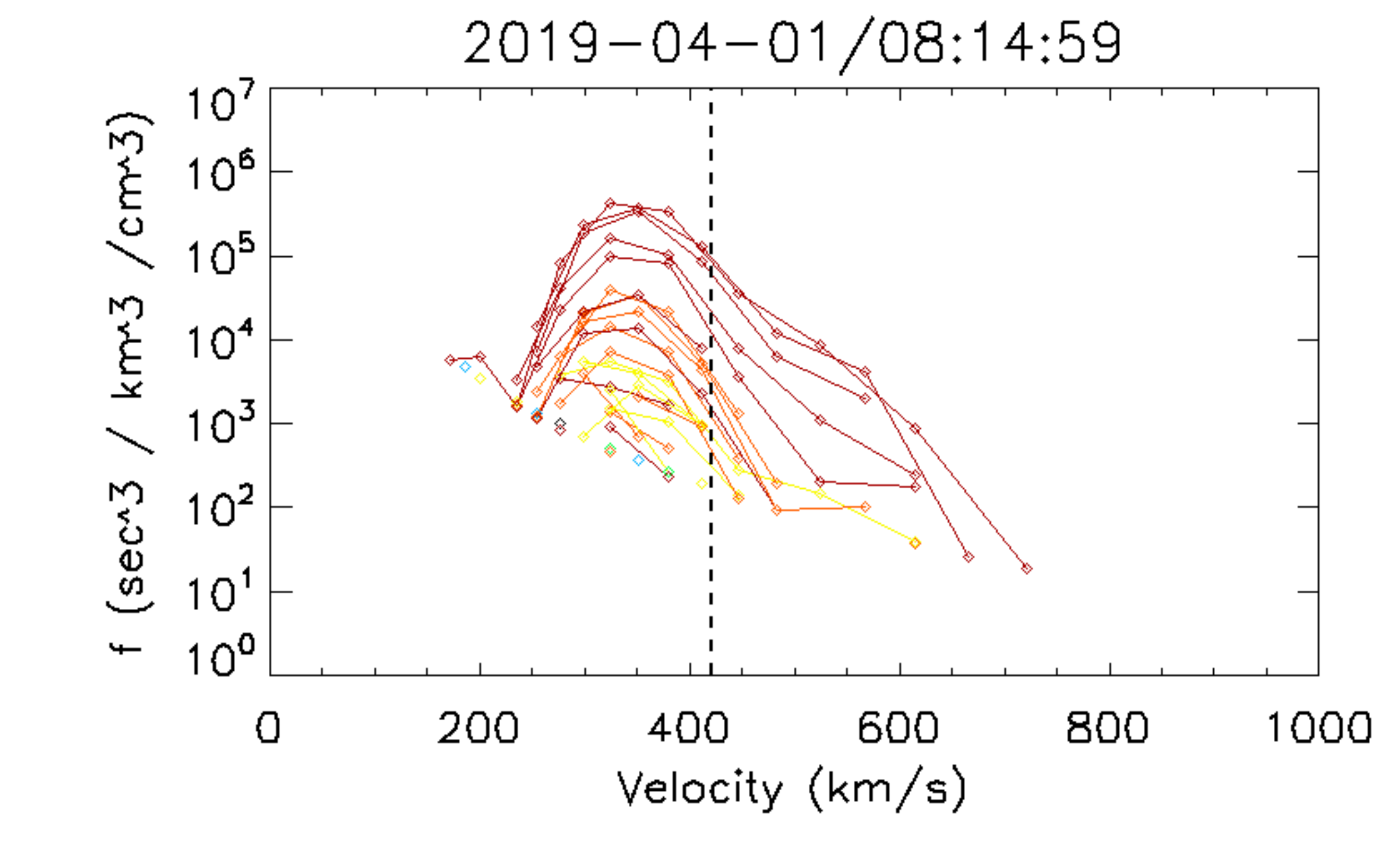} \hspace{.02in}
\includegraphics[trim=30 10 300 52,clip,scale=.17]{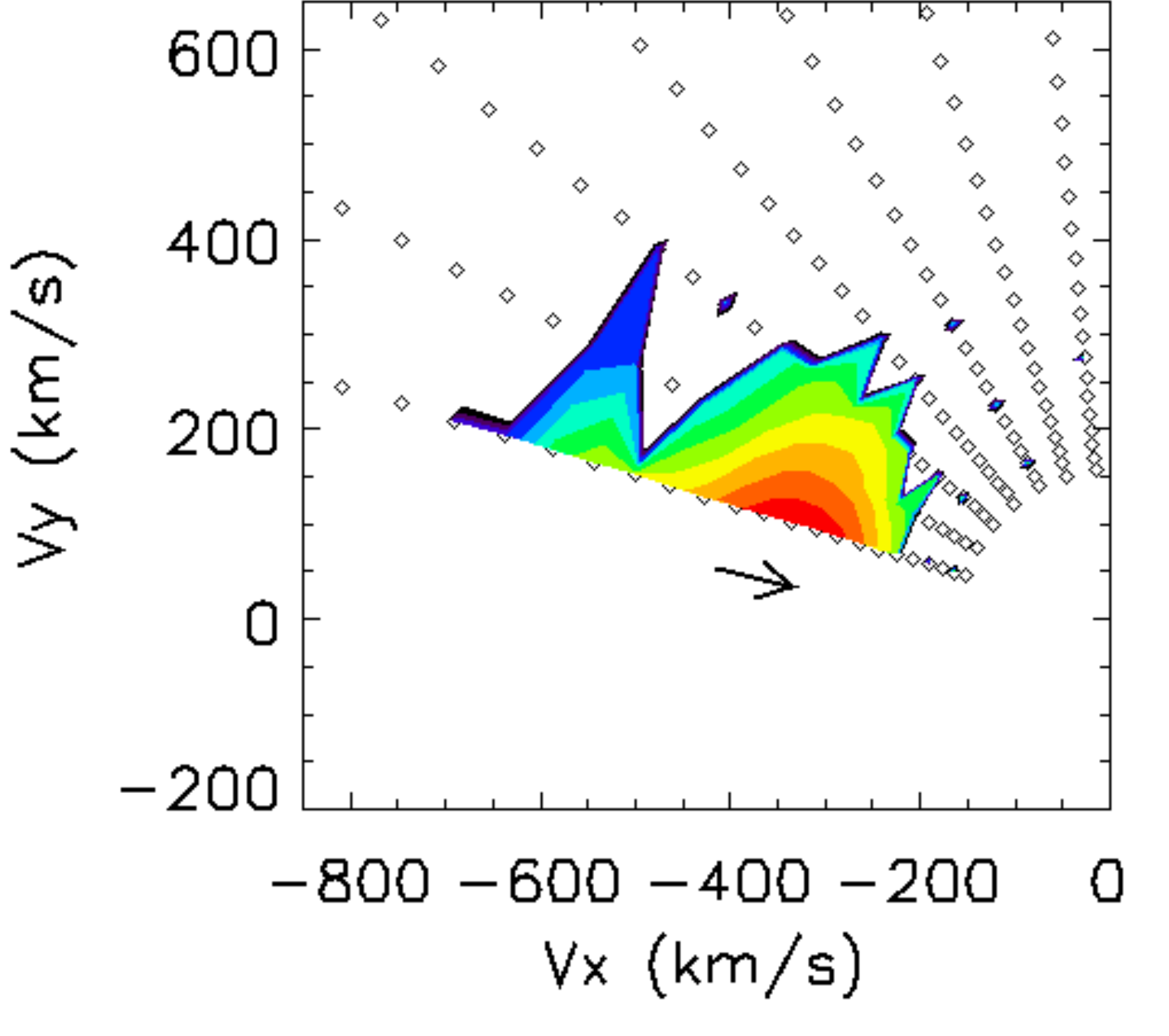}
\hspace{.02in}
\includegraphics[trim=30 10 290 52,clip,scale=.17]{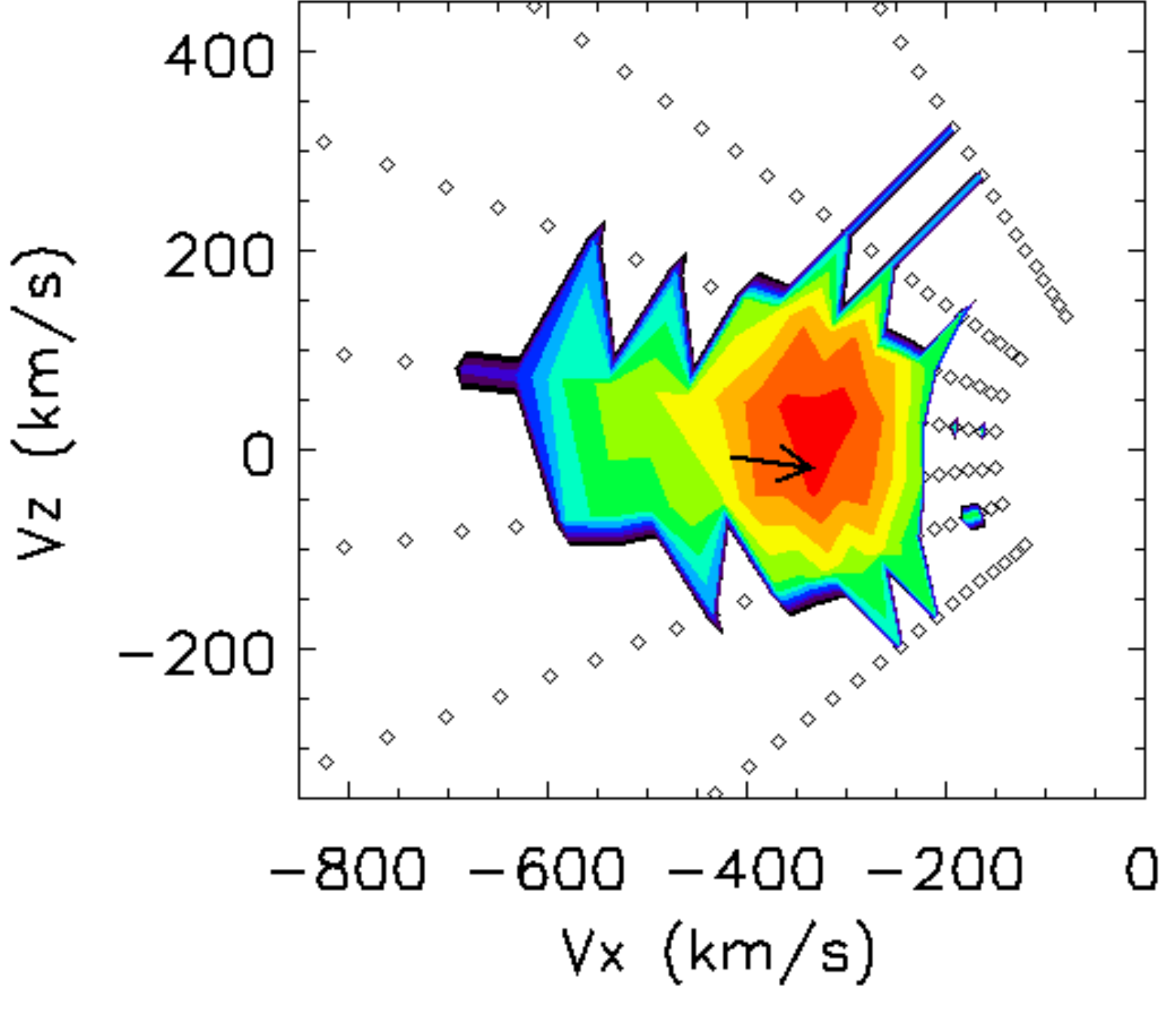}
\includegraphics[scale=.14]{fig5_legend.pdf}
\vfill

\hspace{.05in} (c) t9 = 2019-04-01/08:37:00 
\vfill

\includegraphics[trim=60 10 10 48,clip,scale = .17]{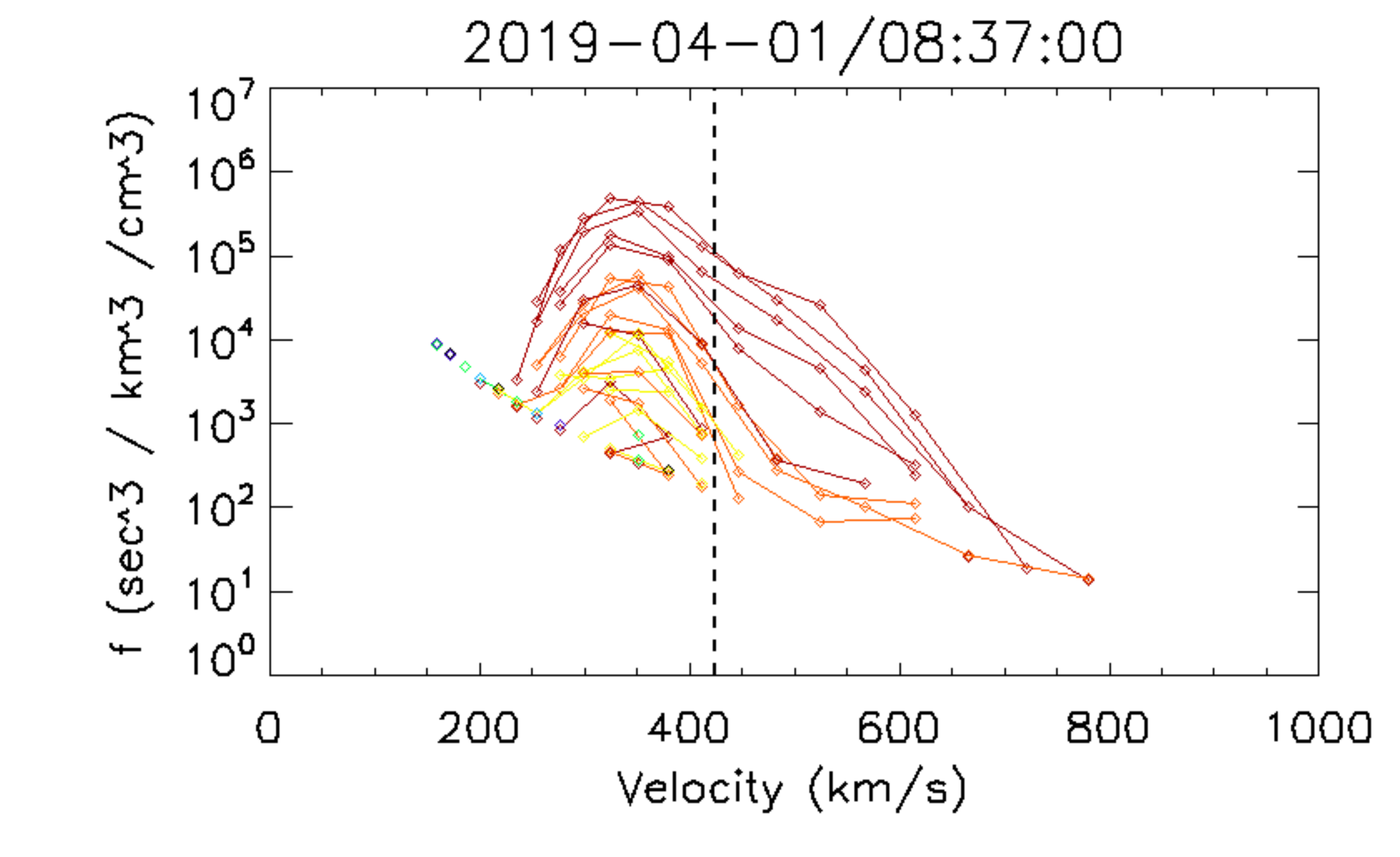} \hspace{.02in}
\includegraphics[trim=30 10 300 52,clip,scale=.17]{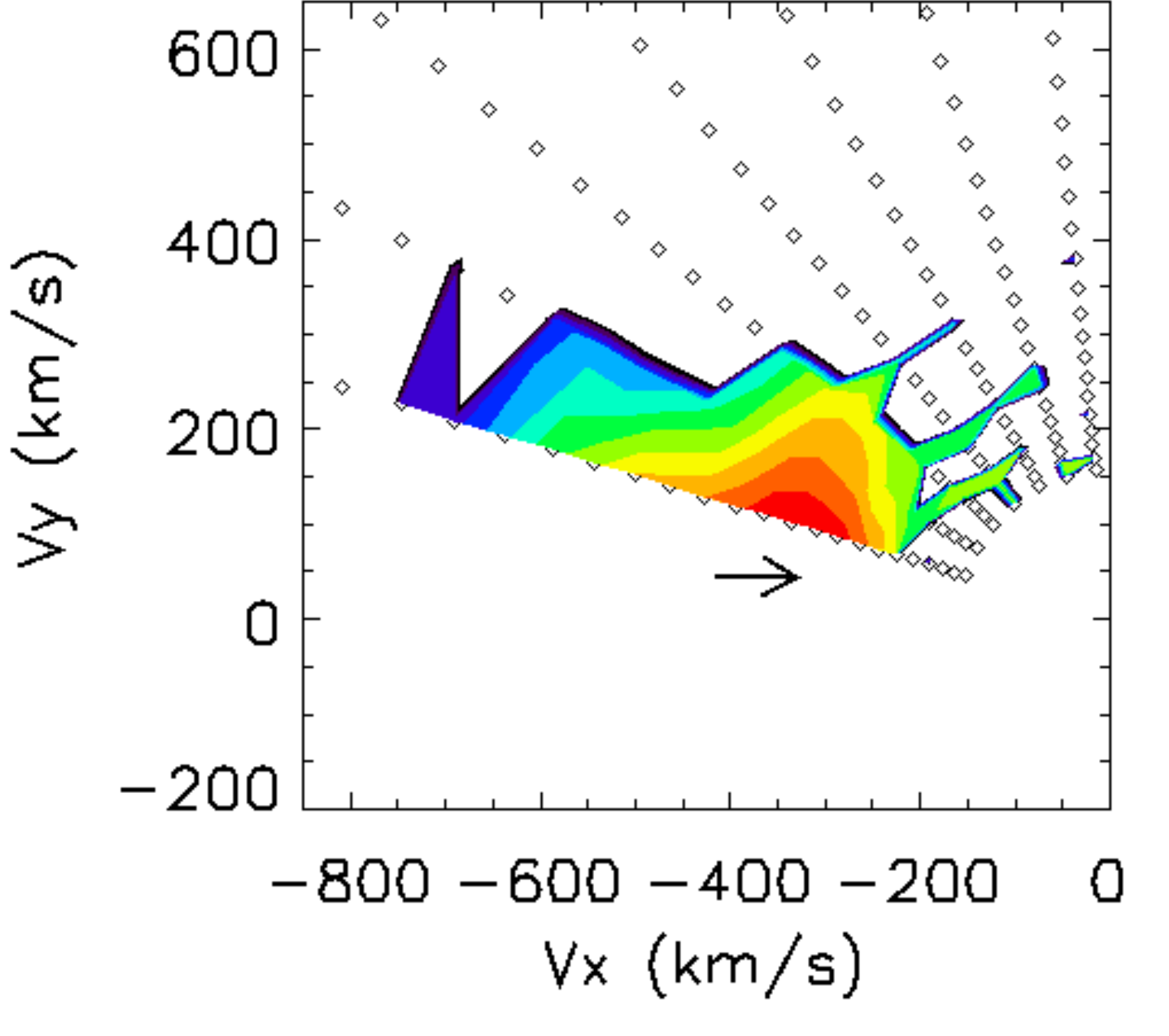}
\hspace{.02in}
\includegraphics[trim=30 10 290 52,clip,scale=.17]{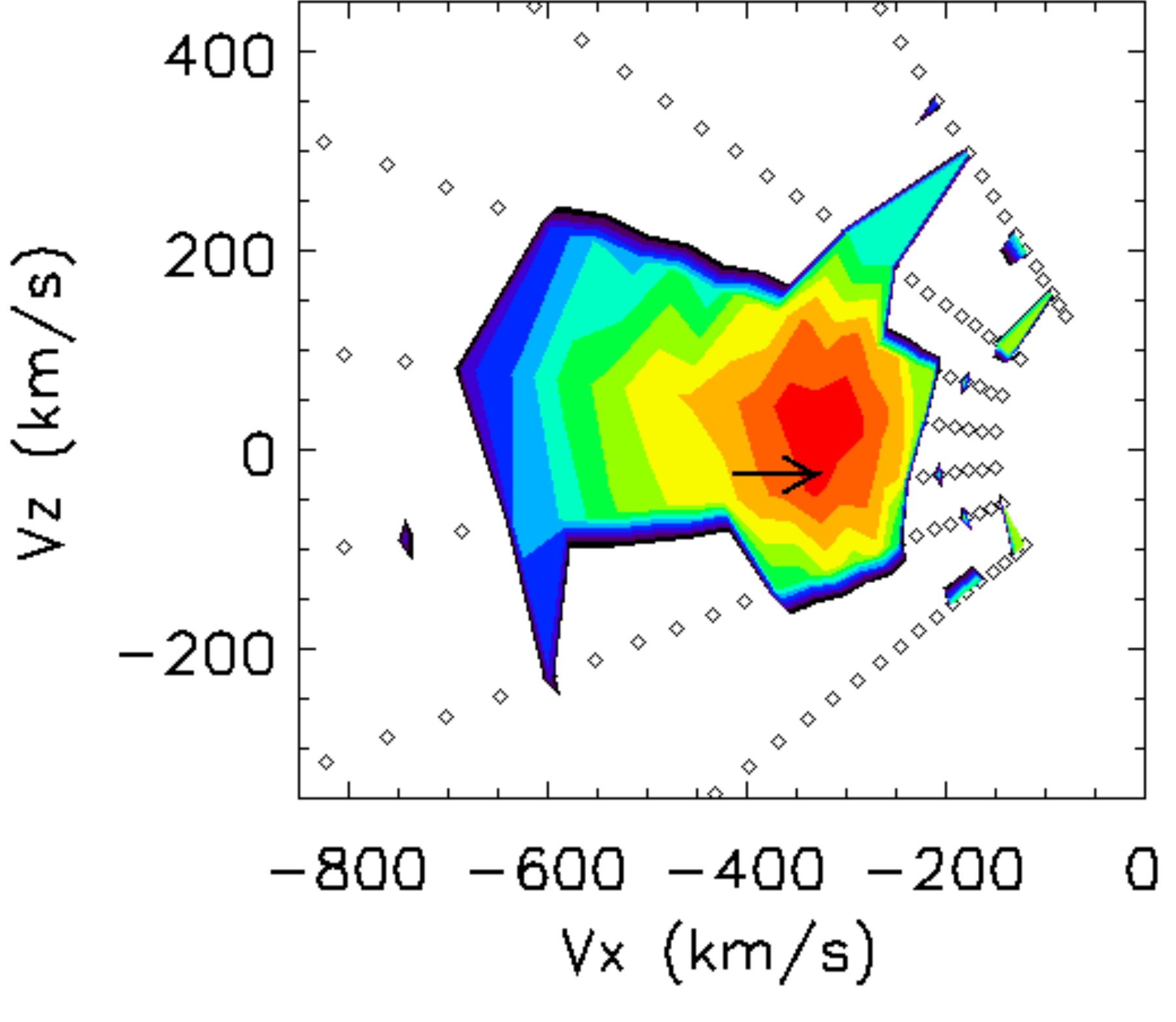}
\includegraphics[scale=.14]{fig5_legend.pdf}
\vfill

\hspace{.05in} (d) t10 = 2019-04-01/10:01:56 
\vfill
\includegraphics[trim=60 10 10 48,clip,scale = .17]{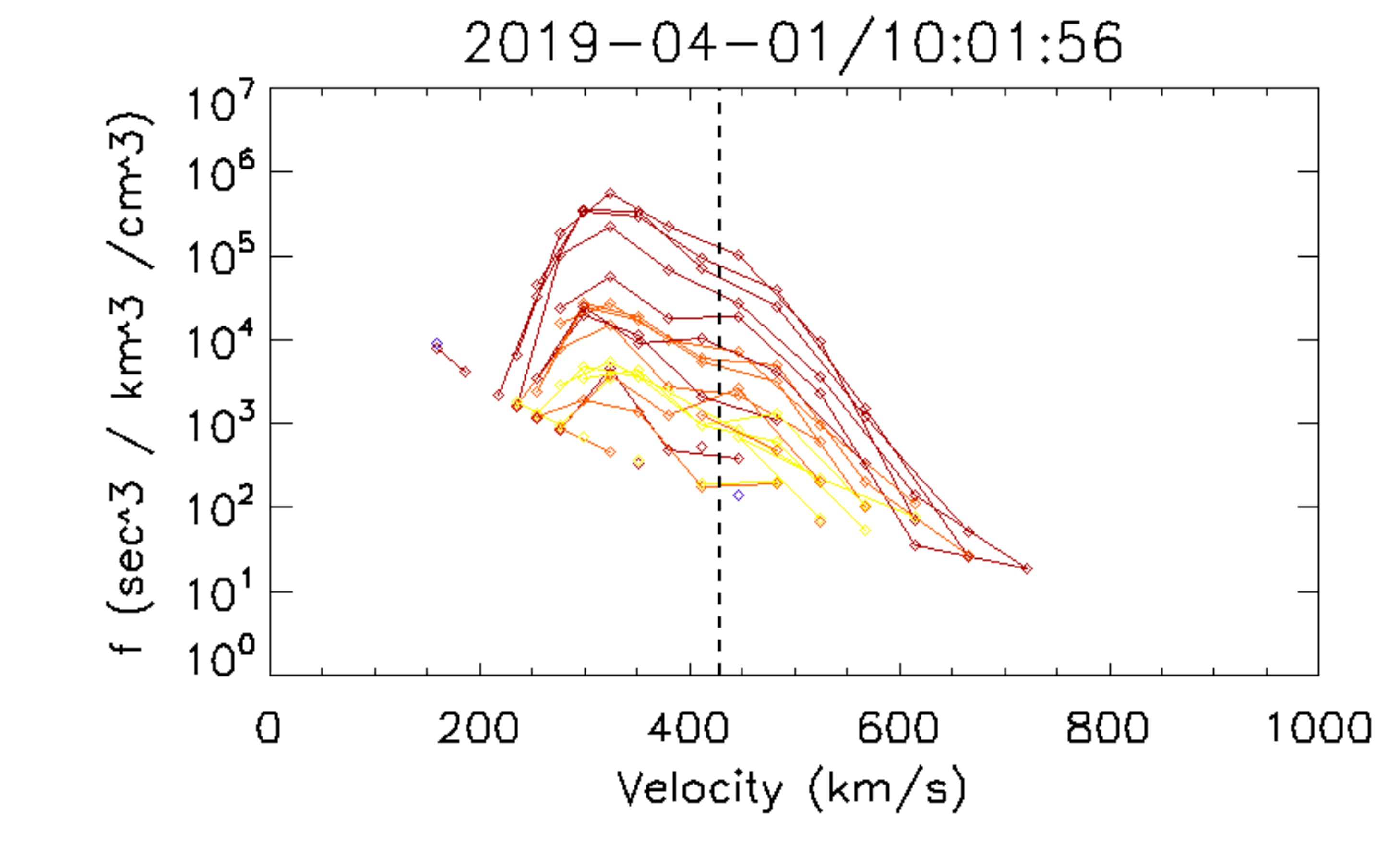} \hspace{.02in}
\includegraphics[trim=30 10 300 52,clip,scale=.17]{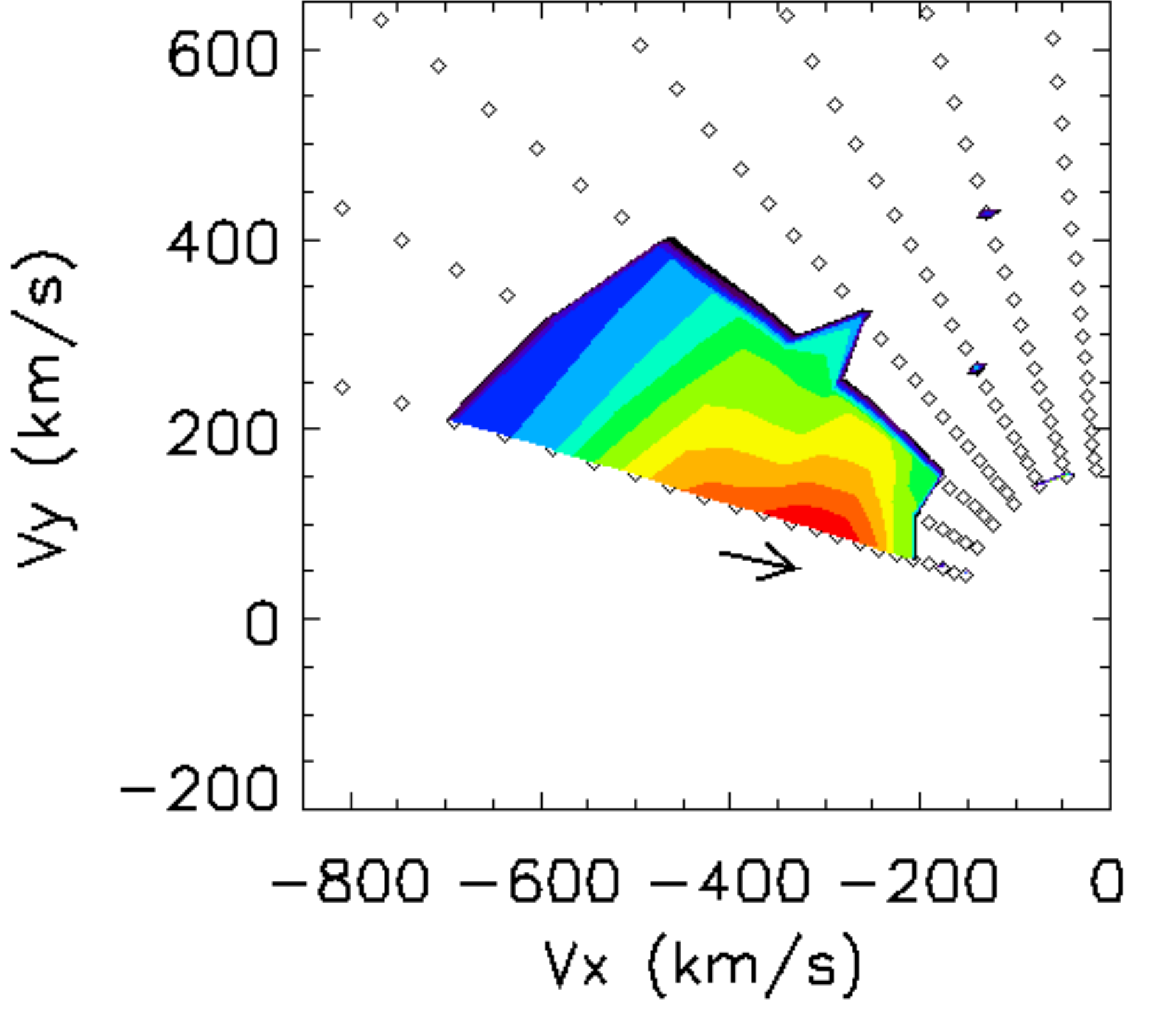}
\hspace{.02in}
\includegraphics[trim=30 10 290 52,clip,scale=.17]{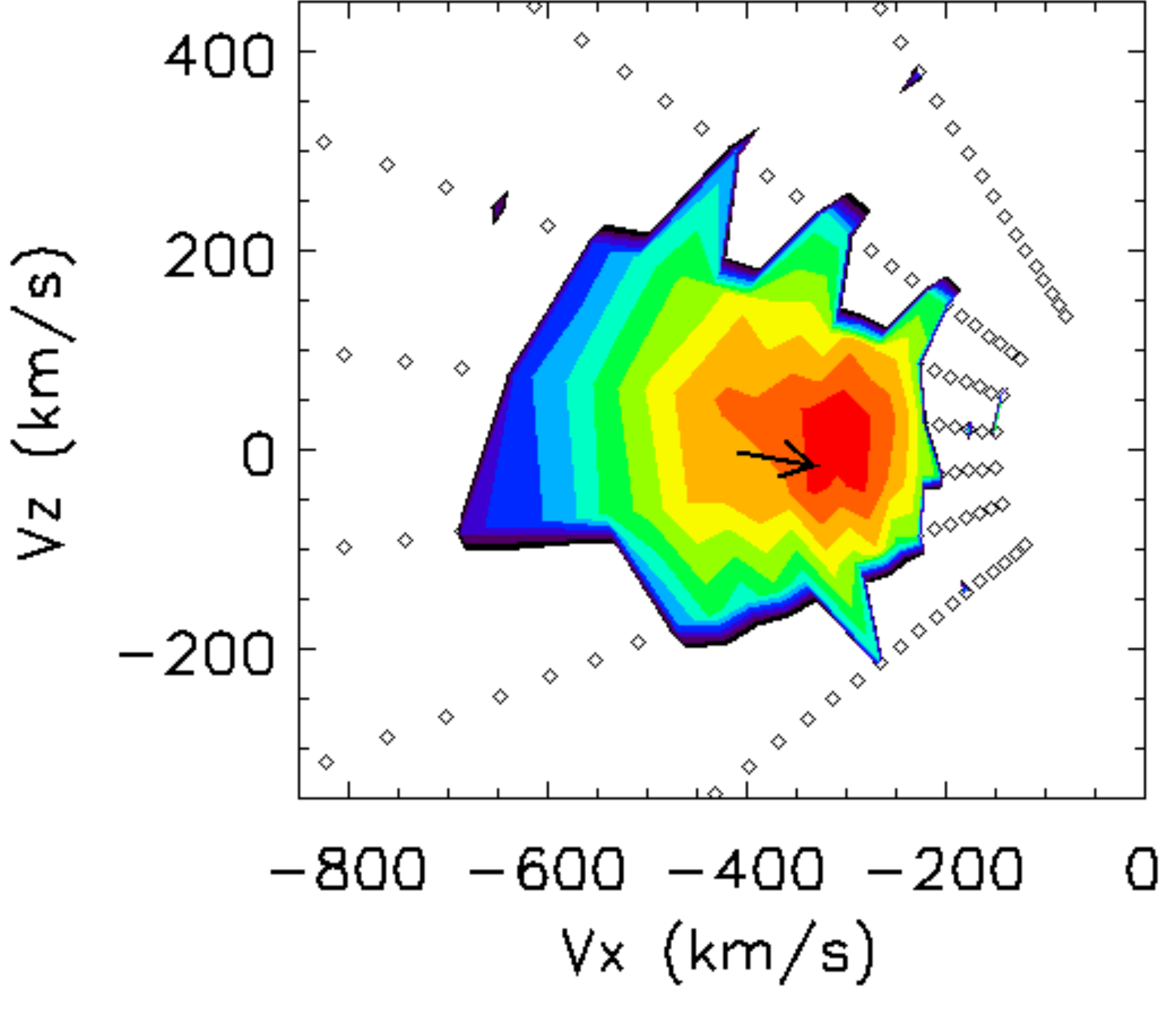}
\includegraphics[scale=.14]{fig5_legend.pdf}
\vfill

\hspace{.05in} (e) t11 = 2019-04-01/10:05:40 
\vfill
\includegraphics[trim=60 10 10 48,clip,scale = .17]{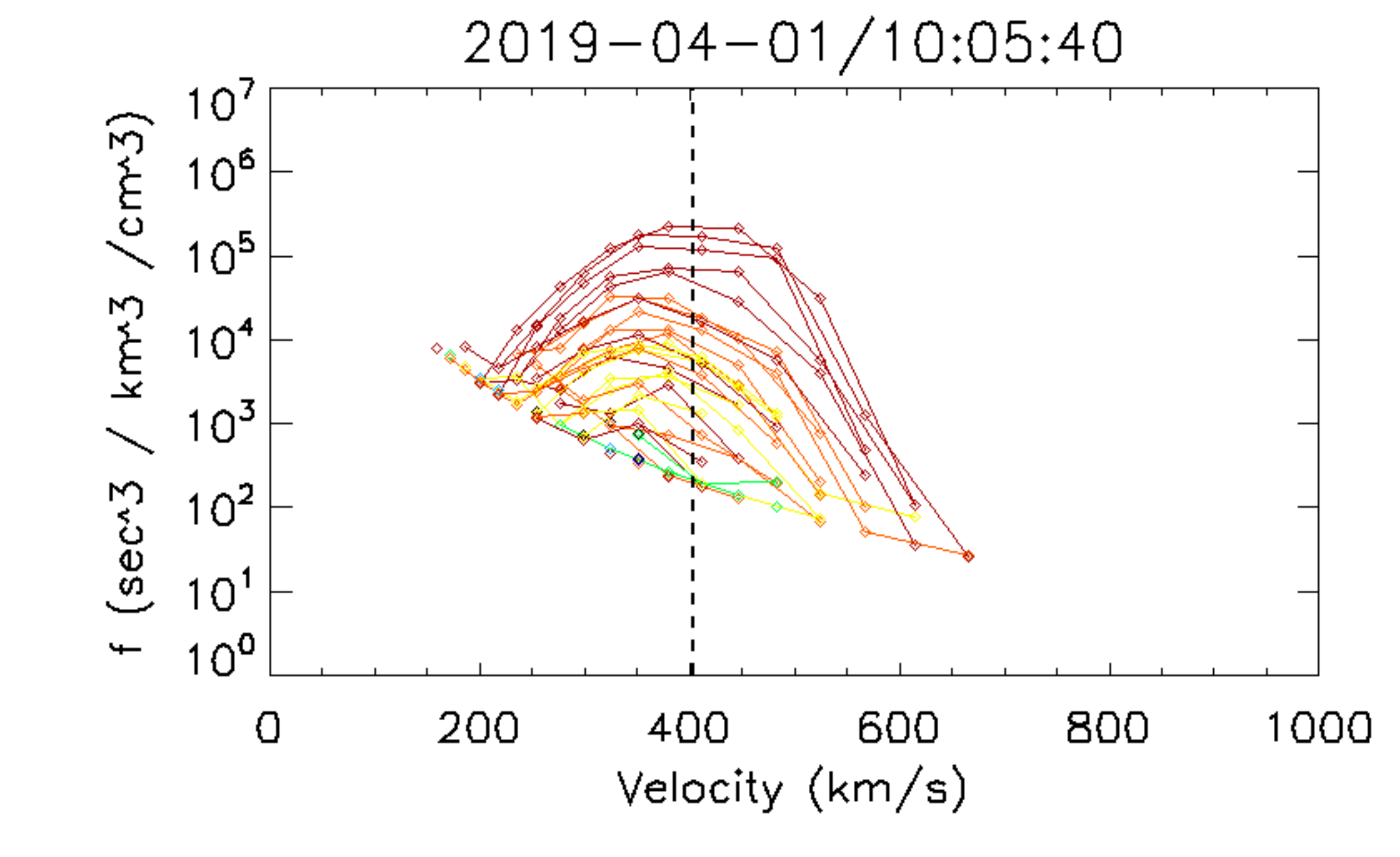} \hspace{.02in}
\includegraphics[trim=30 10 300 52,clip,scale=.17]{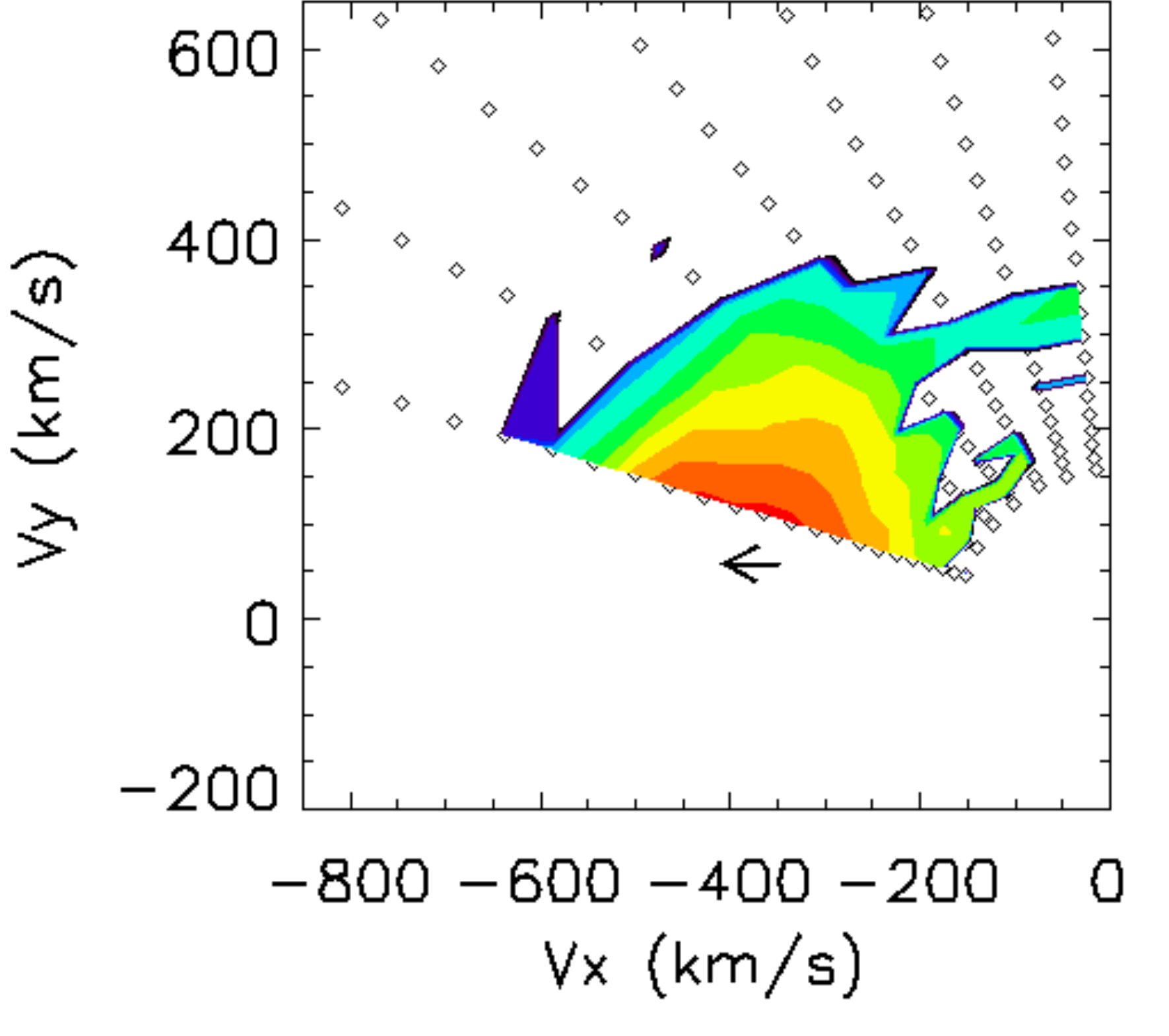}
\hspace{.02in}
\includegraphics[trim=30 10 300 52,clip,scale=.17]{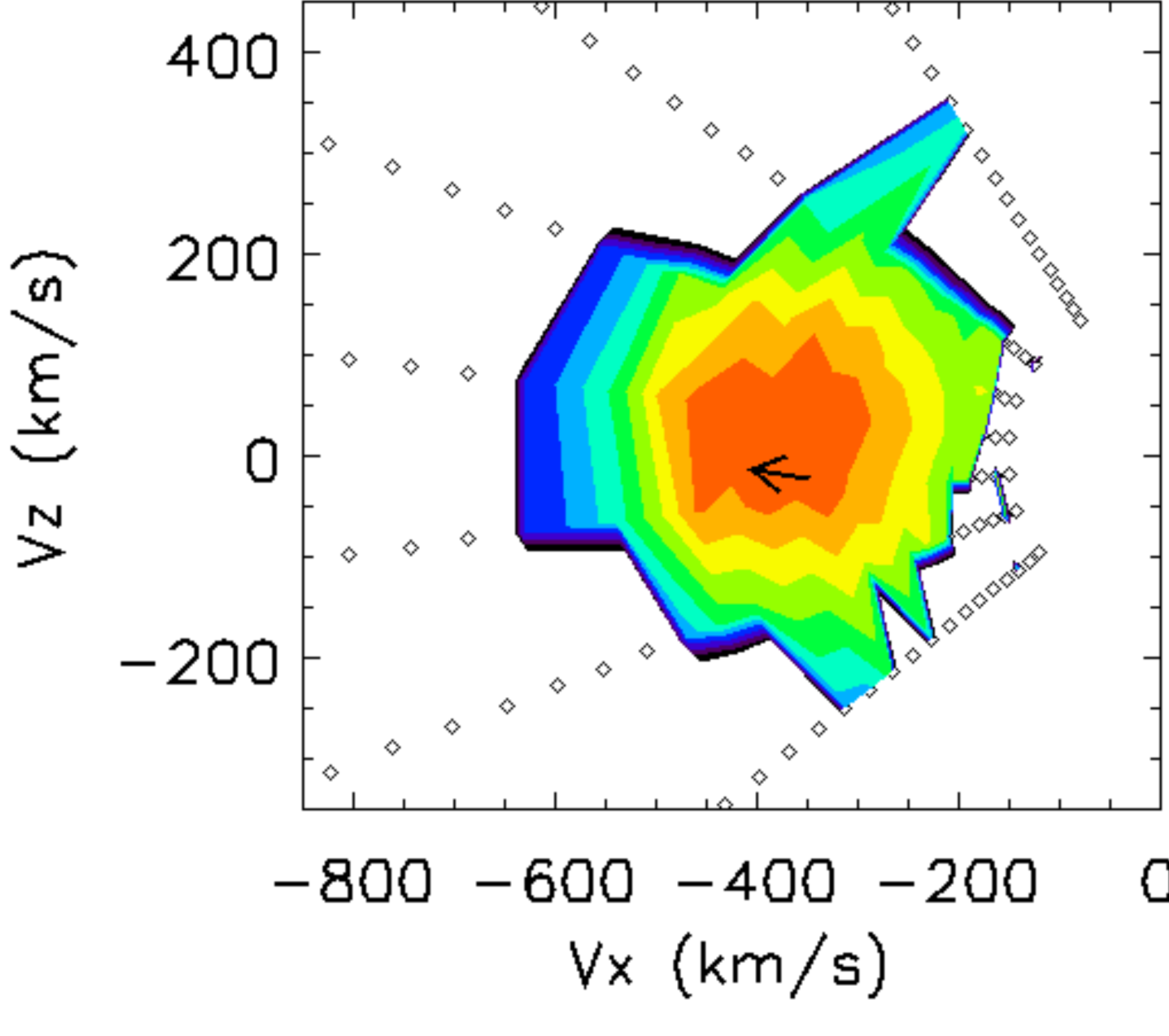}
\includegraphics[scale=.14]{fig5_legend.pdf}
\vfill

\hspace{.05in} (f) t12 = 2019-04-01/10:29:19 
\vfill
\includegraphics[trim=60 10 10 48,clip,scale = .17]{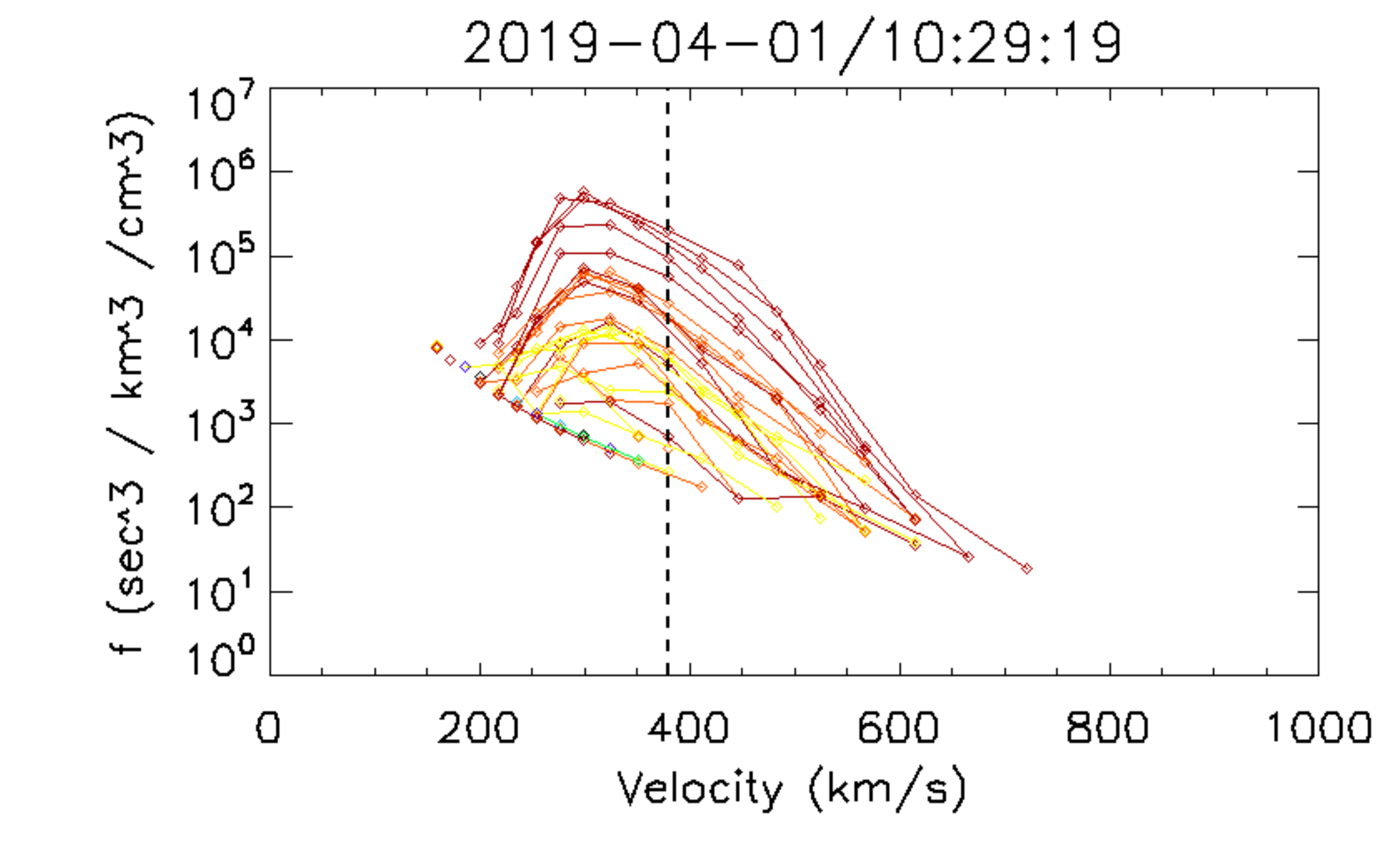} \hspace{.02in}
\includegraphics[trim=30 10 300 52,clip,scale=.17]{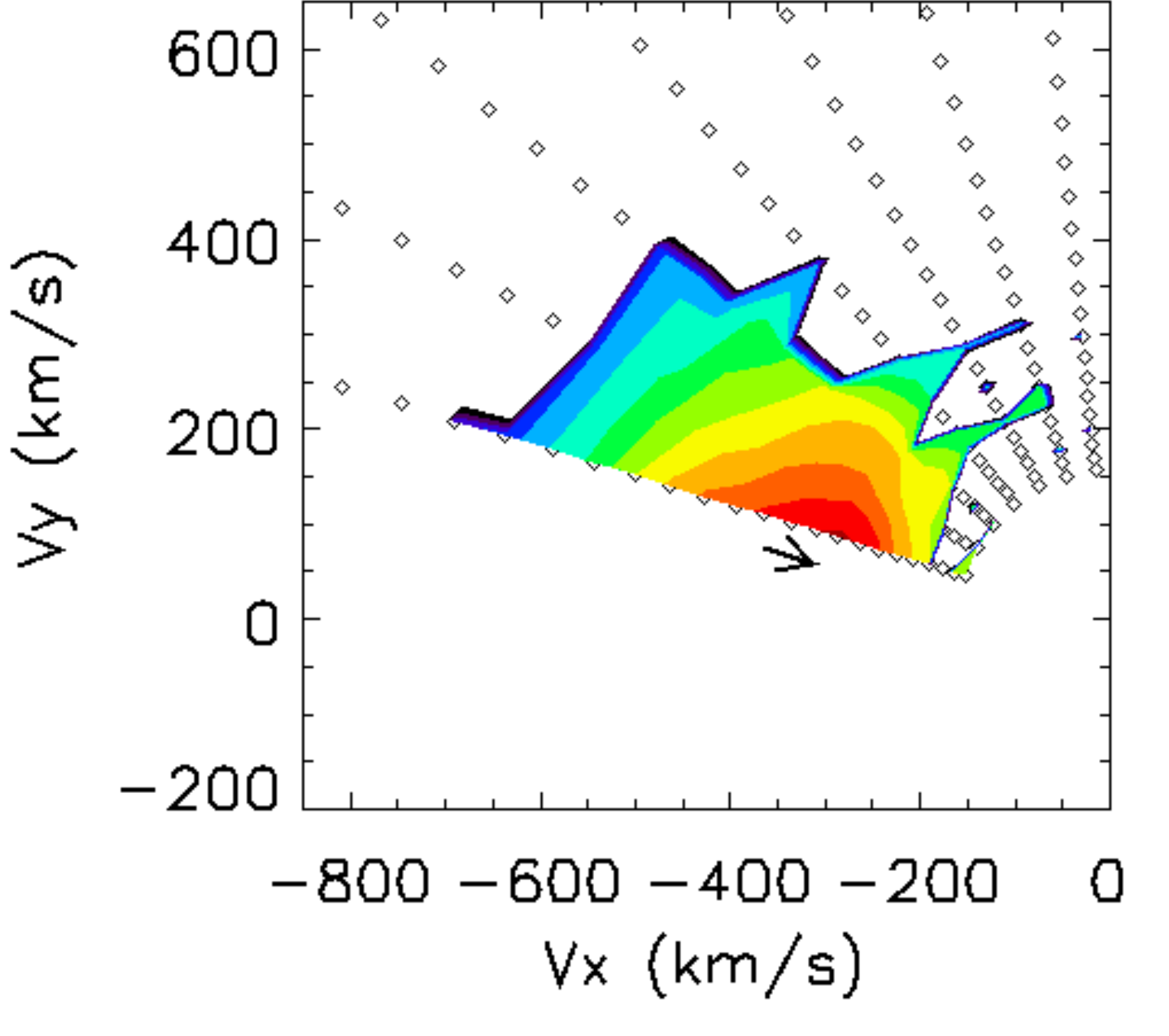}
\hspace{.02in}
\includegraphics[trim=30 10 290 52,clip,scale=.17]{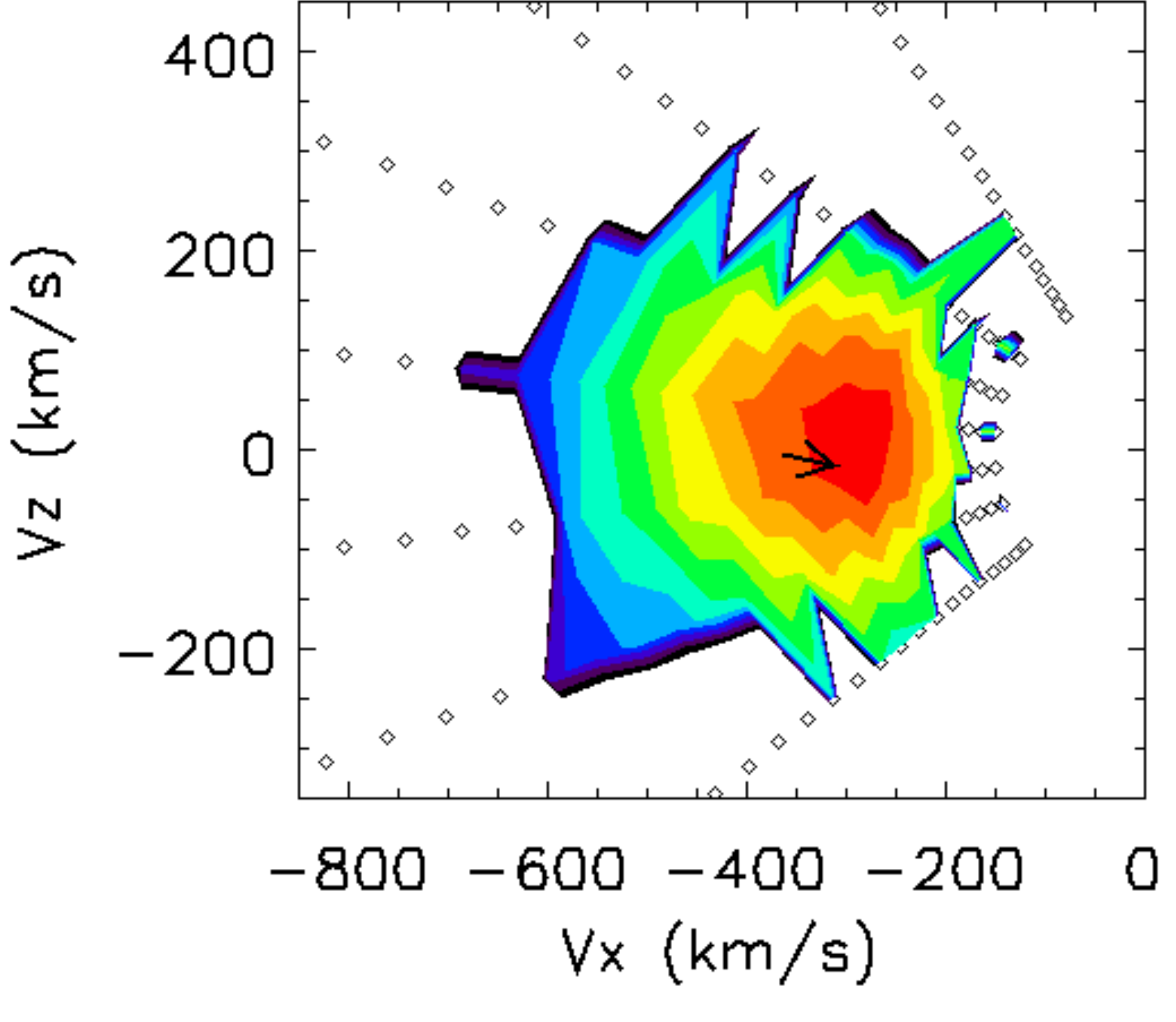}
\includegraphics[scale=.14]{fig5_legend.pdf}
    \caption{VDF evolution for times indicated by the dashed black lines in \figref{fig:4_1_allvars}, with coordinates defined in \secref{sec:coord}. Left: Proton VDF where each line refers to an energy sweep at different elevations. The dashed line represents \Alfven speed. Middle: VDF contour elevations which are summed and collapsed onto azimuthal plane. Right: VDF contour elevations which are summed and collapsed onto $\theta$ plane. The black arrow represents magnetic field direction in SPAN-I coordinates, where the head is at the solar wind velocity (measured by SPC) and the length is the \Alfven speed.
}
    \label{fig:2019_04_01_vdf}
\end{figure*}
 This event is illustrated in \figref{fig:4_1_allvars}, where each panel depicts the same variables as \figref{fig:4_5_allvars}. \figref{fig:4_1_allvars}(a) again shows a quiet, radial magnetic field (with the exception of a 30 minute duration magnetic field reversal lurking at 10:00), (b) shows that the angle $\theta_{kb}$ between the wave normal vector, $\V{k}$, and $\V{B}$ is $< 10^\circ$ during times where the wave power is sufficiently dominant over the background magnetic field, and (c) depicts results from MVA showing circular polarization by two overlapping eigenvalues (red and green), accompanied by a clear subdominant eigenvalue (blue). Wavelet analysis of the magnetic field is shown in \figref{fig:4_1_allvars}(d), depicting the wave power as a function of frequency, and the corresponding spacecraft frame polarizations are shown in (e). Similar to Event $\#$1, the wave power and polarization frequencies in \figref{fig:4_1_allvars}(d) and (e), respectively, do not include the Doppler-shifted correction to the plasma frame (see \secref{sec:stab} for explanation), and the dashed-dotted white line shown in \figref{fig:4_1_allvars}(d) and (e) is the local proton gyrofrequency at $\approx$0.88 Hz.

Contrasting with \figref{fig:4_5_allvars}(d) and (e) from Event $\#$1, in \figref{fig:4_1_allvars}, we notice that the wave power suddenly increased to a higher frequency in the 20 minute segment centered around the dashed black vertical line at 08:14:59, hereafter called t8. The subsequent polarization is also significantly broader in frequency. During the time-averaged period of 06:21:20-07:46:20, the peak wave power was 2.44 times larger than than the background at 2.74 Hz (2.41$\Omega_i$), while the peak wave power jumped to 12.44 times the background field at 3.56 Hz (3.13 $\Omega_i$) during the period 08:04:20-08:24:40.

At times before 08:24:40, the polarization is mostly right-handed (red) and after 08:24:40, it is mostly left-handed (blue). However, it is still not clear what the intrinsic polarization is in the plasma frame. \figref{fig:4_1_allvars}(f) shows the differential energy flux measured from SPAN-I and one can notice that during the 20 minute segment centered at t8, there was a slight depletion in differential energy flux. Also during this time, \figref{fig:4_1_allvars}(g)-(k) show distinguishing features indicative of energy transfer. Specifically, \figref{fig:4_1_allvars} shows (g) a decrease in beam-to-core density ratio, with a significant increase afterwards, (h) an increase in beam temperature (blue), (i) a decrease in $\alpha$-proton differential speed (green) and slight increase in $v_D^*$ (blue), (j) temperature isotropization, with $T_\parallel/T_\perp < 1$ directly before and after this time period, and (k) a slight decrease in $\alpha$ density. 

\figref{fig:2019_04_01_vdf} shows the time series of the proton VDF (left) and contour elevations (middle and right) during this event, where coordinates are defined in \secref{sec:coord}. The evolution of a super-\Alfvenic beam/high-energy shoulder is demonstrated at the times indicated by the dashed black vertical lines in \figref{fig:4_1_allvars}. \figref{fig:2019_04_01_vdf}(a) shows an example VDF at time t7 from \figref{fig:4_1_allvars} when the polarization is right-handed in the spacecraft frame. \figref{fig:2019_04_01_vdf}(b) shows the VDF during the time of intense RH polarized wave power. After this time period, \figref{fig:2019_04_01_vdf}(c) shows the VDF at a time of concurrent RH and LH polarities, revealing a significant elongated beam component from the right panel. From \figref{fig:4_1_allvars}(e), we observe that at t9, the RH (red) polarization is at lower frequencies than the simultaneously observed LH (blue) polarization. This agrees with previous observations of concurrent polarities in the spacecraft frame \citep{Jian:2014,Zhao:2019}. 

The VDFs shown in \figref{fig:2019_04_01_vdf}(d)-(f) correspond to times before, during, and after the magnetic field reversal shown in between t10 and t12 in \figref{fig:4_1_allvars}(a). These magnetic field reversals, also known as ``switchbacks", have been observed to be pervasive in PSP's initial orbits, but their origin and physical nature remains a mystery \citep{Kasper:2019,Bale:2019,McManus:2020a,Tenerani:2020,Dudok_de_Wit:2020,Horbury:2020}.
Directly before this 30 minute-duration switchback, \figref{fig:2019_04_01_vdf}(d) reveals from the middle and right panels that the magnetic field arrow is pointing towards the left, with the small beam component aligned anti-parallel to $\V{B}$ in the right panel. Approximately 4 seconds later, the plasma parcel entered the switchback at t11, represented by \figref{fig:2019_04_01_vdf}(e). Note that from the right panel, the beam component did not change positions, but the magnetic field vector reversed to now point to the right, meaning that the beam is now aligned parallel to $\V{B}$. After the switchback at t11, \figref{fig:2019_04_01_vdf}(f) shows that the plasma reverted back to a state similar to t12.

The behavior of the plasma parcel as it traveled through the switchback in \figref{fig:2019_04_01_vdf}(d)-(f) means that inside the switchback, the definition of the beam and core switched in the plasma fitting routine, since the field-aligned beam was moving backwards from the core because of the field reversal. In between t11 and t12 in \figref{fig:4_1_allvars}(h), the most noticeable effect of this choice is the sudden swap of the beam (blue) and core (red) temperature. Future Encounters will look for similar observations of switchbacks and further analyze this choice of switching the beam and core for the plasma parameter fits.

\subsubsection{An Unexplained Nonlinear Phenomenon} \label{sec:static}

\begin{figure*}
\centering
\includegraphics[scale=1]{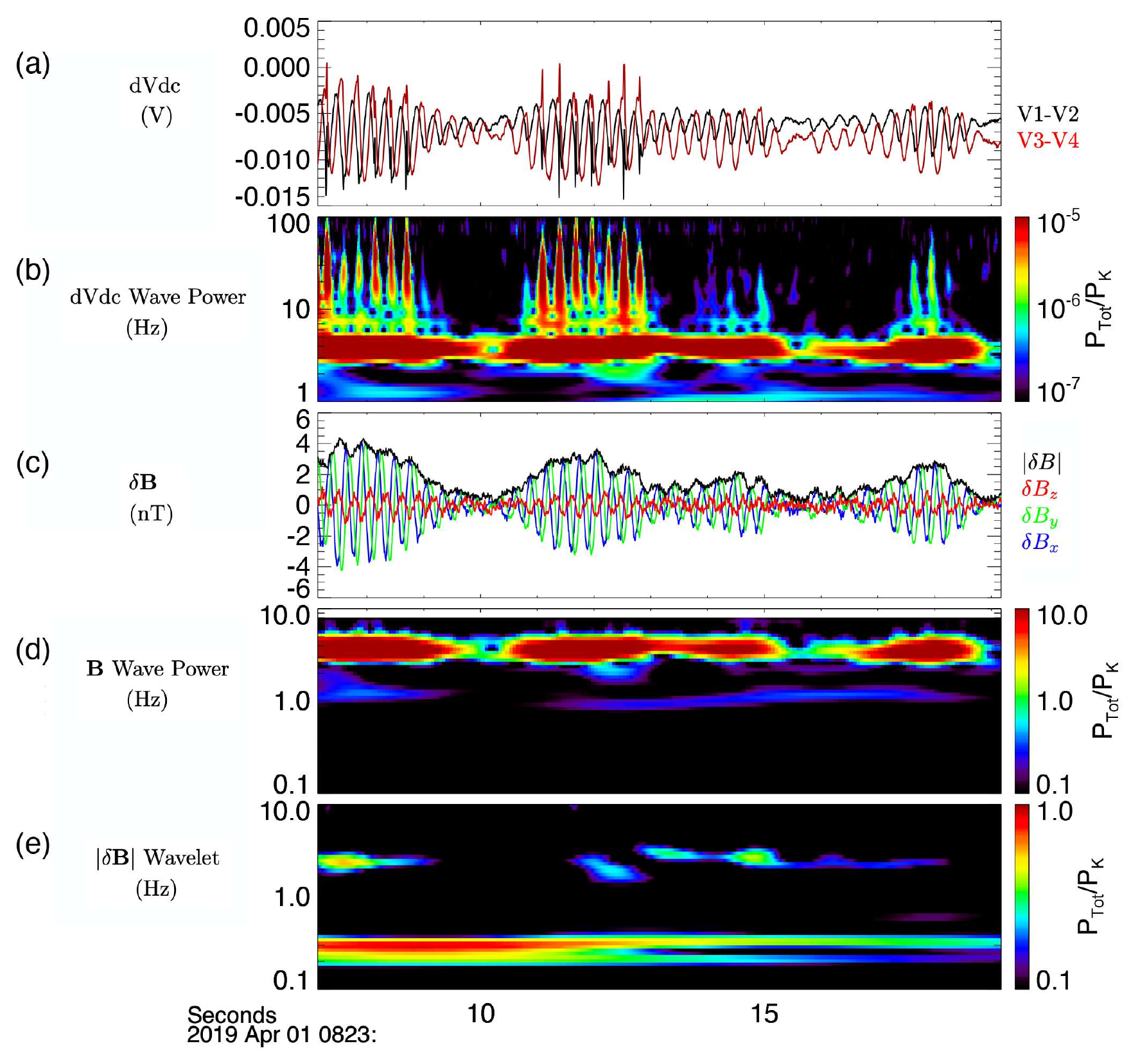}
\caption{Unexplained nonlinear wave mode observed in electric field data (unnormalized by effective antenna length) during period of intense wave power about t8 in \figref{fig:4_1_allvars}. Panel (a) shows the FIELDS DFB DC differential voltage waveform data, (b) is the wavelet transform of (a), (c) shows $\delta \V{B}$, (d) is the wavelet transform of $\V{B}$, and (e) is the wavelet transform of $|\delta B|$. Note that both the differential voltage and $\delta \V{B}$ wave forms are shown in spacecraft coordinates.}
\label{fig:zoom}
\end{figure*}

During the 20 minute period of intense wave power centered at t8 in \figref{fig:4_1_allvars}(d), an unexplained nonlinear structure was observed in the FIELDS Digital Fields Board (DFB) DC differential voltage waveform data, which is essentially the electric field without normalizing by the effective antenna length, i.e. the voltage difference is proportional to the electric field. We used this data since the normalized electric field data product is not yet available. \figref{fig:zoom} shows electric and magnetic field information at shorter timescales within the region of intense wave power centered at t8 from \figref{fig:4_1_allvars}. \figref{fig:zoom}(a) shows the voltage difference measured by FIELDS and (b) shows the wavelet transform of (a). The higher resolution (140 Hz) magnetic field data was employed in \figref{fig:zoom}(c), where we subtracted the mean magnetic field (averaged over 1 second) to obtain the waveform of $\delta \V{B}$. This high-pass filter was performed to compare small-scale fluctuations with the electric field data. \figref{fig:zoom}(d) shows the wavelet transform of $\V{B}$, which corresponds to \figref{fig:4_1_allvars}(d) on a shorter timescale. Finally, the wavelet transform of $|\delta B|$ is shown in \figref{fig:zoom}(e). Note that both the electric and magnetic field data are in spacecraft coordinates.

Note from the wavelet spectra in \figref{fig:zoom}(b) and (d), that the maximum wave power in both the electric and magnetic field appears at the same frequency of 3.56 Hz. However, \figref{fig:zoom}(b) shows that only the electric field has wave power above this frequency, exhibiting sharp, possibly nonlinear structures. The specific shape of these structures is due to the wavelet response to the impulsive features in the electric field data from \figref{fig:zoom}(a). Plotted in \figref{fig:zoom}(c) is $\delta B_x$ (blue), $\delta B_y$ (red), $\delta B_z$ (green), and $|\delta B|$. By visual inspection of both \figref{fig:zoom}(c) and (d), the magnetic field does not have the same impulsive signatures as the electric field in \figref{fig:zoom}(a) and (b). We therefore conclude that these higher frequency structures may be electrostatic. Note that the authors have ruled out various instrumental effects, such as instrumental cross-talk, reaction wheel contamination, and dust impacts. This is indeed a physically meaningful signal.

Furthermore, we see that from \figref{fig:zoom}(c) that $|\delta B|$ assumes a waveform with higher frequency oscillations inside an envelope of lower frequency oscillations. These two frequencies are evident in \figref{fig:zoom}(e), where the wavelet transform shows power as a narrowband frequency centered at $\approx$2 Hz, and another at $\approx$0.25 Hz. By comparing \figref{fig:zoom}(b) and \figref{fig:zoom}, it appears that the lower frequency envelope of $|\delta B|$ is correlated with the electrostatic wave power. This explains why the $\V{B}$ wave power in \figref{fig:zoom}(d) appears as discrete packets. Note that in the particular time interval shown in \figref{fig:zoom}(e), it appears that the frequency band centered about 0.25 Hz has larger wave power than the 2 Hz band. However, other time periods within the 20 minute interval centered at t8 reveals times where the 2 Hz frequency band has significantly more power (not shown), suggesting that the coupled waves at two different frequencies may be undergoing continuous damping and wave growth.

The existence of the unexplained phenomenon may be key to discovering plasma instabilities capable of producing such a nonlinear wave mode or electrostatic structure at ion-scales. This phenomenon occurred during the exact time interval of intense wave power and RH polarity, centered around t8 in \figref{fig:4_1_allvars}. Any plasma instability shown to exist there will be imperative to understanding the nature this complex wave-particle interaction event candidate.
\newline
\newline
\section{Initial Instability Analysis} \label{sec:stab}

Particle velocity distribution functions that depart from local thermodynamic equilibrium, as shown in the previous sections, are prone to plasma instabilities, which may contain enough free energy to contribute to wave growth or decay. From the 1D plasma fits from SPAN-I, we perform an instability analysis using a reduced parameter space that does not include separate temperature anisotropies for the individual proton core, proton beam, electron, and $\alpha$-particle components. For this task, 3D plasma fits are required, and not yet available. Even though we are using simplified plasma parameters as input, these preliminary results remain insightful. Also note that in this section, we omit the superscript, *, on the parameter fits. All references to the proton beam and core are from the fits, while the $\alpha$ parameters are from the SPAN-I moments.

\subsection{Doppler-Shift Estimation} \label{sec:dop}

One of the primary reasons to invoke a plasma dispersion solver for this investigation was to accurately Doppler-shift the wave power and polarization frequencies from \secref{sec:obs} to reveal the intrinsic handedness of the waves. We also sought to test whether the frequencies Doppler-shifted approximate to the local gyrofrequency, suggesting that the waves may have been generated locally. The Doppler shift equation is
\begin{equation}
2\pi f_{sc}^\pm=2\pi f_p\pm\V{k\cdot V_{sw}}
\label{eq:dopf}
\end{equation}
where $f_{sc}$ and $f_p$ denotes the spacecraft and plasma frame frequency, respectively, and $\V{V_{sw}}$ is the solar wind velocity. Previous authors \citep{Jian:2009,Jian:2010,Jian:2014,Wicks:2016} have computed this Doppler-shifted frequency by assuming that the phase speed of the wave was approximate to the \Alfven speed. However, we seek to find a more accurate value of $\V{k}$ by using a warm plasma dispersion solver.

As first introduced by \citet{Verscharen:2013b}, the New Hampshire Dispersion Relation Solver (NHDS) is an open source numerical tool for solving the hot-plasma dispersion relation, guided by \citet{Stix:1992}. It assumes a drifting bi-Maxwellian for each particle species and the reader may consult \citet{Verscharen:2018} for further details. From the 1D fits, we inputted plasma parameters from 4 different particle species: proton core, proton beam, electrons, and $\alpha$-particles. Since the waves were shown to be parallel propagating, we set the angle between $\V{k}$ and $\V{B}$ fixed at $1^\circ$. In addition, all calculations were performed in the proton core frame of reference. Although the relevant electron data is not yet available, we set values to obey quasineutrality and zero net current. In general, both the proton beam and $\alpha$-particles had nontrivial drift speed from the core, which was found to be the main driver for unstable modes.

Table \ref{tab:nhds_t1} describes the parameters inputted into NHDS from Event $\#$1 at t1 = 2019-04-05/18:33:22 in \figref{fig:4_5_allvars}, directly before the large enhancement in differential energy flux when $T_\perp/T_\parallel > 1$. Table \ref{tab:nhds_t8} describes the input parameters from Event $\#$2 at t8=2019-04-01/08:14:59 in \figref{fig:4_1_allvars}, during the period of intense wave power that correlated with the observed nonlinear phenomenon described \secref{sec:static}. The proton VDFS at t1 and t8 are seen in \figref{fig:4_5_vdfs}(a) and \figref{fig:2019_04_01_vdf}(b), respectively. As an approximation to the temperature anisotropy, we used a value of $T_\perp/T_\parallel$=2 for all particle species by inspection of \figref{fig:4_5_allvars}(j) at t1. Similarly, we set $T_\perp/T_\parallel$=1 for all species by inspection of \figref{fig:4_1_allvars}(j) at t8. 

\begin{figure*}

\hspace{.3in} (a) Dispersion curves at t1 = 2019-04-05/18:33:22 \hspace{.5in} (b) Growth (decay) rates at t1 = 2019-04-05/18:33:22
\vfill
\includegraphics[scale = .45]{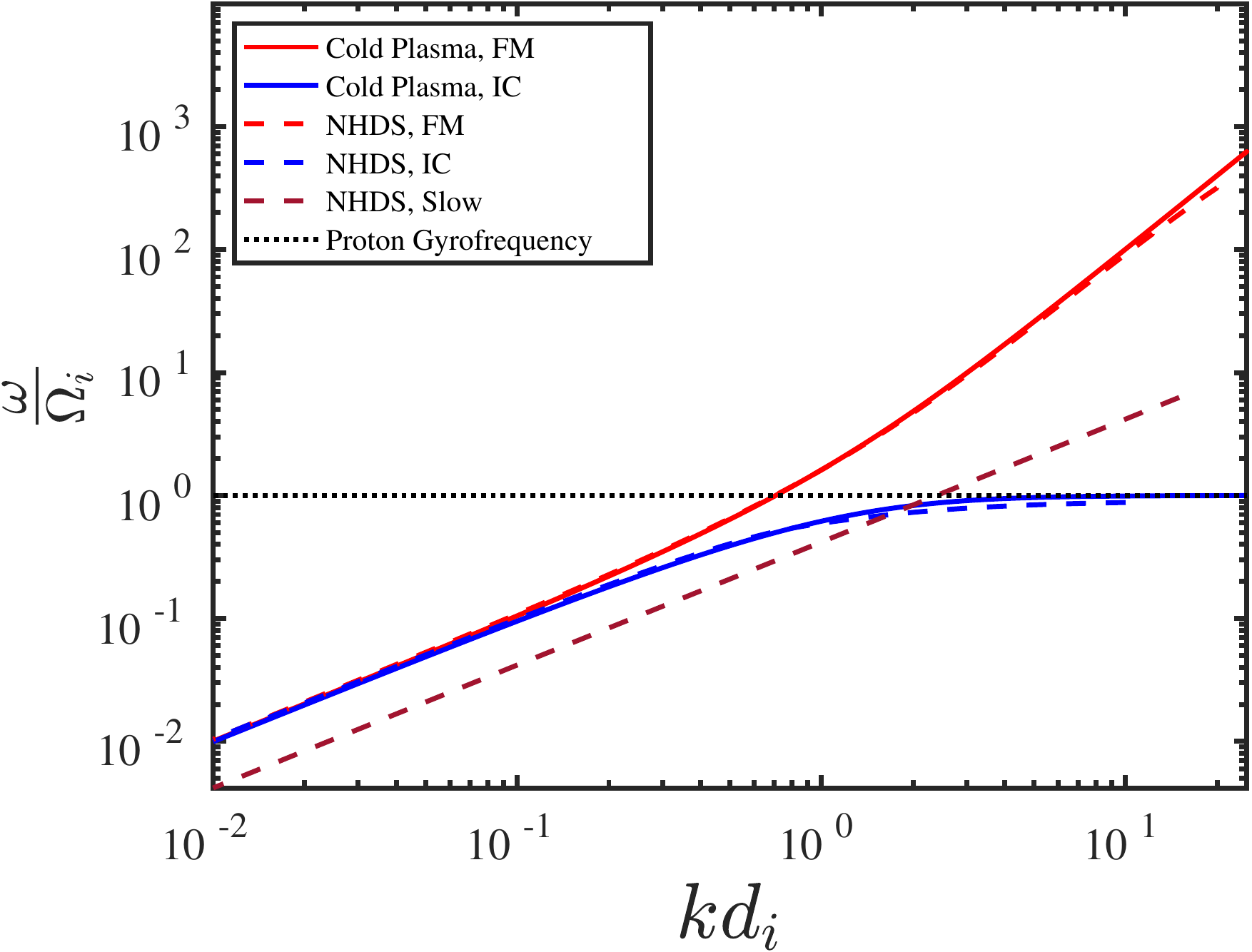} \hspace{.1in}
\includegraphics[scale=.45]{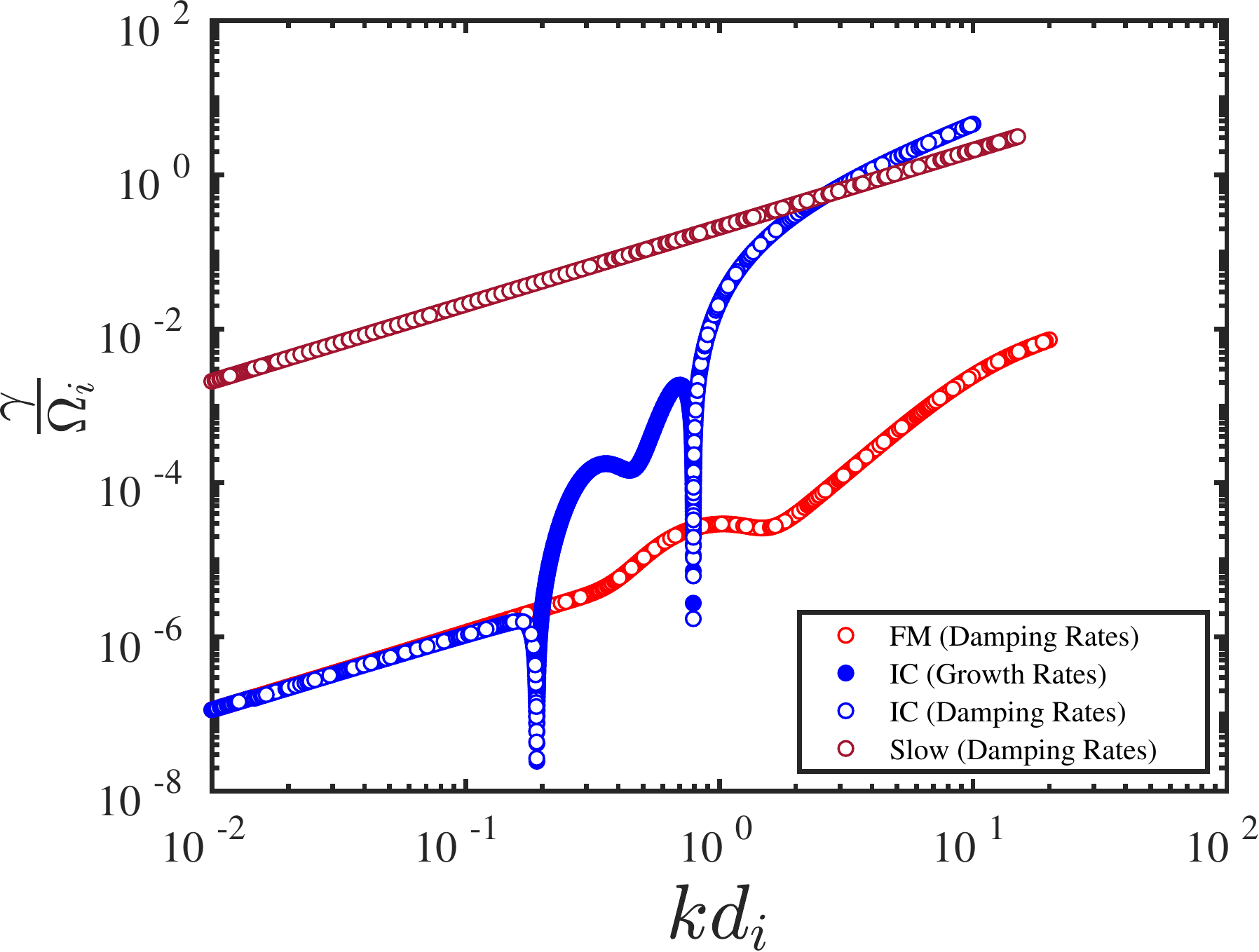}

\hspace{.3in} (c) Dispersion curves at t8 = 2019-04-01/08:14:59 \hspace{.5in} (d) Growth (decay) rates at t8 = 2019-04-01/08:14:59
\vfill
\includegraphics[scale=.45]{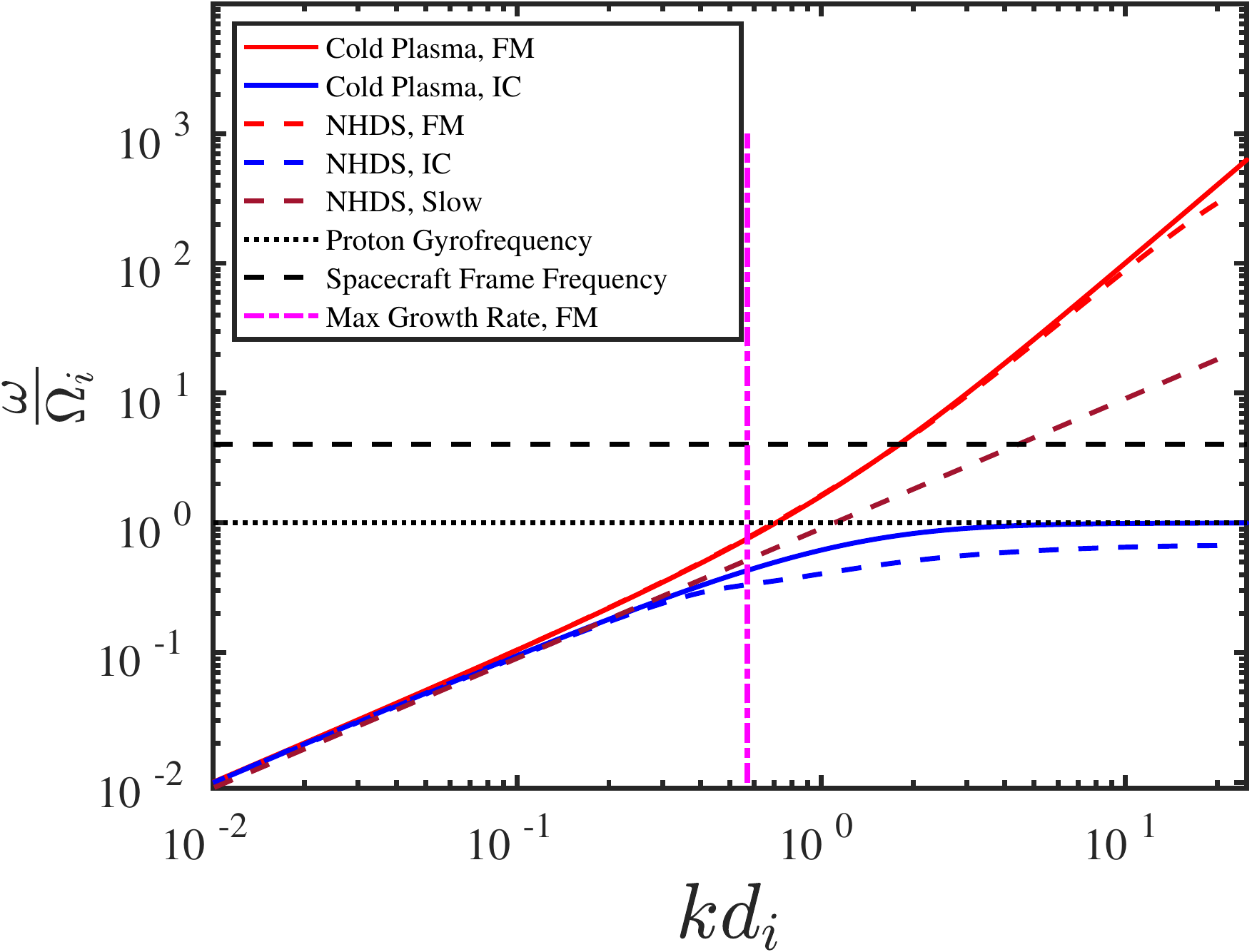} \hspace{.1in}
\includegraphics[scale=.45]{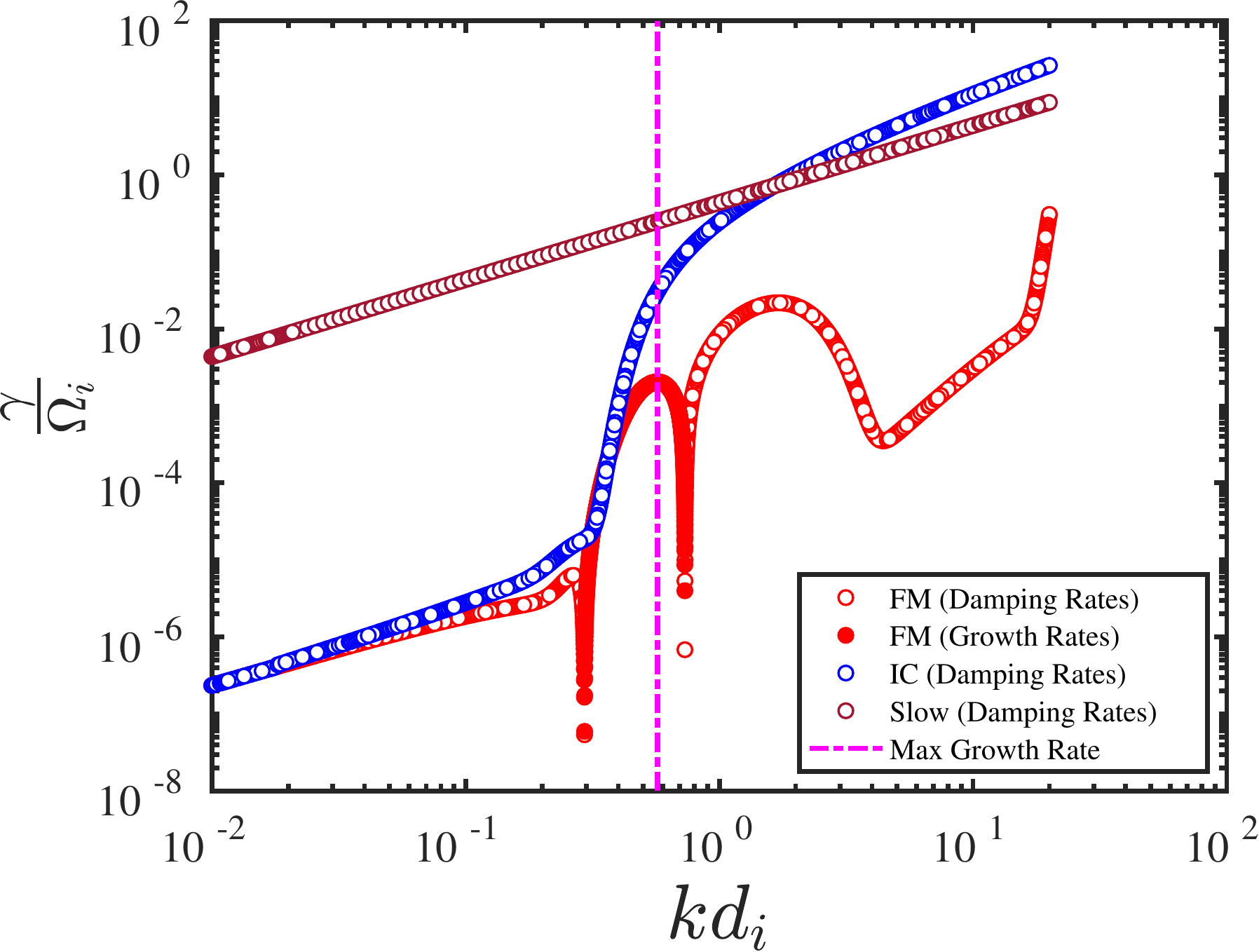}
\caption{Results from New Hampshire Dispersion Solver (NHDS), inputting 1D fits of SPAN-I plasma parameters. Panels (a) and (b) correspond to dispersion curves and growth (decay) rates from Event $\#$1 at t1 in \figref{fig:4_5_allvars}. Panels (c) and (d) correspond to dispersion curves and growth (decay) rates from Event $\#$2 at t8 in \figref{fig:4_1_allvars}. The pink dashed line is located at the maximum growth rate.}
\label{fig:disp}
\end{figure*}

\figref{fig:disp}(a) and (c) show the dispersion relation curves from each time, respectively. Note that the $y$-axis is normalized by the proton gyrofrequency, $\Omega_i$, and the $x$-axis is normalized by the (proton core) ion inertial length, $d_i$. The solid red and blue curves in \figref{fig:disp}(a) and (c) correspond to the cold plasma approximation for the fast magnetosonic (FM) branch and the ion-cyclotron branch (IC), respectively, for comparison to the NHDS solution in dashed red and blue. The brown dashed line represents the slow mode, also obtained by NHDS, to show the three (forward-propagating) possible solutions. The horizontal black dotted line at $\omega/\Omega_i=1$ represents the ion gyrofrequency. \figref{fig:disp}(b) and (d) show the growth rates (solid circles) and decay rates (open circles) for each mode.
\begin{table*}
\caption{NHDS input plasma parameters from 1D SPAN-I fits at t1 = 2019-04-05/18:33:22. The densities are in units of proton core density and the drift speed $v_D$ is respect to the proton core, normalized by the proton core \Alfven speed.}
\centering
\begin{tabular}{| c | c | c | c | c |}
\hline 
Parameter & Proton Core & Electrons & Proton Beam & $\alpha$-Particles \\
\hline 
$\beta$ & 0.0908 & 0.0908 & 0.0374 & 0.0022 \\
$Density$ & 1.0000 & 1.1014 & 0.0992 & 0.0011  \\
$v_D$ & 0 & 0.0406 & 0.4426 & 0.3505\\
\hline
\end{tabular}
\label{tab:nhds_t1}
\end{table*}

\begin{table*}
\caption{NHDS input plasma parameters from 1D SPAN-I fits at t8 = 2019-04-01/08:14:59. The densities are in units of proton core density and the drift speed $v_D$ is respect to the proton core, normalized by the proton core \Alfven speed.}
\centering
\begin{tabular}{| c | c | c | c | c |}
\hline 
Parameter & Proton Core & Electrons & Proton Beam & $\alpha$-Particles \\
\hline 
$\beta$ & 0.4290 & 0.4290 & 0.0916 & 0.0026 \\
$Density$ & 1.0000 & 1.0858 & 0.0845 & 0.0007  \\
$v_D$ & 0 & 0.1146 & 1.4609 & 0.7210\\
\hline
\end{tabular}
\label{tab:nhds_t8}
\end{table*}
Using a low-frequency, massless electron approximation to the dispersion relation found in Eq. (2.5) of \citet{Stix:1992}, we solved for $\omega/\Omega_i$ as a function of $kd_i$ to obtain the cold plasma curves in the same normalized units as NHDS output. Specifically, we get
\[
\left(\frac{\omega}{\Omega_i}\right)^\pm = \frac{kd_i}{2}\sqrt{kd_i^2+4}\pm kd_i
\]
where the (+) and (-) signs correspond to the RH FM mode and LH IC mode, respectively.

From \figref{fig:disp}(b), we see that the only mode with positive growth rates was the IC mode (solid blue circles). This suggests that an ion-cyclotron instability may have occurred at t1, right before the large wave storm in Event $\#$1, indicating that the IC instability was indeed the driver of these waves. Note that from \figref{fig:4_5_allvars}, the wave storm was all left-hand polarized in the spacecraft frame and had a peak wave power at 2.4 Hz, which was only slightly higher that the local gyrofrequency of 1.7 Hz. However, strong wave power was present between 1 and 5 Hz (0.59$\Omega_i$ - 3.34$\Omega_i$). Therefore, these waves could be ion-cyclotron wave modes that were locally generated. 

From \figref{fig:disp}(d), we observe that the mode with the largest growth rate was the FM mode (solid red circles). Therefore, a magnetosonic instability may have existed at t8, in the region of intense wave power and right-handed polarization of \figref{fig:4_1_allvars} from Event $\#$2. The magenta vertical dashed line in \figref{fig:disp}(c) and (d) indicates the value of $kd_i$ associated with this maximum growth rate, $\gamma_{max}$. In contrast to Event $\#$1, the spacecraft-frame frequency of the dominant wave power was 3.6 Hz (4.06 $\Omega_i$), modestly above the local proton gyrofrequency of 0.88 Hz, as indicated by the horizontal black dashed line in \figref{fig:disp}(c). Since the dispersion solver was performed in the plasma frame, one approach to finding the frequency in the plasma frame is to inspect where the magenta dashed line, the line corresponding to the $\V{k}$ value of FM $\gamma_{max}$ ($kd_i=0.55$), intersects with the FM branch (dashed red) in \figref{fig:disp}(c). By inspection, we see that their intersection occurs at 0.66 Hz (0.75$\Omega_i$). However, when the values of $V_{sw}$=333.4 $kms^{-1}$ and $d_i$=17.9 km are inserted into Eq. \ref{eq:dopf}, we get
\[
2\pi(3.56) = 2\pi(0.66)+(0.55/17.9)\times 333.4
\]
But, this yields an inconsistent result of 22.93 $\omega_p$ = 14.39 $\omega_p$. This could mean that either our found value of $kd_i$ was inaccurate, and/or that the frequency of the wave is a mix of modes. We also note that the strong ($\geq$10 times the background) wave power was in the range 3 Hz - 5.5 Hz (2.6$\Omega_i$-16.5$\Omega_i$), with weaker wave power extending to even lower frequencies at 1 Hz (0.88$\Omega_i$). Therefore, if we include the lower and upper end of this frequency range, the lower and upper bounds of the Doppler-shifted frequency is in the range 6.28$\omega_p$ - 31.4$\omega_p$. The computed plasma frame frequency of 14.39$\omega_p$ indeed falls within this range.

\begin{figure*}
\centering
(a) Possible Doppler-shifted FM Branch at t8
\vfill
\includegraphics[scale=.7]{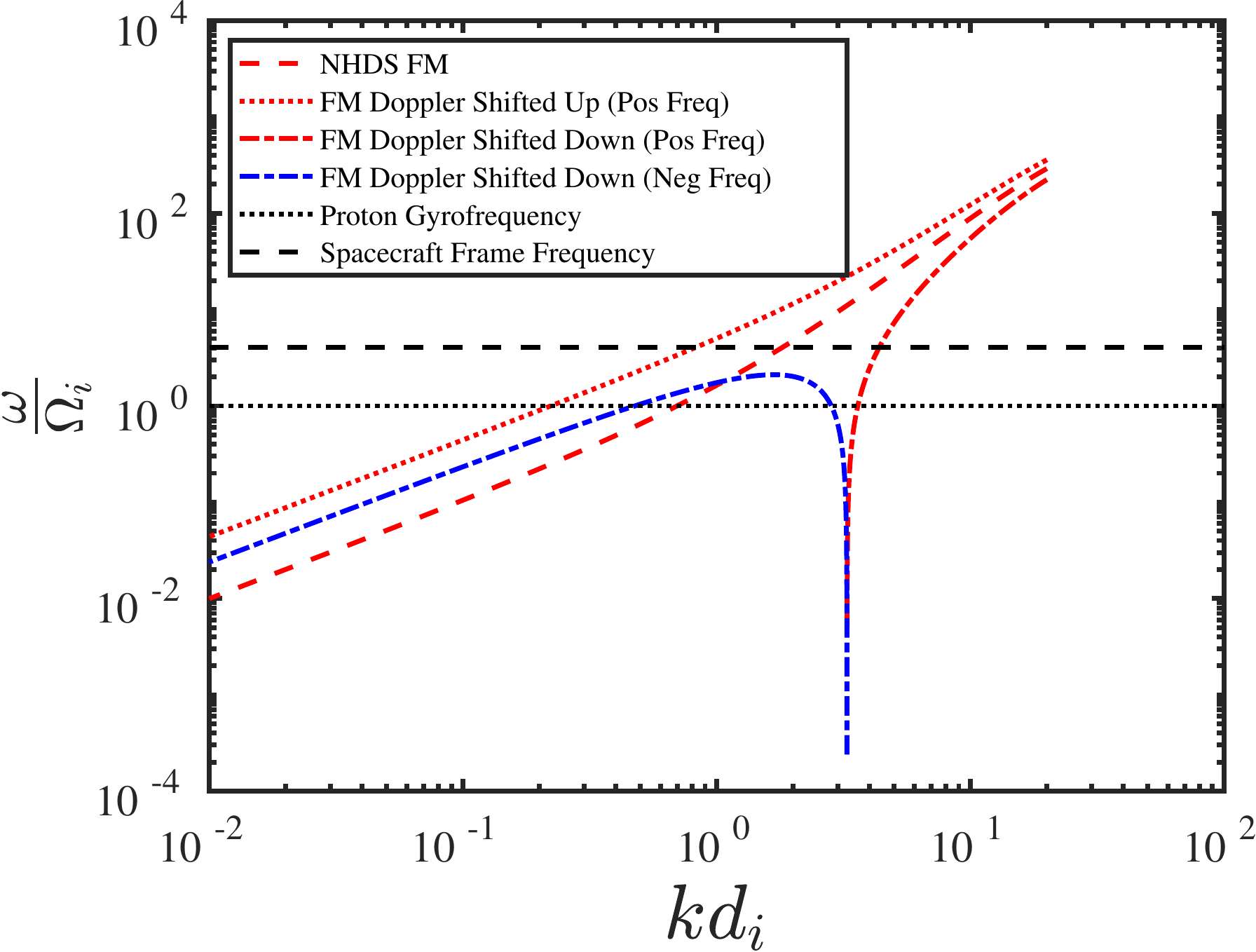}
\vfill
(b) Possible Doppler-shifted IC Branch at t8
\vfill
\includegraphics[scale=.7]{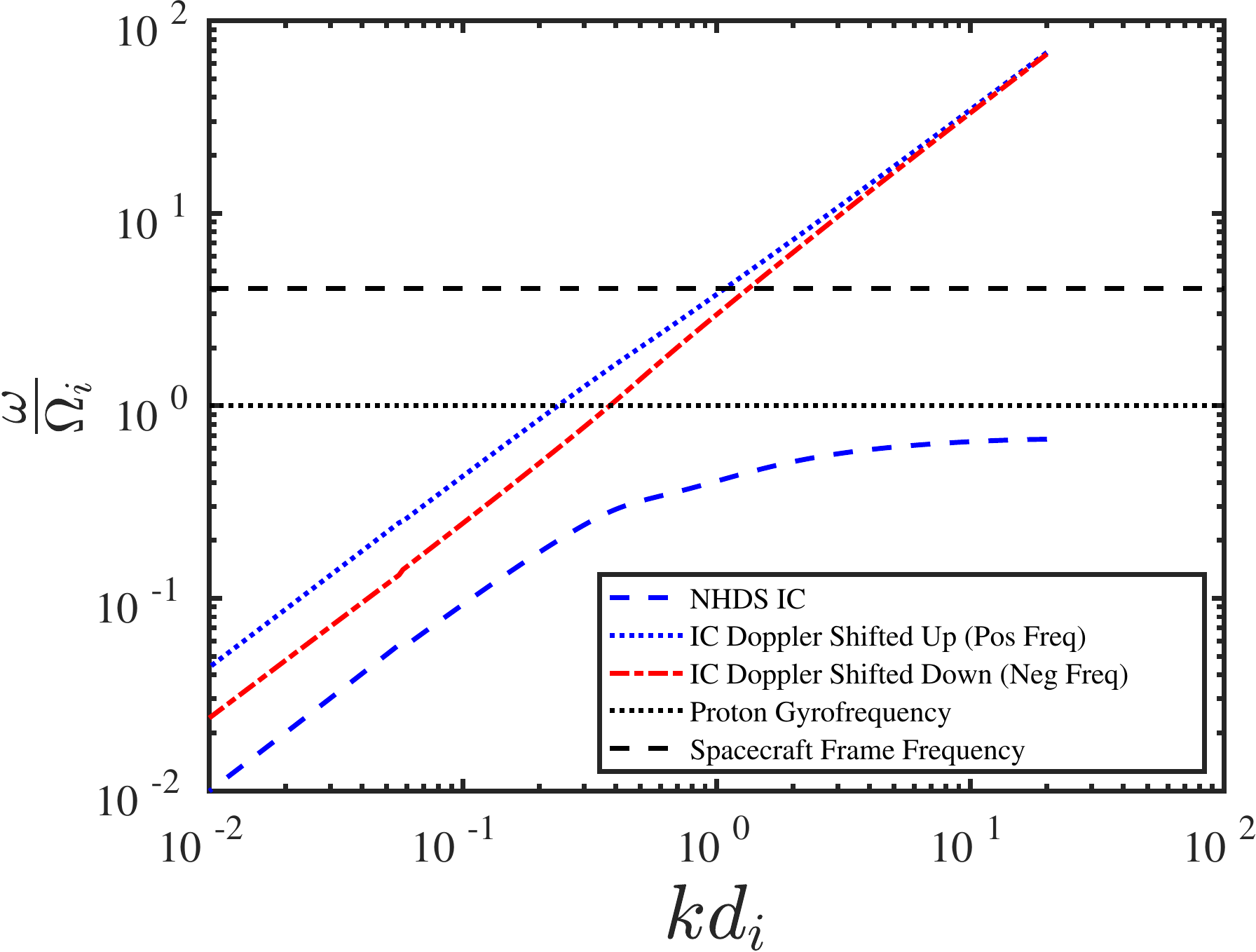}
\caption{Possible method to Doppler-shift the dispersion curves up and down to find the intrinsic handedness of the wave at t8 from Event $\#$2. Panel (a) represents the procedure of Doppler-shifting the FM branch up and down to find the intersection with the spacecraft frame frequency. Panel (b) shows the procedure for the IC branch. }
\label{fig:dop}
\end{figure*}

We also present an alternative method to computing the Doppler shift, without relying on the value of $\gamma_{max}$. A procedure to find the set of all possible values of $kd_i$ is highlighted in \figref{fig:dop}. First, in \figref{fig:dop}(a), we take the FM branch found by NHDS (dashed red), and Doppler-shift the entire curve up and down to see where it intersects with the spacecraft frame frequency (horizontal dashed black line). The resulting Doppler-shifted curves correspond to the spacecraft frame frequency, $\left(\omega/\Omega_i\right)_{sc}$. For example, we compute new curves, based on the NHDS solution of the FM branch in the plasma frame, $\left(\omega/\Omega_i\right)_p$, as
\[
\left(\frac{\omega}{\Omega_i}\right)_{sc} = \left(\frac{\omega}{\Omega_i}\right)_p \pm k\times \frac{V_{sw}}{d_i\Omega_i}
\]
where the (+) or (-) sign signifies shifting up or down, respectively.

The dotted red line and dashed-dotted red line in \figref{fig:dop}(a) represents the FM branch Doppler-shifted up and down to positive frequencies, respectively. The dashed-dotted blue curve in \figref{fig:dop}(a) corresponds to the FM branch Doppler-shifted down with resulting negative frequencies, meaning that the wave polarization changed handedness from RH (positive) to LH (negative). We can now observe where each curve intersects the spacecraft frame frequency. The dotted red curve in \figref{fig:dop}(a) intersects the horizontal black dashed curve at $kd_i$=0.82, mapping back to a plasma-frame frequency of 1.09 Hz (1.24$\Omega_i$). Since the values of the FM positive growth rates in \figref{fig:disp}(d) extend to 0.73 $kd_i$, this case is within a reasonable margin of error based on our estimation of $V_{sw}$, $d_i$, and the inputted plasma fits. If we consider broadening the range of spacecraft-frame frequencies to all wave power above 10 times the background field (2.6$\Omega_i$-16.5$\Omega_i$), then our found plasma frame frequency falls well within this range. 

The dashed-dotted curve in \figref{fig:dop}(a) intersects the spacecraft frame frequency at $kd_i$=4.37, which is too large and therefore this case can be discarded. The dashed-dotted blue curve in \figref{fig:dop}(a) does not intersect the spacecraft frame frequency, but approaches 1.85 Hz (2.11$\Omega_i$) at $kd_i$=1.66 before heavily damping to zero. Since this mode only reaches half the spacecraft frame frequency, this case can also be discarded.

 We then repeat the same procedure in \figref{fig:dop}(b) for the IC branch found by NHDS (dashed blue), and Doppler-shift up to positive frequencies (dotted blue), which intersects the spacecraft frame frequency at $kd_i$=1.08, corresponding to a plasma frame frequency of 0.36 Hz (0.41$\Omega_i$). Doppler-shifting down to negative frequencies (dashed-dotted red) in \figref{fig:dop}(b) also changes sign and therefore from LH to RH. This curve intersects the spacecraft frame frequency at $kd_i$=1.33, similarly mapping to a plasma frame frequency of 0.45 Hz (0.40$\Omega_i$). However, \figref{fig:disp}(d) shows that the IC branch (blue open circles) is strongly damped around $kd_i$=1. We can therefore discard both of these cases.
 
We therefore conclude that the most likely value of $kd_i$ is 0.82, represented by the intersection location of the FM branch Doppler-shifted up to positive frequencies (dotted red curve) with the spacecraft frame frequency (dashed black line) in \figref{fig:dop}(a). The plasma frame frequency would then be approximately 20\% higher than the local proton gyrofrequency, suggesting that the fast magnetosonic wave was generated locally.  

\subsection{Additional Instability Characterizations}
\label{sec:plume}

The overall purpose of this paper is to highlight information gleaned from SPAN-I measurements. In a complementary investigation to \secref{sec:dop}, we have employed the instability analysis tool, \T{PLUMAGE} \citep{Klein:2017c}, which determines unstable modes based on the Nyquist Criterion \citep{Nyquist:1932}. The hot-plasma dispersion relation is solved numerically by \T{PLUME} \citep{Klein:2015a} for an arbitrary number of particle species, each represented by drifting bi-Maxwellians. \T{PLUMAGE} has recently demonstrated success by characterizing instabilities from Helios data, finding that 88\% of the surveyed intervals were linearly unstable \citep{Klein:2019}.

Since separate temperature anisotropy measurements for the different particle components are not yet available, we turn to the reduced parameter space:
\begin{multline}
\mathcal{P} = \bigg\{ \beta_{core}, \frac{w_{core}}{c}, \frac{T_{core}}{T_{beam}}, \frac{T_{core}}{T_{\alpha}}, \frac{n_{beam}}{n_{core}}, \frac{n_\alpha}{n_{core}},\\
\frac{\Delta v_{beam,core}}{vA_{core}}, \frac{\Delta v_{\alpha,core}}{vA_{core}}\bigg\}
\label{eq:param}
\end{multline}
where $\beta_{core}$ is in reference to the proton core and $w_{core}/c$ is the proton core thermal speed (normalized by the speed of light). $T_{core}/T_{beam}$ and $T_{core}/T_{\alpha}$ refers to the proton core-to-beam and core-to-$\alpha$ temperature ratios, respectively. $n_{beam}/n_{core}$ and $n_\alpha/n_{core}$ represents the proton beam-to-core and $\alpha$-to-core density ratios, respectively. Finally, $\Delta v_{beam,core}/vA_{core}$ and $\Delta v_{\alpha,core}/vA_{core}$ signifies the relative drift speeds of the proton beam and $\alpha$-particles with respect to the proton core (normalized by the proton core \Alfven velocity), respectively. SPAN-I produced proton core, proton beam, and $\alpha$-particle parameters for 995 time intervals spanning 2019-04-05/18:30:00 - 2019-04-05/20:30:00, corresponding to the Event $\#$1 times in \figref{fig:4_5_allvars}. The instrument also yielded plasma parameters for 2183 time intervals spanning 2019-04-01/06:30:00 - 2019-04-05/10:30:30, corresponding to the Event $\#$2 times in \figref{fig:4_1_allvars}. We then inputted these parameters for both events into \T{PLUMAGE}. For each time interval, \T{PLUMAGE} determined the fastest growing mode as well as the associated plasma response at length scales spanning $k_\perp \rho_c \in [10^{-3},3]$ and $k_\parallel \rho_c \in[10^{-2},3]$ (see Fig. 1 and the associated test in \cite{Klein:2019} for additional details on the method).

\begin{figure*}
\centering
(a) \T{PLUMAGE} Results from Event $\#$1
\vfill
\includegraphics[scale=1]{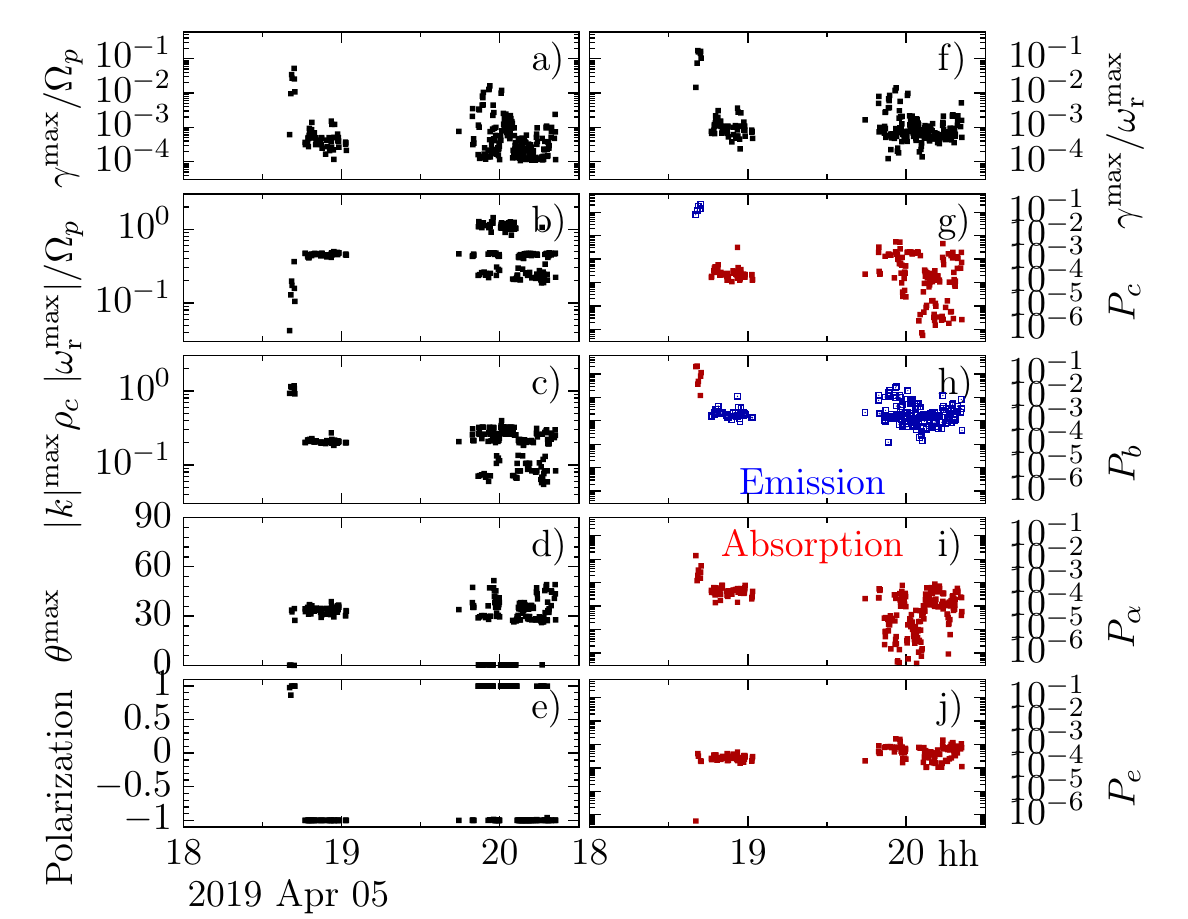}
\vfill
(b) \T{PLUMAGE} Results from Event $\#$2
\vfill
\includegraphics[scale=1]{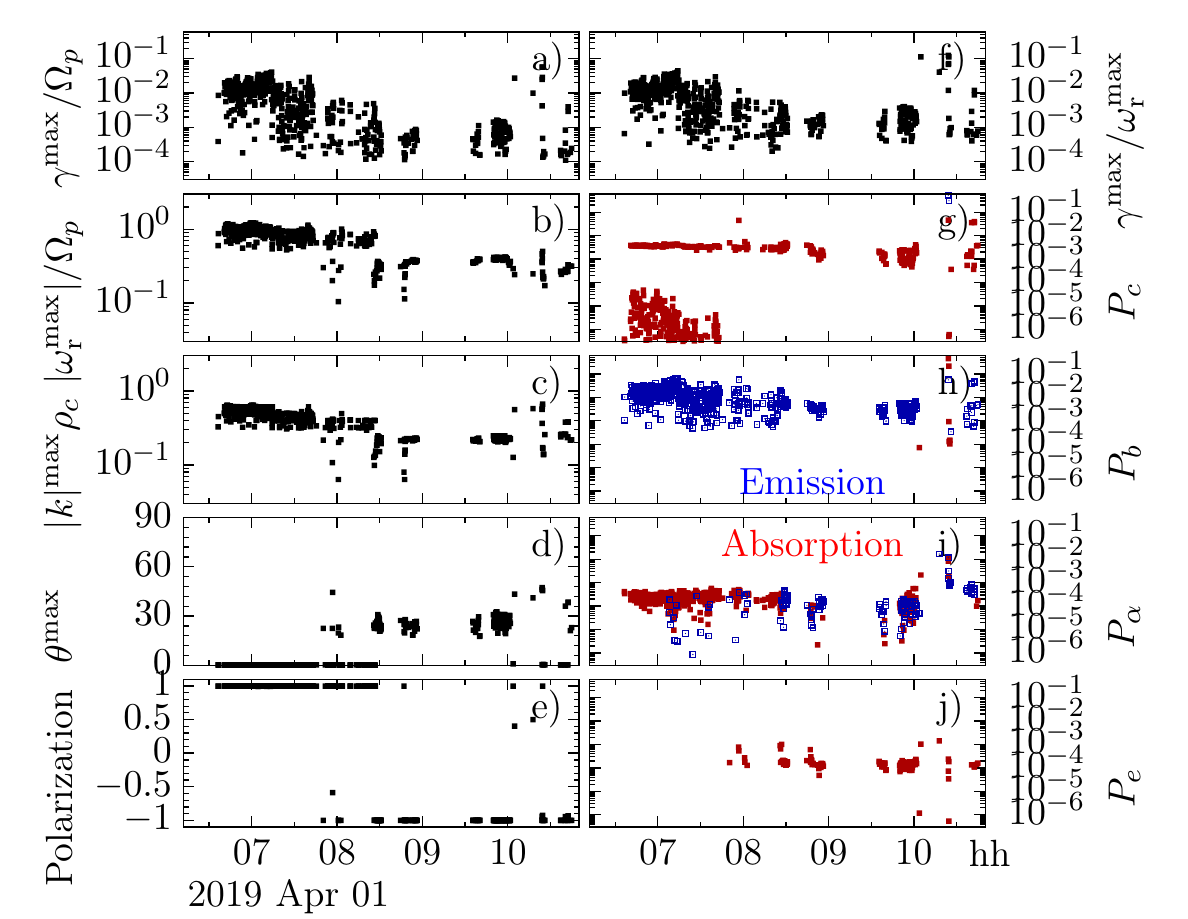}
\caption{Initial instability analysis in reduced parameter space, $T_\perp/T_\parallel$=1, for all wave propagation angles. The power absorbed (red) and emitted (blue) from each particle species is shown for each positive growth rate. Panel (a) shows results from the Event $\#$1 discussed in \secref{sec:e1} and (b) shows results from Event $\#$2 detailed in \secref{sec:e2}. The time axis on panels (a) and (b) corresponds to the same intervals from \figref{fig:4_5_allvars} and \figref{fig:4_1_allvars}, respectively.}
\label{fig:plume}
\end{figure*}

\figref{fig:plume} displays the linear growth rates computed by \T{PLUMAGE}, extracted from the time series data of the 1D SPAN-I proton population fits and $\alpha$-particle moments. The information is organized as follows: a) the maximum growth rate $\gamma^\textrm{max}/\Omega_p$, b) the real frequency solution associated with the maximum growth rate $|\omega^\textrm{max}|/\Omega_p$, c) the magnitude of the wavevector associated with the maximum growth rate $|k|^\textrm{max}| \rho_c$, d) the angle between the wavevector and background magnetic field associated with the maximum growth rate $\theta^\textrm{max}$, e) the magnetic field polarization about $B_0$,
\begin{equation}
    \frac{i \left(\delta B_x \delta B_y^* - \delta B_x^* \delta B_y \right)}{|\delta B_x||\delta B_y|}
\end{equation}
f) the normalized maximum growth rate $\gamma^\textrm{max}/\omega^\textrm{max}$, and g-j) the power absorbed (red) or emitted (blue) by each of the four plasma components, with the subscripts \textit{c,b,$\alpha$,e} corresponding to the proton core, proton beam, $\alpha$, and electron distributions.

Overall, \figref{fig:plume} shows that \T{PLUMAGE} found many unstable modes, generally at the same times of enhanced power at ion scales in the wavelet spectra. From the first row of \figref{fig:plume}(a) and (b), we see there are significant instabilities predicted by linear theory. \figref{fig:plume}(a) reveals growth rates from Event $\#$1, with the proton beam largely (solely, for most of the intervals) responsible for the instabilities. The growth rates from Event $\#$2 are shown in \figref{fig:plume}(b). From the right column, we see that the proton beam is the main source of free energy driving the instabilities, with some contributions from the $\alpha$-particle component (and rare contributions from the proton core). 

In addition, \figref{fig:plume}(b) demonstrates that the plasma is mostly unstable before and after the switchback, corresponding to the time 2019-04-01/10:00:00 in \figref{fig:4_1_allvars}. However, during the switchback, the narrowband wave power signal vanished (\figref{fig:4_1_allvars}(d)). \figref{fig:plume}(b) shows that the plasma still remained unstable (albeit for fewer time intervals) as $\V{B}$ rotated. Furthermore, at the onset of the switchback, the growth rate increased by an order of magnitude. These preliminary results may have implications on the nature of the physical processes that govern switchbacks.

Both \figref{fig:plume}(a) and (b) show that in the reduced parameter space given by (\ref{eq:param}), the unstable modes are mostly oblique in both events. Although this contrasts with the observations from \secref{sec:obs}, that the waves were parallel-propagating, it is possible that if temperature anisotropies were included in the calculation, \T{PLUMAGE} may determine that the parallel-propagating modes were more unstable than the oblique waves.

\section{Discussion and Conclusions} \label{sec:con}

Parker Solar Probe has now completed 4 orbits around the Sun. This paper presented observations from Encounter 2, where we have demonstrated that SPAN-I has the capability to resolve features in the solar wind relevant to PSP's mission goals. More specifically, the proton VDFs measured from SPAN-I is capable of showing ubiquitous non-Maxwellian features with enough free energy capable of driving ion-scale waves.

As discussed in \secref{sec:e1}, Event $\#$1 showed a strong correlation with the proton beam and ion-scale wave storm measured by FIELDS. The magnetic field information depicted in \figref{fig:4_5_allvars} shows that this event satisfies all criteria necessary to classify it as an IC wave storm propagating (anti) parallel to the magnetic field, at a frequency approximate to the ion gyrofrequency, with left-handed polarization. The plasma parameter information shows that the proton beam played a dominant role in the likely energy transfer between the waves and the particles. Notice from the left panel in \figref{fig:4_5_vdfs}(b) that the \Alfven speed, $v_A$, (vertical dashed black line) is located at a region of the VDF where the slope of the beam component is positive. According to plasma theory \citep{Nicholson:1983}, this is the criteria for particles moving slightly slower or faster than $v_A$ to participate in a resonant wave-particle interaction, such as a bump-on-tail instability. Directly after the event, the middle panel of \figref{fig:4_5_vdfs}(f) reveals that the proton core almost traveled completely into SPAN-I's FOV. By inspecting this same time (t6) in \figref{fig:4_5_allvars}(f), we see that at energies above 1000 eV ($y$-axis) the differential energy flux ($z$-axis) disappeared and increased by several orders of magnitude at energies between 300 eV and 800 eV. This is most likely due to a slight change in magnetic field direction that resulted in the core portion of the distribution moving into SPAN-I's FOV, while the beam moved out of the FOV. This illustrates the difficulties in analyzing partial distributions.

The evolution of the beam showed in \figref{fig:4_5_vdfs} revealed unprecedented rates of beam density growth (justified by \figref{fig:fits}) and showcased which regions of velocity-space can be measured with SPAN-I.
In between times t2 and t3, the magnetic field was completely quiet and radial (\figref{fig:4_5_vdfs}(a)) and the beam population indeed grew to comparable height to the core, where it appears as a single distribution in \figref{fig:4_5_vdfs}(c). Consider t3 and t4 from the right panels in \figref{fig:4_5_vdfs}(c) and (d), respectively. During both these times, \figref{fig:4_5_allvars}(g) shows that $\left(n_b/n_c\right)^*$ grew larger than 1, a result that refutes the commonly accepted notions of ``beam" and ``core."
The dark red circle in the contour plots of \figref{fig:4_5_vdfs} represents the peak phase-space density. If we require $\left(n_b/n_c\right)^* < 1$, then the population centered around this peak would be labeled as the ``beam,” contradicting our particle population definitions outlined in \secref{sec:anal}. Note that a peak in phase-space density does not imply a peak in number density. There are indeed more particles concentrated in this circle than any other region in the VDF, but the ``tail" or ``beam" of the VDF could collectively have more particles. If one were to switch the ``beam" and ``core" fit parameters whenever $\left(n_b/n_c\right)^* > 1$, then there would exist unphysical jumps in the temperature and the velocity components. The VDFs at times not specified by \figref{fig:4_5_vdfs}, indicate that the two populations did not physically switch in velocity-space (not shown). Furthermore, the physics does not depend upon the words used to label the different portions of the distribution function.  As an example, the plasma dispersion solver \T{PLUMAGE}, discussed in \secref{sec:plume}, found no changes in the instability analysis when reassigning ``beam" and ``core" fit parameters. If we were to perform 3D fits, it may turn out that there are indeed more particles in the ``core" population than in the ``beam." Nevertheless, the 1D fits yield valuable qualitative information about the evolution of the VDF during simultaneous periods of narrow-band ion-scale wave power. Future work will perform 3D fits to make this analysis more quantitative.

Event $\#$2 discussed in \secref{sec:e2} showed an example of an ion-scale wave storm with both LH and RH polarities. Unlike previous findings \citep{Jian:2009}, the 20 minute duration wave mode with RH spacecraft frame polarity, centered at t8 in \figref{fig:4_1_allvars}, had both higher wave power and frequencies than the LH modes. This revealed itself as a possible FM mode with intrinsic plasma frame right-handedness. A change in all plasma parameters, shown in \figref{fig:4_1_allvars}(f)-(k), were correlated with this interval, suggesting that the plasma was undergoing a wave-particle interaction via an unknown instability. \figref{fig:2019_04_01_vdf}(a)-(c) revealed that the shape of VDFs differed before, during, and after this wave mode centered at t8. We speculate that SPAN-I may have flown through a different kind of plasma than the one that was measured before 2019-04-01/08:04:20.

Unlike Event $\#$1, the magnetic field in Event $\#$2 was not completely quiet and radial, especially for times after t9, shown in \figref{fig:4_1_allvars}(a). Therefore, extra precautions need to be considered in plasma parameter fitting before assigning which population is the beam and the core. Defining the ``core" as the population with the highest phase-space density is consistent in scenarios where the plasma is traversing through a non-radial magnetic field. \figref{fig:2019_04_01_vdf}(d)-(f) displayed the evolution of the VDF through the switchback. Considering that \figref{fig:2019_04_01_vdf}(d) reverted to a similar plasma state after the switchback in \figref{fig:2019_04_01_vdf}, suggested that this specific plasma parcel that SPAN-I observed possessed features that were invariant under magnetic field reversals. Characterizing the properties of the observed plasma during long switchback intervals may be important for understanding the nature of switchbacks themselves, which has been an outstanding question for PSP observations thus far \citep{Kasper:2019,Bale:2019,McManus:2020a,Tenerani:2020,Dudok_de_Wit:2020,Horbury:2020}.

The phenomenon discussed in \secref{sec:static} showed strong evidence of a nonlinear electrostatic structure associated with the 20 minute interval of intense wave power centered at t8 from Event $\#$2. \figref{fig:zoom} exhibits features suggesting strong coupling between the electrostatic structures in \figref{fig:zoom}(b) and the $|\delta B|$ modulation in \figref{fig:zoom}(c). One possibility is that the wavelet of $|\delta B|$ shown in \figref{fig:zoom}(e) could be revealing wave modes at the $\alpha$ or other heavy ion gyrofrequencies. Since the RH wave polarization in this time period (\figref{fig:4_1_allvars}(e)) extended from 0.5 Hz-5 Hz, the magnetic field was in an ordered state for a broad range of ion-scale frequencies. This suggests that a resonant multi-ion species coupling occurred. Further work is necessary to prove this claim, such as testing for coherent three wave interactions that are locked in phase. Perhaps there are undiscovered mechanisms that prove to be more plausible. In future orbits, we will look for similar features in the FIELDS DFB DC differential voltage data that may yield more insight into what kind of phenomenon is being observed and to what extent plasma instabilities play a role in associated solar wind thermalization.    

From the initial results from NHDS, we speculate that the wave storm in Event $\#$1 was generated locally and triggered by an IC instability. The wave storm in Event $\#$2 was possibly driven by a magnetosonic instability and a mix of wave modes may have been present. Complementary results from \T{PLUMAGE} in \figref{fig:plume} show that the initial 1D fits from SPAN-I produced a large number of unstable modes at time intervals correlated with the wave events. The positive growth rates were mostly due to the proton beam, and do not include the effects of temperature anisotropies. If separate proton core, proton beam, and $\alpha$-particle temperature anisotropies were included, then perhaps parallel-propagating and circularly polarized modes would have higher growth rates. These results show that even just the relative drift speeds and density ratios of each particle species produce VDFs with sufficient free energy to drive the plasma unstable.

Further work is needed to both characterize the instabilities for each event and to determine an accurate value of $kd_i$ to properly compute the Doppler-shifted frequencies. In \secref{sec:stab}, we described a procedure for finding a range of possible values. However, our results remain speculative. To definitively determine $\V{k}$, we should also consider solutions for backward-propagating modes. More importantly, extending the plasma parameters fits to 3D will enable fits for separate core and beam temperature anisotropies as input into plasma dispersion solvers, such as NHDS and \T{PLUMAGE}. We recognize that progress has been made by \citet{Huang:2020} for determining temperature anisotropies for the core population, but their work exclusively used data from SPC. Future work will address how to properly combine information from both SPAN-I and SPC for an accurate, complete picture.
 
Although we have shown strong evidence of wave-particle interaction events, we do not have enough information at the present time to conclude definitively that direct energy transfer transpired between the waves and particles. Nevertheless, the 1D fits generated from SPAN-I data have been shown to produce promising results for characterizing instabilities, which will be more relevant closer to the Sun, where we expect the particle VDFs to behave further away from local thermodynamic equilibrium. By virtue of SPAN-I's design, we also expect higher fidelity data products at closest approach to the Sun towards the end of the mission.

This study lays the initial groundwork for understanding what types of wave-particle interaction events PSP can observe in order to implement data analysis from the onboard wave-particle correlator. This future project may be able to definitively characterize the energy transfer mechanisms for each event, or even discover new ones. Since PSP is continuously flying through uncharted territory, as remarked by \citet{Daughton:1998}, ``we cannot guarantee that another, yet undiscovered, branch does not lurk somewhere in an obscure corner of parameter space." 

\section{Acknowledgements}
This work was funded through work on the NASA contract NNN06AA01C. K.G.K was supported by NASA grant 80NSSC19K0912. The authors wish to thank K. Goodrich, T. S. Horbury, and C. C. Chaston for helpful conversations, in addition to the entire SWEAP and FIELDS teams.

\newpage
\bibliography{ic_beams_arx}

\begin{thebibliography}{}
\expandafter\ifx\csname natexlab\endcsname\relax\def\natexlab#1{#1}\fi
\providecommand{\url}[1]{\href{#1}{#1}}

\bibitem[{{Alterman} {et~al.}(2018){Alterman}, {Kasper}, {Stevens}, \&
  {Koval}}]{Alterman:2018}
{Alterman}, B.~L., {Kasper}, J.~C., {Stevens}, M.~L., \& {Koval}, A. 2018,
  \apj, 864, 112

\bibitem[{{Araneda} {et~al.}(2002){Araneda}, {Vi{\~n}As}, \&
  {Astudillo}}]{Araneda:2002}
{Araneda}, J.~A., {Vi{\~n}As}, A.~F., \& {Astudillo}, H.~F. 2002,
  J.~Geophys.~Res., 107, 1453

\bibitem[{{Bale} {et~al.}(2009){Bale}, {Kasper}, {Howes}, {Quataert}, {Salem},
  \& {Sundkvist}}]{Bale:2009}
{Bale}, S.~D., {Kasper}, J.~C., {Howes}, G.~G., {et~al.} 2009,
  Phys.~Rev.~Lett., 103, 211101

\bibitem[{{Bale} {et~al.}(2016){Bale}, {Goetz}, {Harvey}, {Turin}, {Bonnell},
  {Dudok de Wit}, {Ergun}, {MacDowall}, {Pulupa}, {Andre}, {Bolton},
  {Bougeret}, {Bowen}, {Burgess}, {Cattell}, {Chandran}, {Chaston}, {Chen},
  {Choi}, {Connerney}, {Cranmer}, {Diaz-Aguado}, {Donakowski}, {Drake},
  {Farrell}, {Fergeau}, {Fermin}, {Fischer}, {Fox}, {Glaser}, {Goldstein},
  {Gordon}, {Hanson}, {Harris}, {Hayes}, {Hinze}, {Hollweg}, {Horbury},
  {Howard}, {Hoxie}, {Jannet}, {Karlsson}, {Kasper}, {Kellogg}, {Kien},
  {Klimchuk}, {Krasnoselskikh}, {Krucker}, {Lynch}, {Maksimovic}, {Malaspina},
  {Marker}, {Martin}, {Martinez-Oliveros}, {McCauley}, {McComas}, {McDonald},
  {Meyer-Vernet}, {Moncuquet}, {Monson}, {Mozer}, {Murphy}, {Odom},
  {Oliverson}, {Olson}, {Parker}, {Pankow}, {Phan}, {Quataert}, {Quinn},
  {Ruplin}, {Salem}, {Seitz}, {Sheppard}, {Siy}, {Stevens}, {Summers}, {Szabo},
  {Timofeeva}, {Vaivads}, {Velli}, {Yehle}, {Werthimer}, \&
  {Wygant}}]{Bale:2016}
{Bale}, S.~D., {Goetz}, K., {Harvey}, P.~R., {et~al.} 2016, Space Sci.~Rev.,
  204, 49

\bibitem[{Bale {et~al.}(2019)Bale, Badman, Bonnell, Bowen, Burgess, Case,
  Cattell, Chandran, Chaston, Chen, Drake, de~Wit, Eastwood, Ergun, Farrell,
  Fong, Goetz, Goldstein, Goodrich, Harvey, Horbury, Howes, Kasper, Kellogg,
  Klimchuk, Korreck, Krasnoselskikh, Krucker, Laker, Larson, MacDowall,
  Maksimovic, Malaspina, Martinez-Oliveros, McComas, Meyer-Vernet, Moncuquet,
  Mozer, Phan, Pulupa, Raouafi, Salem, Stansby, Stevens, Szabo, Velli, Woolley,
  \& Wygant}]{Bale:2019}
Bale, S.~D., Badman, S.~T., Bonnell, J.~W., {et~al.} 2019, Nature, 576, 237.
\newblock \url{https://doi.org/10.1038/s41586-019-1818-7}

\bibitem[{{Belcher} {et~al.}(1969){Belcher}, {Davis}, \&
  {Smith}}]{Belcher:1969}
{Belcher}, J.~W., {Davis}, Leverett, J., \& {Smith}, E.~J. 1969,
  J.~Geophys.~Res., 74, 2302

\bibitem[{{Belcher} \& {Davis}(1971)}]{Belcher:1971}
{Belcher}, J.~W., \& {Davis}, L. 1971, J.~Geophys.~Res., 76, 3534

\bibitem[{{Bowen} {et~al.}(2020){Bowen}, {Mallet}, {Huang}, {Klein},
  {Malaspina}, {Stevens}, {Bale}, {Bonnell}, {Case}, {Chandran}, {Chaston},
  {Chen}, {Dudok de Wit}, {Goetz}, {Harvey}, {Howes}, {Kasper}, {Korreck},
  {Larson}, {Livi}, {MacDowall}, {McManus}, {Pulupa}, {Verniero}, \&
  {Whittlesey}}]{Bowen:2020inpress}
{Bowen}, T.~A., {Mallet}, A., {Huang}, J., {et~al.} 2020, Astrophys.~J., in
  press.

\bibitem[{{Brain} {et~al.}(2002){Brain}, {Bagenal}, {Acu{\~n}a}, {Connerney},
  {Crider}, {Mazelle}, {Mitchell}, \& {Ness}}]{Brain:2002}
{Brain}, D.~A., {Bagenal}, F., {Acu{\~n}a}, M.~H., {et~al.} 2002, Journal of
  Geophysical Research (Space Physics), 107, 1076

\bibitem[{Case {et~al.}(2019)Case, Kasper, Stevens, Korreck, Paulson, Daigneau,
  Caldwell, Freeman, Henry, Klingensmith, Robinson, Berg, Tiu, Jr., Curtis,
  Ludlam, Larson, Whittlesey, Livi, Klein, \& Martinović}]{Case:2019inpress}
Case, A.~W., Kasper, J.~C., Stevens, M.~L., {et~al.} 2019, The Astrophysical
  Journal

\bibitem[{{Chandran} {et~al.}(2010){Chandran}, {Li}, {Rogers}, {Quataert}, \&
  {Germaschewski}}]{Chandran:2010a}
{Chandran}, B.~D.~G., {Li}, B., {Rogers}, B.~N., {Quataert}, E., \&
  {Germaschewski}, K. 2010, Astrophys.~J., 720, 503

\bibitem[{{Chandran} {et~al.}(2013){Chandran}, {Verscharen}, {Quataert},
  {Kasper}, {Isenberg}, \& {Bourouaine}}]{Chandran:2013}
{Chandran}, B.~D.~G., {Verscharen}, D., {Quataert}, E., {et~al.} 2013, \apj,
  776, 45

\bibitem[{{Daughton} \& {Gary}(1998)}]{Daughton:1998}
{Daughton}, W., \& {Gary}, P.~S. 1998, J.~Geophys.~Res., 103, 20613.
\newblock
  \url{https://agupubs.onlinelibrary.wiley.com/doi/abs/10.1029/98JA01385}

\bibitem[{de~Wit {et~al.}(2020)de~Wit, Krasnoselskikh, Bale, Bonnell, Bowen,
  Chen, Froment, Goetz, Harvey, Jagarlamudi, Larosa, MacDowall, Malaspina,
  Matthaeus, Pulupa, Velli, \& Whittlesey}]{Dudok_de_Wit:2020}
de~Wit, T.~D., Krasnoselskikh, V.~V., Bale, S.~D., {et~al.} 2020, The
  Astrophysical Journal Supplement Series, 246, 39.
\newblock \url{https://doi.org/10.3847%2F1538-4365%2Fab5853}

\bibitem[{{Delva} {et~al.}(2011){Delva}, {Mazelle}, \& {Bertucci}}]{Delva:2011}
{Delva}, M., {Mazelle}, C., \& {Bertucci}, C. 2011, \ssr, 162, 5

\bibitem[{{Dunlop} {et~al.}(1995){Dunlop}, {Woodward}, \&
  {Farrugia}}]{Dunlop:1995}
{Dunlop}, M.~W., {Woodward}, T.~I., \& {Farrugia}, C.~J. 1995, in ESA Special
  Publication, Vol. 371, Proceedings of the Cluster Workshops, Data Analysis
  Tools and Physical Measurements and Mission-Oriented Theory, ed. K.~H.
  {Glassmeier}, U.~{Motschmann}, \& R.~{Schmidt}, 33

\bibitem[{Farge(1992)}]{Farge:1992}
Farge, M. 1992, Annual Review of Fluid Mechanics, 24, 395.
\newblock \url{https://doi.org/10.1146/annurev.fl.24.010192.002143}

\bibitem[{Farge \& Schneider(2015)}]{Farge:2015}
Farge, M., \& Schneider, K. 2015, J.~Plasma Phys., 81, 435810602

\bibitem[{{Feldman} {et~al.}(1973){Feldman}, {Asbridge}, {Bame}, \&
  {Montgomery}}]{Feldman:1973}
{Feldman}, W.~C., {Asbridge}, J.~R., {Bame}, S.~J., \& {Montgomery}, M.~D.
  1973, J.~Geophys.~Res., 78, 2017

\bibitem[{{Feldman} {et~al.}(1974){Feldman}, {Asbridge}, {Bame}, \&
  {Montgomery}}]{Feldman:1974}
---. 1974, Reviews of Geophysics and Space Physics, 12, 715

\bibitem[{{Fox} {et~al.}(2016){Fox}, {Velli}, {Bale}, {Decker}, {Driesman},
  {Howard}, {Kasper}, {Kinnison}, {Kusterer}, {Lario}, {Lockwood}, {McComas},
  {Raouafi}, \& {Szabo}}]{Fox:2016}
{Fox}, N.~J., {Velli}, M.~C., {Bale}, S.~D., {et~al.} 2016, Space Sci.~Rev.,
  204, 7

\bibitem[{{Gary}(1993)}]{Gary:1993}
{Gary}, S.~P. 1993, {Theory of Space Plasma Microinstabilities} (Theory of
  Space Plasma Microinstabilities, by S.~Peter Gary, pp.~193.~ISBN
  0521431670.~Cambridge, UK: Cambridge University Press, September 1993.)

\bibitem[{{Gary} {et~al.}(2016){Gary}, {Jian}, {Broiles}, {Stevens}, {Podesta},
  \& {Kasper}}]{Gary:2016}
{Gary}, S.~P., {Jian}, L.~K., {Broiles}, T.~W., {et~al.} 2016,
  J.~Geophys.~Res., 121, 30

\bibitem[{{Gary} {et~al.}(2003){Gary}, {Yin}, {Winske}, {Ofman}, {Goldstein},
  \& {Neugebauer}}]{Gary:2003}
{Gary}, S.~P., {Yin}, L., {Winske}, D., {et~al.} 2003, J.~Geophys.~Res., 108,
  1068

\bibitem[{{Gary} {et~al.}(2000){Gary}, {Yin}, {Winske}, \&
  {Reisenfeld}}]{Gary:2000}
{Gary}, S.~P., {Yin}, L., {Winske}, D., \& {Reisenfeld}, D.~B. 2000,
  Geophys.~Res.~Lett., 27, 1355

\bibitem[{{Hellinger} {et~al.}(2006){Hellinger}, {Tr{\'a}vn{\'{\i}}{\v c}ek},
  {Kasper}, \& {Lazarus}}]{Hellinger:2006}
{Hellinger}, P., {Tr{\'a}vn{\'{\i}}{\v c}ek}, P., {Kasper}, J.~C., \&
  {Lazarus}, A.~J. 2006, Geophys.~Res.~Lett., 33, 9101

\bibitem[{{Hellinger} {et~al.}(2003){Hellinger}, {Tr{\'a}vn{\'{\i}}{\v c}ek},
  {Mangeney}, \& {Grappin}}]{Hellinger:2003}
{Hellinger}, P., {Tr{\'a}vn{\'{\i}}{\v c}ek}, P., {Mangeney}, A., \& {Grappin},
  R. 2003, Geophys.~Res.~Lett., 30, 1959

\bibitem[{Horbury {et~al.}(2020)Horbury, Woolley, Laker, Matteini, Eastwood,
  Bale, Velli, Chandran, Phan, Raouafi, Goetz, Harvey, Pulupa, Klein, Wit,
  Kasper, Korreck, Case, Stevens, \& Livi}]{Horbury:2020}
Horbury, T., Woolley, T., Laker, R., {et~al.} 2020, The Astrophysical Journal
  Supplement Series, 246, 45

\bibitem[{{Hu} \& {Habbal}(1999)}]{Hu:1999}
{Hu}, Y.~Q., \& {Habbal}, S.~R. 1999, J.~Geophys.~Res., 104, 17045

\bibitem[{Huang {et~al.}(2020)Huang, Kasper, Vech, Klein, Stevens,
  Martinovi{\'{c}}, Alterman, {\v{D}}urovcov{\'{a}}, Paulson, Maruca, Qudsi,
  Case, Korreck, Jian, Velli, Lavraud, Hegedus, Bert, Holmes, Bale, Larson,
  Livi, Whittlesey, Pulupa, MacDowall, Malaspina, Bonnell, Harvey, Goetz, \&
  de~Wit}]{Huang:2020}
Huang, J., Kasper, J.~C., Vech, D., {et~al.} 2020, The Astrophysical Journal
  Supplement Series, 246, 70.
\newblock \url{https://doi.org/10.3847%2F1538-4365%2Fab74e0}

\bibitem[{Isenberg \& Vasquez(2007)}]{Isenberg:2007}
Isenberg, P.~A., \& Vasquez, B.~J. 2007, The Astrophysical Journal, 668, 546.
\newblock \url{https://doi.org/10.1086%2F521220}

\bibitem[{Isenberg \& Vasquez(2009)}]{Isenberg:2009}
---. 2009, The Astrophysical Journal, 696, 591.
\newblock \url{https://doi.org/10.1088%2F0004-637x%2F696%2F1%2F591}

\bibitem[{Isenberg \& Vasquez(2011)}]{Isenberg:2011}
---. 2011, The Astrophysical Journal, 731, 88.
\newblock \url{https://doi.org/10.1088%2F0004-637x%2F731%2F2%2F88}

\bibitem[{{Jian} {et~al.}(2010){Jian}, {Russell}, {Luhmann}, {Anderson},
  {Boardsen}, {Strangeway}, {Cowee}, \& {Wennmacher}}]{Jian:2010}
{Jian}, L.~K., {Russell}, C.~T., {Luhmann}, J.~G., {et~al.} 2010,
  J.~Geophys.~Res., 115, A12115

\bibitem[{{Jian} {et~al.}(2009){Jian}, {Russell}, {Luhmann}, {Strangeway},
  {Leisner}, \& {Galvin}}]{Jian:2009}
---. 2009, apjl, 701, L105

\bibitem[{{Jian} {et~al.}(2014){Jian}, {Wei}, {Russell}, {Luhmann}, {Klecker},
  {Omidi}, {Isenberg}, {Goldstein}, {Figueroa-Vi{\~{n}}as}, \&
  {Blanco-Cano}}]{Jian:2014}
{Jian}, L.~K., {Wei}, H.~K., {Russell}, C.~T., {et~al.} 2014, Astrophys.~J.,
  786, 123.
\newblock \url{https://doi.org/10.1088%2F0004-637x%2F786%2F2%2F123}

\bibitem[{{Kasper} {et~al.}(2002){Kasper}, {Lazarus}, \& {Gary}}]{Kasper:2002}
{Kasper}, J.~C., {Lazarus}, A.~J., \& {Gary}, S.~P. 2002, Geophys.~Res.~Lett.,
  29, 20

\bibitem[{{Kasper} {et~al.}(2008){Kasper}, {Lazarus}, \& {Gary}}]{Kasper:2008}
---. 2008, Phys.~Rev.~Lett., 101, 261103

\bibitem[{{Kasper} {et~al.}(2006){Kasper}, {Lazarus}, {Steinberg}, {Ogilvie},
  \& {Szabo}}]{Kasper:2006}
{Kasper}, J.~C., {Lazarus}, A.~J., {Steinberg}, J.~T., {Ogilvie}, K.~W., \&
  {Szabo}, A. 2006, Journal of Geophysical Research (Space Physics), 111,
  A03105

\bibitem[{{Kasper} {et~al.}(2013){Kasper}, {Maruca}, {Stevens}, \&
  {Zaslavsky}}]{Kasper:2013}
{Kasper}, J.~C., {Maruca}, B.~A., {Stevens}, M.~L., \& {Zaslavsky}, A. 2013,
  Phys.~Rev.~Lett., 110, 091102

\bibitem[{{Kasper} {et~al.}(2015){Kasper}, {Abiad}, {Austin}, {Balat-Pichelin},
  {Bale}, {Belcher}, {Berg}, {Bergner}, {Berthomier}, {Bookbinder}, {Brodu},
  {Caldwell}, {Case}, {Chandran}, {Cheimets}, {Cirtain}, {Cranmer}, {Curtis},
  {Daigneau}, {Dalton}, {Dasgupta}, {DeTomaso}, {Diaz-Aguado}, {Djordjevic},
  {Donaskowski}, {Effinger}, {Florinski}, {Fox}, {Freeman}, {Gallagher},
  {Gary}, {Gauron}, {Gates}, {Goldstein}, {Golub}, {Gordon}, {Gurnee}, {Guth},
  {Halekas}, {Hatch}, {Heerikuisen}, {Ho}, {Hu}, {Johnson}, {Jordan},
  {Korreck}, {Larson}, {Lazarus}, {Li}, {Livi}, {Ludlam}, {Maksimovic},
  {McFadden}, {Marchant}, {Maruca}, {McComas}, {Messina}, {Mercer}, {Park},
  {Peddie}, {Pogorelov}, {Reinhart}, {Richardson}, {Robinson}, {Rosen},
  {Skoug}, {Slagle}, {Steinberg}, {Stevens}, {Szabo}, {Taylor}, {Tiu}, {Turin},
  {Velli}, {Webb}, {Whittlesey}, {Wright}, {Wu}, \& {Zank}}]{Kasper:2015}
{Kasper}, J.~C., {Abiad}, R., {Austin}, G., {et~al.} 2015, Space Sci.~Rev.,
  204, 131

\bibitem[{{Kasper} {et~al.}(2016){Kasper}, {Abiad}, {Austin}, {Balat-Pichelin},
  {Bale}, {Belcher}, {Berg}, {Bergner}, {Berthomier}, {Bookbinder}, {Brodu},
  {Caldwell}, {Case}, {Chandran}, {Cheimets}, {Cirtain}, {Cranmer}, {Curtis},
  {Daigneau}, {Dalton}, {Dasgupta}, {DeTomaso}, {Diaz-Aguado}, {Djordjevic},
  {Donaskowski}, {Effinger}, {Florinski}, {Fox}, {Freeman}, {Gallagher},
  {Gary}, {Gauron}, {Gates}, {Goldstein}, {Golub}, {Gordon}, {Gurnee}, {Guth},
  {Halekas}, {Hatch}, {Heerikuisen}, {Ho}, {Hu}, {Johnson}, {Jordan},
  {Korreck}, {Larson}, {Lazarus}, {Li}, {Livi}, {Ludlam}, {Maksimovic},
  {McFadden}, {Marchant}, {Maruca}, {McComas}, {Messina}, {Mercer}, {Park},
  {Peddie}, {Pogorelov}, {Reinhart}, {Richardson}, {Robinson}, {Rosen},
  {Skoug}, {Slagle}, {Steinberg}, {Stevens}, {Szabo}, {Taylor}, {Tiu}, {Turin},
  {Velli}, {Webb}, {Whittlesey}, {Wright}, {Wu}, \& {Zank}}]{Kasper:2016}
---. 2016, Space Sci.~Rev., 204, 131

\bibitem[{Kasper {et~al.}(2019)Kasper, Bale, Belcher, Berthomier, Case,
  Chandran, Curtis, Gallagher, Gary, Golub, Halekas, Ho, Horbury, Hu, Huang,
  Klein, Korreck, Larson, Livi, Maruca, Lavraud, Louarn, Maksimovic,
  Martinovic, McGinnis, Pogorelov, Richardson, Skoug, Steinberg, Stevens,
  Szabo, Velli, Whittlesey, Wright, Zank, MacDowall, McComas, McNutt, Pulupa,
  Raouafi, \& Schwadron}]{Kasper:2019}
Kasper, J.~C., Bale, S.~D., Belcher, J.~W., {et~al.} 2019, Nature, 576, 228.
\newblock \url{https://doi.org/10.1038/s41586-019-1813-z}

\bibitem[{{Kennel}(1966)}]{Kennel:1966}
{Kennel}, C. 1966, Phys.~Fluids, 9, 2190

\bibitem[{{Kivelson} {et~al.}(1996){Kivelson}, {Khurana}, {Walker}, {Russell},
  {Linker}, {Southwood}, \& {Polanskey}}]{Kivelson:1996}
{Kivelson}, M.~G., {Khurana}, K.~K., {Walker}, R.~J., {et~al.} 1996, Science,
  273, 337

\bibitem[{{Klein} {et~al.}(2018){Klein}, {Alterman}, {Stevens}, {Vech}, \&
  {Kasper}}]{Klein:2018}
{Klein}, K.~G., {Alterman}, B.~L., {Stevens}, M.~L., {Vech}, D., \& {Kasper},
  J.~C. 2018, \prl, 120, 205102

\bibitem[{{Klein} \& {Howes}(2015)}]{Klein:2015a}
{Klein}, K.~G., \& {Howes}, G.~G. 2015, Phys.~Plasmas, 22, 032903

\bibitem[{{Klein} {et~al.}(2017){Klein}, {Kasper}, {Korreck}, \&
  {Stevens}}]{Klein:2017c}
{Klein}, K.~G., {Kasper}, J.~C., {Korreck}, K.~E., \& {Stevens}, M.~L. 2017,
  J.~Geophys.~Res., doi:10.1002/2017JA024486.
\newblock \url{http://dx.doi.org/10.1002/2017JA024486}

\bibitem[{Klein {et~al.}(2019)Klein, Martinovi{\'{c}}, Stansby, \&
  Horbury}]{Klein:2019}
Klein, K.~G., Martinovi{\'{c}}, M., Stansby, D., \& Horbury, T.~S. 2019, The
  Astrophysical Journal, 887, 234.
\newblock \url{https://doi.org/10.3847%2F1538-4357%2Fab5802}

\bibitem[{{Leubner} \& {Vinas}(1986)}]{Leubner:1986}
{Leubner}, M.~P., \& {Vinas}, A.~F. 1986, J.~Geophys.~Res., 91, 13366

\bibitem[{Levenberg(1944)}]{Levenberg:1944}
Levenberg, K. 1944, Quarterly of Applied Mathematics, 2, 164.
\newblock \url{http://www.jstor.org/stable/43633451}

\bibitem[{{Li} \& {Habbal}(2000)}]{Li:2000}
{Li}, X., \& {Habbal}, S.~R. 2000, J.~Geophys.~Res., 105, 7483

\bibitem[{Livi {et~al.}(2014)Livi, Goldstein, Burch, Crary, Rymer, Mitchell, \&
  Persoon}]{Livi:2014}
Livi, R., Goldstein, J., Burch, J.~L., {et~al.} 2014, Journal of Geophysical
  Research: Space Physics, 119, 3683

\bibitem[{{Livi} {et~al.}(2020){Livi}, {Larson}, {Kasper}, {Abiad}, {Case},
  {Klein}, {Curtis}, {Dalton}, {Stevens}, {Korreck}, {Robinson}, {Tiu},
  {Whittlesey}, {Verniero}, {Halekas}, {McFadden}, {Marckwordt}, {Slagle},
  {Abatcha}, \& {Rahmati}}]{Livi:2020inpress}
{Livi}, R., {Larson}, D.~E., {Kasper}, J.~C., {et~al.} 2020, Astrophys.~J., {In
  preparation.}

\bibitem[{{Maneva} {et~al.}(2013){Maneva}, {Vi{\~n}As}, \&
  {Ofman}}]{Maneva:2013}
{Maneva}, Y.~G., {Vi{\~n}As}, A.~F., \& {Ofman}, L. 2013, J.~Geophys.~Res.,
  118, 2842

\bibitem[{Marquardt(1963)}]{Marquardt:1963}
Marquardt, D.~W. 1963, Journal of the Society for Industrial and Applied
  Mathematics, 11, 431.
\newblock \url{http://www.jstor.org/stable/2098941}

\bibitem[{Marsch(2006)}]{Marsch:2006}
Marsch, E. 2006, Living Rev.~Solar Phys., 3, 1.
\newblock \url{http://www.livingreviews.org/lrsp-2006-1}

\bibitem[{{Marsch} \& {Livi}(1987)}]{Marsch:1987}
{Marsch}, E., \& {Livi}, S. 1987, J.~Geophys.~Res., 92, 7263.
\newblock
  \url{https://agupubs.onlinelibrary.wiley.com/doi/abs/10.1029/JA092iA07p07263}

\bibitem[{{Marsch} {et~al.}(1982){Marsch}, {Schwenn}, {Rosenbauer},
  {Muehlhaeuser}, {Pilipp}, \& {Neubauer}}]{Marsch:1982}
{Marsch}, E., {Schwenn}, R., {Rosenbauer}, H., {et~al.} 1982, J.~Geophys.~Res.,
  87, 52

\bibitem[{{Maruca} {et~al.}(2011){Maruca}, {Kasper}, \& {Bale}}]{Maruca:2011}
{Maruca}, B.~A., {Kasper}, J.~C., \& {Bale}, S.~D. 2011, Phys.~Rev.~Lett., 107,
  201101

\bibitem[{{McManus} {et~al.}(2020){McManus}, {Larson}, {Livi}, {Rahmati},
  {Whittlesey}, {Verniero}, {Bowen}, {Klein}, {Chandran}, {Kasper}, {Case},
  {Korreck}, {Stevens}, {Bale}, {Pulupa}, {Malaspina}, {Bonnell}, {Harvey},
  {Goetz}, \& {MacDowall}}]{McManus:2020inprep}
{McManus}, M.~D., {Larson}, D.~E., {Livi}, R., {et~al.} 2020, Astrophys.~J.,
  {In preparation.}

\bibitem[{McManus {et~al.}(2020)McManus, Bowen, Mallet, Chen, Chandran, Bale,
  Larson, de~Wit, Kasper, Stevens, Whittlesey, Livi, Korreck, Goetz, Harvey,
  Pulupa, MacDowall, Malaspina, Case, \& Bonnell}]{McManus:2020a}
McManus, M.~D., Bowen, T.~A., Mallet, A., {et~al.} 2020, The Astrophysical
  Journal Supplement Series, 246, 67.
\newblock \url{https://doi.org/10.3847%2F1538-4365%2Fab6dce}

\bibitem[{Means(1972)}]{Means:1972}
Means, J.~D. 1972, J.~Geophys.~Res., 77, 5551.
\newblock
  \url{https://agupubs.onlinelibrary.wiley.com/doi/abs/10.1029/JA077i028p05551}

\bibitem[{{Montgomery} {et~al.}(1976){Montgomery}, {Gary}, {Feldman}, \&
  {Forslund}}]{Montgomery:1976}
{Montgomery}, M.~D., {Gary}, S.~P., {Feldman}, W.~C., \& {Forslund}, D.~W.
  1976, J.~Geophys.~Res., 81, 2743

\bibitem[{{Montgomery} {et~al.}(1975){Montgomery}, {Gary}, {Forslund}, \&
  {Feldman}}]{Montgomery:1975}
{Montgomery}, M.~D., {Gary}, S.~P., {Forslund}, D.~W., \& {Feldman}, W.~C.
  1975, Phys.~Rev.~Lett., 35, 667

\bibitem[{Murphy {et~al.}(1995)Murphy, Smith, Tsurutani, Balogh, \&
  Southwood}]{Murphy:1995}
Murphy, N., Smith, E.~J., Tsurutani, B.~T., Balogh, A., \& Southwood, D.~J.
  1995, Space Science Reviews, 72, 447.
\newblock \url{https://doi.org/10.1007/BF00768819}

\bibitem[{{Neugebauer} {et~al.}(1996){Neugebauer}, {Goldstein}, {Smith}, \&
  {Feldman}}]{Neugebauer:1996}
{Neugebauer}, M., {Goldstein}, B.~E., {Smith}, E.~J., \& {Feldman}, W.~C. 1996,
  J.~Geophys.~Res., 101, 17047

\bibitem[{Nicholson(1983)}]{Nicholson:1983}
Nicholson, D. 1983, Introduction to plasma theory, Wiley series in plasma
  physics (Wiley).
\newblock \url{https://books.google.com/books?id=fyRRAAAAMAAJ}

\bibitem[{Nyquist(1932)}]{Nyquist:1932}
Nyquist, H. 1932, Bell System Technical Journal, 11, 126.
\newblock
  \url{https://onlinelibrary.wiley.com/doi/abs/10.1002/j.1538-7305.1932.tb02344.x}

\bibitem[{{Podesta} \& {Gary}(2011{\natexlab{a}})}]{Podesta:2011a}
{Podesta}, J.~J., \& {Gary}, S.~P. 2011{\natexlab{a}}, Astrophys.~J., 734, 15

\bibitem[{{Podesta} \& {Gary}(2011{\natexlab{b}})}]{Podesta:2011b}
---. 2011{\natexlab{b}}, Astrophys.~J., 742, 41

\bibitem[{{Quest} \& {Shapiro}(1996)}]{Quest:1996}
{Quest}, K.~B., \& {Shapiro}, V.~D. 1996, J.~Geophys.~Res., 101, 24457

\bibitem[{{Rosin} {et~al.}(2011){Rosin}, {Schekochihin}, {Rincon}, \&
  {Cowley}}]{Rosin:2011}
{Rosin}, M.~S., {Schekochihin}, A.~A., {Rincon}, F., \& {Cowley}, S.~C. 2011,
  Mon.~Not.~Roy.~Astron.~Soc., 413, 7

\bibitem[{Russell {et~al.}(1990)Russell, Luhmann, Schwingenschuh, Riedler, \&
  Yeroshenko}]{Russell:1990}
Russell, C.~T., Luhmann, J.~G., Schwingenschuh, K., Riedler, W., \& Yeroshenko,
  Y. 1990, Geophysical Research Letters, 17, 897.
\newblock
  \url{https://agupubs.onlinelibrary.wiley.com/doi/abs/10.1029/GL017i006p00897}

\bibitem[{{Sonnerup} \& {Cahill}(1967)}]{Sonnerup:1967}
{Sonnerup}, B.~U.~O., \& {Cahill}, Jr., L.~J. 1967, J.~Geophys.~Res., 72, 171

\bibitem[{{Steinberg} {et~al.}(1996){Steinberg}, {Lazarus}, {Ogilvie},
  {Lepping}, \& {Byrnes}}]{Steinberg:1996}
{Steinberg}, J.~T., {Lazarus}, A.~J., {Ogilvie}, K.~W., {Lepping}, R., \&
  {Byrnes}, J. 1996, Geophys.~Res.~Lett., 23, 1183

\bibitem[{{Stix}(1962)}]{Stix:1962}
{Stix}, T.~H. 1962, {The Theory of Plasma Waves} (McGraw-Hill)

\bibitem[{{Stix}(1992)}]{Stix:1992}
---. 1992, {Waves in Plasmas} (New York: American Institute of Physics)

\bibitem[{Tenerani {et~al.}(2020)Tenerani, Velli, Matteini, R{\'{e}}ville, Shi,
  Bale, Kasper, Bonnell, Case, de~Wit, Goetz, Harvey, Klein, Korreck, Larson,
  Livi, MacDowall, Malaspina, Pulupa, Stevens, \& Whittlesey}]{Tenerani:2020}
Tenerani, A., Velli, M., Matteini, L., {et~al.} 2020, The Astrophysical Journal
  Supplement Series, 246, 32.
\newblock \url{https://doi.org/10.3847%2F1538-4365%2Fab53e1}

\bibitem[{Torrence \& Compo(1998)}]{Torrence:1998}
Torrence, C., \& Compo, G.~P. 1998, Bulletin of the American Meteorological
  Society, 79, 61.
\newblock \url{https://doi.org/10.1175/1520-0477(1998)079<0061:APGTWA>2.0.CO;2}

\bibitem[{{Tsurutani} {et~al.}(1994){Tsurutani}, {Arballo}, {Mok}, {Smith},
  {Mason}, \& {Tan}}]{Tsurutani:1994}
{Tsurutani}, B.~T., {Arballo}, J.~K., {Mok}, J., {et~al.} 1994, \grl, 21, 633

\bibitem[{{Tu} \& {Marsch}(2002)}]{Tu:2002}
{Tu}, C.-Y., \& {Marsch}, E. 2002, J.~Geophys.~Res., 107, 1249

\bibitem[{{Verscharen} {et~al.}(2013{\natexlab{a}}){Verscharen}, {Bourouaine},
  \& {Chand ran}}]{Verscharen:2013c}
{Verscharen}, D., {Bourouaine}, S., \& {Chand ran}, B. D.~G.
  2013{\natexlab{a}}, Astrophys.~J., 773, 163

\bibitem[{{Verscharen} {et~al.}(2013{\natexlab{b}}){Verscharen}, {Bourouaine},
  {Chandran}, \& {Maruca}}]{Verscharen:2013b}
{Verscharen}, D., {Bourouaine}, S., {Chandran}, B.~D.~G., \& {Maruca}, B.~A.
  2013{\natexlab{b}}, Astrophys.~J., 773, 8

\bibitem[{{Verscharen} \& {Chandran}(2013)}]{Verscharen:2013a}
{Verscharen}, D., \& {Chandran}, B.~D.~G. 2013, Astrophys.~J., 764, 88

\bibitem[{Verscharen \& Chandran(2018)}]{Verscharen:2018}
Verscharen, D., \& Chandran, B. D.~G. 2018, Research Notes of the {AAS}, 2, 13.
\newblock \url{https://doi.org/10.3847%2F2515-5172%2Faabfe3}

\bibitem[{{Verscharen} {et~al.}(2019){Verscharen}, {Klein}, \&
  {Maruca}}]{Verscharen:2019}
{Verscharen}, D., {Klein}, K.~G., \& {Maruca}, B.~A. 2019, Living Reviews in
  Solar Physics, 16, 5

\bibitem[{Whittlesey {et~al.}(2020)Whittlesey, Larson, Kasper, Halekas,
  Abatcha, Abiad, Berthomier, Case, Chen, Curtis, Dalton, Klein, Korreck, Livi,
  Ludlam, Marckwordt, Rahmati, Robinson, Slagle, Stevens, Tiu, \&
  Verniero}]{Whittlesey:2020}
Whittlesey, P.~L., Larson, D.~E., Kasper, J.~C., {et~al.} 2020, The
  Astrophysical Journal Supplement Series, 246, 74.
\newblock \url{https://doi.org/10.3847%2F1538-4365%2Fab7370}

\bibitem[{{Wicks} {et~al.}(2016){Wicks}, {Alexander}, {Stevens}, {Wilson}~III,
  {Moya}, {Vi{\~{n}}as}, {Jian}, {Roberts}, {O'Modhrain}, {Gilbert}, \&
  {Zurbuchen}}]{Wicks:2016}
{Wicks}, R.~T., {Alexander}, R.~L., {Stevens}, M., {et~al.} 2016,
  Astrophys.~J., 819, 6.
\newblock \url{https://doi.org/10.3847%2F0004-637x%2F819%2F1%2F6}

\bibitem[{Woodham {et~al.}(2019)Woodham, Wicks, Verscharen, Owen, Maruca, \&
  Alterman}]{Woodham:2019}
Woodham, L.~D., Wicks, R.~T., Verscharen, D., {et~al.} 2019, The Astrophysical
  Journal, 884, L53.
\newblock \url{https://doi.org/10.3847%2F2041-8213%2Fab4adc}

\bibitem[{Zhao {et~al.}(2018)Zhao, Feng, Wu, Liu, Zhao, Zhao, \&
  Huang}]{Zhao:2018}
Zhao, G.~Q., Feng, H.~Q., Wu, D.~J., {et~al.} 2018, Journal of Geophysical
  Research: Space Physics, 123, 1715.
\newblock
  \url{https://agupubs.onlinelibrary.wiley.com/doi/abs/10.1002/2017JA024979}

\bibitem[{Zhao {et~al.}(2019)Zhao, Feng, Wu, Pi, \& Huang}]{Zhao:2019}
Zhao, G.~Q., Feng, H.~Q., Wu, D.~J., Pi, G., \& Huang, J. 2019, The
  Astrophysical Journal, 871, 175.
\newblock \url{https://doi.org/10.3847%2F1538-4357%2Faaf8b8}

\end{thebibliography}

\end{document}